\documentclass[%
 reprint,
 superscriptaddress,
 amsmath,amssymb,
 aps,
 prx
]{revtex4-2}
\usepackage[T1]{fontenc}
\usepackage{graphicx}
\usepackage{dcolumn}
\usepackage{bm}
\usepackage[colorlinks=true, allcolors=blue]{hyperref}
\usepackage{braket}
\usepackage[table]{xcolor}
\usepackage{xfrac}
\usepackage{comment}
\usepackage{ragged2e}
\usepackage[normalem]{ulem}
\usepackage{mathtools}
\usepackage{gensymb}
\usepackage{layouts}
\usepackage{subcaption}
\usepackage{overpic}
\usepackage{quantikz}
\usepackage{tabularx}
\usepackage{multirow}
\usepackage[section]{placeins}
\usepackage{esint}
\usepackage{enumitem}
\usepackage{booktabs}
\usepackage{amsthm}
\usepackage[table]{xcolor}
\usepackage{array}

\usepackage[capitalise]{cleveref}
\usepackage{tikz}
\usetikzlibrary{shapes.geometric, arrows.meta, positioning,shadings, fadings}

\usepackage{adjustbox} 

\tikzstyle{process} = [
  rectangle,
  rounded corners,
  text width=2cm,
  minimum height=1.4cm,
  align=center,
  draw=black,
  font=\footnotesize,
  text=black,
  top color=blue!20,
  bottom color=purple!20
]

\tikzstyle{arrow} = [thick,->,>=Stealth]

\begin{document}
\title{Color it, Code it, Cancel it: $k$-local dynamical decoupling from classical additive codes}
\author{Minh T. P. Nguyen}
\email{m.t.phamnguyen@tudelft.nl}
\affiliation{QuTech and Kavli Institute of Nanoscience, Delft University of Technology, Lorentzweg 1, 2628 CJ Delft, The Netherlands}
 \author{Maximilian Rimbach-Russ}
 \affiliation{QuTech and Kavli Institute of Nanoscience, Delft University of Technology, Lorentzweg 1, 2628 CJ Delft, The Netherlands}
\author{Stefano Bosco}
\email{s.bosco@tudelft.nl}
\affiliation{QuTech and Kavli Institute of Nanoscience, Delft University of Technology, Lorentzweg 1, 2628 CJ Delft, The Netherlands}

\newtheorem{theorem}{Theorem}[section]
\theoremstyle{definition}
\newtheorem{assumption}{Assumption}
\theoremstyle{remark}
\newtheorem*{remark}{Remark}
\theoremstyle{definition}
\newtheorem{definition}[theorem]{Definition}
\newtheorem{corollary}[theorem]{Corollary}
\newtheorem{lemma}[theorem]{Lemma}
\newtheorem{proposition}[theorem]{Proposition}
\newtheorem{criterion}{Criterion}
\newtheorem{appendixcriterion}{Criterion}[section]
\Crefname{criterion}{Criterion}{Criteria}
\crefname{criterion}{Crit.}{Crit.}

\begin{abstract}
Dynamical decoupling is a central technique in quantum computing for actively suppressing decoherence and systematic imperfections through sequences of single-qubit operations. Conventional sequences  typically aim to completely freeze system dynamics, often resulting in long protocols whose length scales exponentially with system size. In this work, we introduce a general framework for constructing time-optimal, selectively-tailored sequences that remove only specific local interactions.
By combining techniques from graph coloring and classical coding theory, our approach enables compact and hardware-tailored sequences across diverse qubit platforms, efficiently canceling undesired Hamiltonian terms while preserving target interactions. This opens up broad applications in quantum computing and simulation. 
At the core of our method is a mapping between dynamical decoupling sequence design and error-detecting codes, which allows us to leverage powerful coding-theoretic tools to construct customized sequences. To overcome exponential overheads, we exploit symmetries in colored interaction hypergraphs, extending graph-coloring strategies to arbitrary many-body Hamiltonians.
We demonstrate the effectiveness of our framework through concrete examples, including compact sequences that suppress residual ZZ and ZZZ interactions in superconducting qubits and Heisenberg exchange coupling in spin qubits. We also show how it enables Hamiltonian engineering by simulating the anisotropic Kitaev honeycomb model using only isotropic Heisenberg interactions.
\end{abstract}
\maketitle

\section{Introduction}

Dynamical decoupling (DD), originally inspired by nuclear magnetic resonance~\cite{Hahn1950,Meiboom1958,maudsley1986modified,brinkmann2016introduction,BRINKMANN2025100191}, has become a powerful technique in quantum information processing. Traditionally, DD has been employed to mitigate imperfections in near-term quantum devices by using sequences of single-qubit operations to isolate quantum systems from environmental noise~\cite{Vitali1999,viola1998bang,viola1999dynamical,duan1999suppressing}. Sequences that suppress on-site decoherence to high order in time include for example universal single-qubit protocols~\cite{viola1999dynamical,viola1998bang}, XY family~\cite{GULLION1990479,Wang2012,ZhiHui2012}, concatenated sequences~\cite{West2010,pasini2010optimized}, and the Uhrig family~\cite{Uhrig2007,wang2011protection,yang2008universality}. DD has also found applications in quantum sensing, such as frequency-selective spectroscopy~\cite{szankowski2017environmental,Alvarez2011}, and in quantum error correction~\cite{paz2013optimally,vezvaee2025demonstrationhighfidelityentangledlogical}.

Beyond decoherence, DD protocols have been effective in suppressing systematic errors originating from high-order qubit interactions, such as residual crosstalk \cite{Tripathi2022,qiu2021suppressing,evert2024syncopated} from two-body couplings, leading to advanced sequences such as WAHUHA~\cite{WAHUHA1968} and fractal decoupling~\cite{agarwal2020dynamical}. However, these methods were designed primarily for small systems with limited connectivity. Recent advances aim to develop time-efficient DD sequences suppressing 2-local interactions for large devices, with sequence length depending only on the topology of the connectivity graphs of the device~\cite{brown2024efficient,Coote2025,evert2024syncopated,hickman2025crosstalkrobustdynamicaldecouplingbipartitetopology}. Moreover, while conventional DD sequences typically freeze the system dynamics entirely, making them difficult to adapt for designing specific quantum evolutions, recent developments have proposed DD protocols for quantum simulation, including the engineering of target Hamiltonians using average Hamiltonian theory~\cite{Choi2020}.

\begin{figure}
    \centering
    \includegraphics[width=0.95\linewidth]{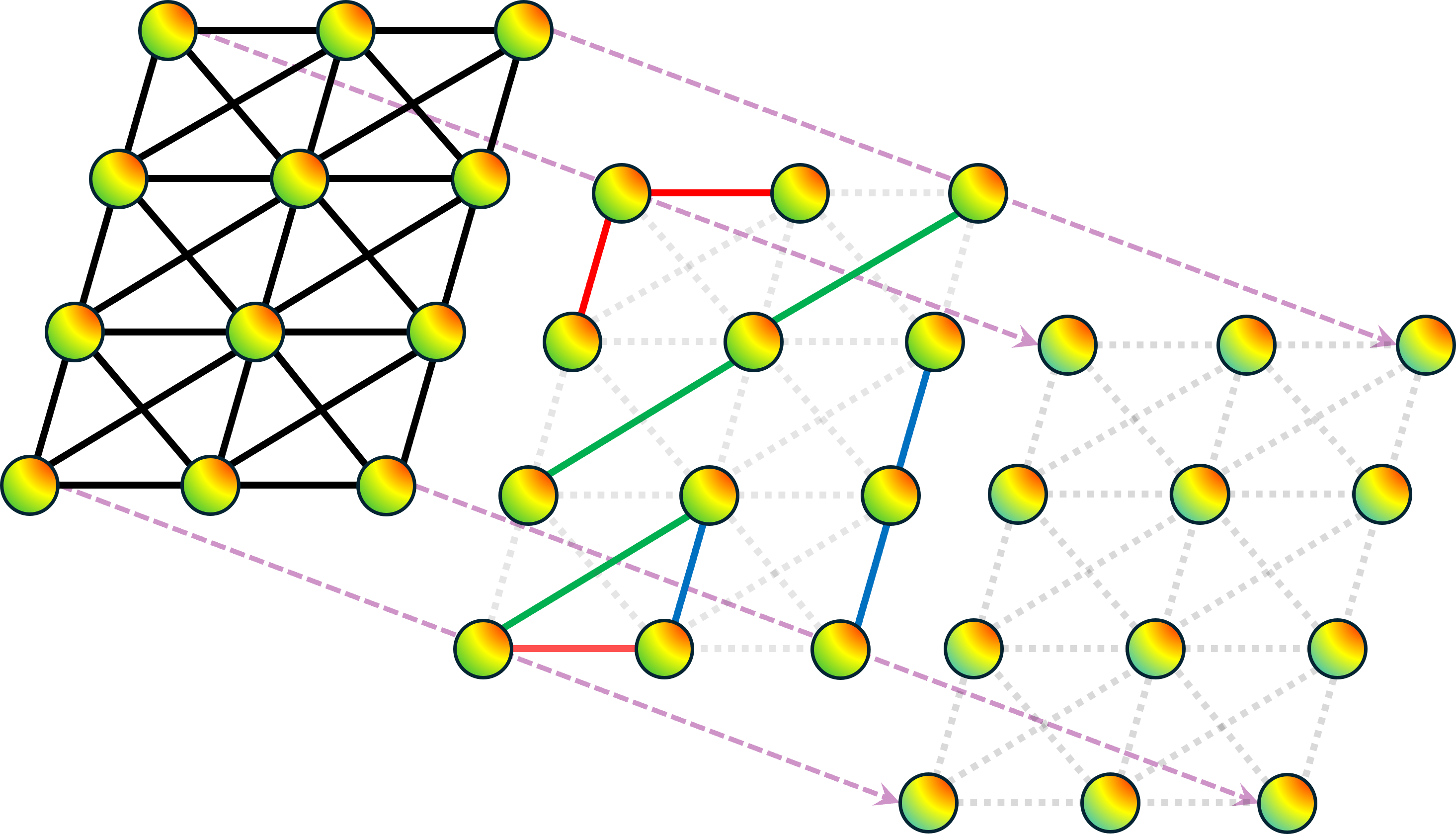}
    \caption{\justifying \textbf{Canceling local interactions by dynamical decoupling.} Our framework permits an efficient and selective suppression of local interactions. For example, by applying tailored DD sequences to a device with dense connectivity (top), we can separate the system into non-interacting clusters, preserving only certain interactions within each cluster (middle), or completely freezing the system's dynamics (bottom). Our framework also applies to general interactions graphs going beyond nearest neighbor and is not restricted to instantaneous control pulses.  }
    \label{fig: 3-layer summary}
\end{figure}

In this work, we present a unified framework that enables the systematic construction of time-efficient and robust DD sequences capable of selectively suppressing any chosen $k$-local interactions, as sketched in Fig.~\ref{fig: 3-layer summary}. Our framework is built on three key ideas. First, we show that any $k$-local Hamiltonian induces a hypergraph symmetry that partitions physical qubits into equivalence color classes. This symmetry enables the design of DD sequences that suppress $k$-local interactions with sequence lengths determined only by the device topology, rather than by system size. This generalizes prior work \cite{brown2024efficient,Coote2025,evert2024syncopated,hickman2025crosstalkrobustdynamicaldecouplingbipartitetopology}, which was limited to 2-local interactions. Second, we exploit the established connection between DD and classical additive codes \cite{Rotteler_2006,Bookatz_2016}. This correspondence allows us to identify high-performance DD sequences directly from code parameters and allow us to design hardware-tailored sequences for both superconducting and spin qubit platforms. Furthermore, it opens paths for using DD as a tool in Hamiltonian engineering \cite{Choi2020,votto2024universal,rajabi2019dynamical}. Finally, we show how to construct DD sequences based on the underlying group structure that are intrinsically robust to pulse imperfections. Our framework naturally supports finite-width pulses with bounded-strength control \cite{Viola2003}, eliminating the need for idealized instantaneous (bang–bang) operations \cite{viola1998bang}. Moreover, the mapping between DD sequences and classical detection codes can potentially enables the use of machine-learning approaches \cite{tong2024empirical,rahman2024learning,zhang2025learningsteerquantummanybody} by dramatically reducing the search space of candidate sequences and providing an efficient route to generating large, task-specific training datasets.

The flowchart of the framework proposed in this work is summarized in Fig.~\ref{fig: flowchart-summary}. The manuscript is organized as follows. In Sec.~\ref{section: DD_background}, we briefly review bang-bang dynamical decoupling, emphasizing the connection between DD sequences and group theory. In Sec.~\ref{section: symmetry and quotient graph}, we reinterpret the results of Refs.~\cite{brown2024efficient, evert2024syncopated} by uncovering a symmetry structure in colored interaction graphs, and extend this perspective to general $k$-local Hamiltonians. We show that by embedding these symmetries into the group structure of DD sequences, one can construct sequences with lengths depending only on the device topology. In Sec.~\ref{section: decoupling group from additive code}, we introduce the correspondence between DD sequences and classical additive codes, which allows us to characterize the properties of DD sequences directly from code parameters. In Sec.~\ref{section: application to QC}, we present two applications that explicitly demonstrate the power of our framework: (i) a general time-efficient decoupling protocol for two- and three-local interactions tailored to superconducting and spin qubits, and (ii) a DD sequence permitting quantum simulation of the Kitaev honeycomb model, which strongly relies on anisotropic interactions, using only isotropic Heisenberg exchange interactions. Finally, in Sec.~\ref{section: robust pulse sequences}, we explain how to construct robust implementations of DD sequences using their group structure, before offering concluding remarks in Sec.~\ref{section: outlook}. While significant room for future improvement remains, our work provides the first systematic framework for constructing scalable DD sequences suitable for near-term quantum devices.

\section{Decoupling sequences from group theory}
\label{section: DD_background}
In this section, we review the basic of dynamical decoupling theory and its formulation in terms of group theory. This section is organized as follows. In Sec.~\ref{section: background theory}, we introduce DD from Floquet theory and the first-order Magnus expansion. In Sec.~\ref{section: bang-bang sequence theory}, we formulate dynamical decoupling sequences within a group-theoretic framework.

\subsection{Dynamical decoupling}
\label{section: background theory}

DD offers a general framework for suppressing unwanted interactions arising from arbitrary Hamiltonians by applying sequences of quantum gates.
This approach is traditionally employed in scenarios where a quantum system is weakly coupled to an environment, with the objective that, after the DD sequence, the influence of the environment on the system is effectively canceled.
In particular, we consider the Hamiltonian \cite{viola1998bang}
\begin{equation}
    H_{tot} = H_{S} + H_{E} + H_{S  E} \ , 
\end{equation}
that couples a qubit system with Hamiltonian $H_{S}$ to an environment Hamiltonian $H_{E}$ by the interaction $H_{S  E}$. Without loss of generality, we assume that
\begin{equation}
    H_{S  E} = \sum_i P_i \otimes O_i,
\end{equation}
where $P_i$ is a string of Pauli operators acting on the system and $O_i$ is an operator acting on the environment.

The DD problem involves designing a control Hamiltonian $H_{c}(t)$ acting solely on the system that suppresses the interactions with the environment. This means that after applying $H_{c}(t)$, the joint system-environment dynamics are governed by the decoupled Hamiltonian
\begin{equation}
    \overline{H}_{tot} = \overline{H}_S + \overline{H}_{E},
\end{equation}
where $\overline{H}_S$ ($\overline{H}_E$) denotes a (possibly) modified system (environment) Hamiltonian.

We restrict ourselves to the analysis of control Hamiltonians comprising sequences  that alternate between fast single-qubit gates and free evolution intervals, where all free-evolution intervals have the same duration. These admit a natural group-theoretic description and are especially advantageous for compilation, as they simplify both the implementation and analysis of control protocols.
In these cases, the effective Hamiltonian $\overline{H}_{tot}$ can be systematically derived using tools such as Floquet theory and Magnus expansion, assuming that the expansion converges \cite{Casas_2007}.

To do so, we denote the time-evolution operator generated by the control Hamiltonian $U_{c}(t)$ associated with $H_{c}(t)$ as 
\begin{equation}
    U_{c}(t) = \mathcal{T} \exp\Big[-i \int_0^{t} H_c(t') dt' \Big],
\end{equation}
where $\mathcal{T}$ is the time ordering operator and we set $\hbar = 1$. 
If the control Hamiltonian is cyclic  with period $T_c$, i.e., $H_c(t) = H_c(t+T_c)$, and the initial time is at $t=0$, the stroboscopic time evolution of the system and bath at times $n T_c$ with $n \in \mathbb{N}$  is described by~\cite{mori2023floquet,iserles2002expansions} 
\begin{equation}
    U(n T_c) = \exp\Big(-in T_c  \overline{H}_{tot} \Big)
\end{equation}
where $\overline{H}_{tot}$ is a time-independent stroboscopic Hamiltonian.

When the frequency $1/T_c$ is larger than the largest energy scale~\cite{Krodjasteh2008}, the effective Hamiltonian $\overline{H}_{tot}$ is well approximated by the first-order Magnus expansion
\begin{equation}
    \label{eq: average Hamiltonian}
    \overline{H}_{tot} \approx \frac{1}{T_c} \int_{0}^{T_c} U_{c}^{\dagger}(t) H_{tot} U_c(t)dt.
\end{equation}
Higher-order corrections in the Magnus expansion can be systematically bounded using Lemma 4 of Ref.~\cite{Krodjasteh2008}. 

In this work, we aim to generalize this idea. Rather than focusing solely on suppressing interactions with the environment, our objective is to design DD sequences that can also selectively suppress certain interactions within the system itself. For example, we either completely freeze the system dynamics (i.e., $\overline{H}_{S} = 0$) or we selectively retain only some desired interactions in $H_S$, thereby engineering specific target Hamiltonians. This capability is particularly useful for digital-analog quantum simulations~\cite{Parra-Rodriguez2020}.

In the next section, we will briefly review the bang-bang control scheme \cite{viola1998bang,viola1999dynamical}, which assumes instantaneous single-qubit pulses. In Appendix~\ref{section: bounded sequence theory}, we also review the bounded-strength control scheme \cite{Viola2003}, which provides a more realistic model assuming that single-qubit gates have a finite duration. Importantly, we emphasize that the framework we introduce in this work applies to both control schemes.

\subsection{Bang-bang control sequence}
\label{section: bang-bang sequence theory}

The bang-bang control scheme \cite{viola1998bang,viola1999dynamical} assumes that single-qubit pulses are instantaneous. The central idea is to define a decoupling group $\mathcal{G} = \{ g_i \}$ \cite{viola1998bang,Viola2003,Viola2003,evert2024syncopated}, which acts on the system Hilbert space via a unitary representation $  g  \to U_g $.
The bang-bang DD sequence is then specified by an ordered list of group elements $(g_1, g_2, \dots, g_L)$, with  total length of the sequence $L = \lambda |\mathcal{G}|$, where $|\mathcal{G}|$ is the order of the group and $\lambda$ is a positive integer, typically $\lambda=1$~\cite{Viola2003}. In general, each group element $g_i \in \mathcal{G}$ appears exactly $\lambda$ times in the sequence. If the group $\mathcal{G}$ is generated by a set of generators $\{ \gamma_1, \dots \gamma_n \}$, we write the group as
\begin{equation}
    \mathcal{G} = \langle \gamma_1 ,\dots, \gamma_n \rangle
\end{equation}
to explicitly denote its generating set.

 A single cycle of the bang-bang control unitary $U_c(t)$ has a total duration $T_c$ and is defined as a step function of time, divided into $L$ equally-spaced intervals of length $\Delta = {T_c}/{L}$. During each segment, lasting for a time  $\tau \in [0,\Delta)$, the control is a constant unitary operator
\begin{equation}
  U_{c}[(j-1)\Delta + \tau,(j-1)\Delta] = U_{g_j} \ . 
\end{equation}
This control unitary is abruptly switched from $U_{g_{j-1}}$ to $U_{g_{j}}$, corresponding to an ideal instantaneous control pulse~\cite{viola1998bang,Viola2003}.

Under the action of the DD group $\mathcal{G}$, the effective Hamiltonian $\overline{H}_{tot}$ in Eq.~\eqref{eq: average Hamiltonian}, is given by the group twirling operation
\begin{equation}
    \overline{H}_{tot} = \frac{1}{|\mathcal{G}|} \sum_{g \in \mathcal{G}} U_{g}^{\dagger} H_{tot} U_{g}:= \Pi_{\mathcal{G}}(H_{tot}) \ . 
\end{equation}
The time-evolution operator of $\overline{H}_{tot}$, corresponding to the time-evolution of $H_{tot}$ at the stroboscopic times $nT_c$, is explicitly given by
\begin{equation}
 U(nT_c)\approx\left(\prod_{g \in \mathcal{G}} U_{g}^{\dagger} U_\Delta U_{g}\right)^n \ , 
\end{equation}
with the free evolution operator defined as $U_\Delta=\mathcal{T}\text{exp}\left(-i \int_0^\Delta H_{tot}(\tau) d\tau\right)$. 

By Schur's lemma, the twirling channel $\Pi_{\mathcal{G}}(\cdot)$ acts as a projection and retains only the components of $H_{tot}$ that commute with all elements of the group $\mathcal{G}$.
When the map $g  \to U_g$ defines an irreducible representation of $\mathcal{G}$, the projection takes the form \cite{viola1998bang,viola1999dynamical,Viola2003,Rotteler_2006}
\begin{equation}
    \label{eq: Schur lemma}
    \Pi_{\mathcal{G}}(O) = \frac{\rm{Tr}(O)}{d} I_d. 
\end{equation}
where $d$ is the dimension of the system Hilbert space. In this special case, the DD sequence completely averages out all nontrivial system operators and $\bar H_{\rm S}$ becomes proportional to the identity. This corresponds to a trivial global phase, meaning that the system dynamics are completely frozen.

In this work, we restrict ourselves to the analysis of $(\pi)$ and/or $(-\pi)$-pulses generating $X, Y, Z$, and  $I$ gates on a system of $N$ qubits.
Consequently, we choose the decoupling group $\mathcal{G}$ to be a subgroup of the $N$-qubit projective Pauli group, i.e. $\mathcal{G} \subseteq \mathcal{P}^{N}/\{\pm i,\pm 1\}$. We remark that although the Pauli group is not Albelian (i.e $g_1 g_2 \neq g_2 g_1$), the projective Pauli group is. The corresponding unitary representation maps each group element $g \in \mathcal{G}$ to its associated standard Pauli operator acting on the system Hilbert space.  

As an example, we consider one cycle of the standard single-qubit $\rm XY4$ sequence, defined by the unitary time-evolution operator 

\begin{equation}
U(T_c)=U_{\Delta} Y U_{\Delta} X U_{\Delta} Y U_{\Delta} X \ ,
\end{equation}
where $U_\Delta$ is a single-qubit free evolution operator for a time $\Delta$.
This sequence has a decoupling group $\mathcal{G}$ corresponding to the full single-qubit Pauli group. This can be seen explicitly by expressing the sequence as a product of Pauli operators applied before and after each free evolution period and using Pauli matrix identities ($XY=Z$ and cyclic permutations)
\begin{equation}
U(T_c) \equiv (I U_{\Delta} I)(Y U_{\Delta} Y)(Z U_{\Delta} Z)  (X U_{\Delta} X) \ .
\end{equation}
This decomposition shows that each interval is conjugated by a Pauli operator and the full sequence effectively performs a twirl over the single-qubit Pauli group. Consequently, the $\rm XY4$ sequence removes all single-qubit terms, resulting in universal noise suppression. 

In bounded-control schemes \cite{Viola2003, Bookatz_2016}, the requirement of applying arbitrarily strong and instantaneous control pulses $U_{g_{j+1}} U_{g_j}^{\dagger}$ at discrete times $t = j \Delta$ is relaxed. Instead, the control strategy involves smoothly interpolating between successive group elements, implementing a continuous transition from $U_{g_j}$ to $U_{g_{j+1}}$ over each subinterval. More details on bounded-strength control scheme can be found in Appendix~\ref{section: bounded sequence theory}.

\section{Permutation-invariant decoupling group from hypergraph symmetry}
\label{section: symmetry and quotient graph}
\begin{figure*}
    \centering
    \begin{subfigure}[t!]{0.95\linewidth}
        \centering
        \begin{adjustbox}{center}
        \begin{tikzpicture}[node distance=3cm, every node/.style={process}, text width=3.8cm]
            \node (step1) {\justifying \textbf{Step 1}: Define interaction hypergraph $I_{\rm Dev}$};
            \node (step2) [right of=step1] {\justifying \textbf{Step 2}:  Color $I_{\rm Dev}$ with $\mathcal{C}[I_{\rm Dev}]$};
            \node (step3) [right of=step2] {\justifying \textbf{Step 3}:  Construct symmetric hypergraph $I'_{\rm Dev}$};
            \node (step4) [right of=step3] {\justifying \textbf{Step 4}:  Form quotient hypergraph $I'_{\rm Dev} / \mathcal{C}[I_{\rm Dev}]$};
            \node (step5) [right of=step4] {\justifying \textbf{Step 5}: Generate universal sequence via additive code};
            \node (step6) [right of=step5] {\justifying \textbf{Step 6}: Tailored sequence or Hamiltonian engineering};

            \draw [arrow] (step1) -- (step2);
            \draw [arrow] (step2) -- (step3);
            \draw [arrow] (step3) -- (step4);
            \draw [arrow] (step4) -- (step5);
            \draw [arrow] (step5) -- (step6);
        \end{tikzpicture}
        \end{adjustbox}
        \caption{Schematic overview of our decoupling framework.}
        \label{fig: flowchart-a}
    \end{subfigure}
    \hfill
    \begin{subfigure}[b]{0.95\linewidth}
        \centering
        \includegraphics[width=\linewidth]{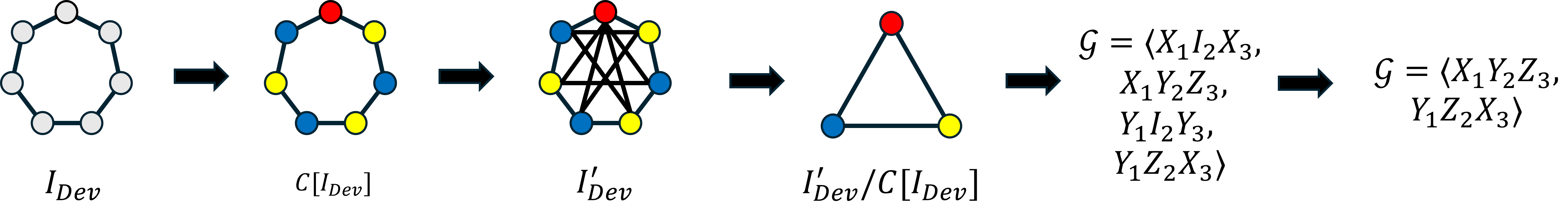}
        \caption{Example illustrating each step of the flowchart.}
        \label{fig: flowchart-b}
    \end{subfigure}
    \caption{\justifying \textbf{Framework for constructing efficient dynamical decoupling sequences.} 
    (a) Flowchart of our approach. Starting from an interaction hypergraph, we color it, generate a symmetric variant, and construct its quotient. Additive codes are then used to produce efficient universal sequences, which are tailored to suppress specific Hamiltonian terms. (b) Example of each step in the protocol. We start from a simple 7-qubit loop (step 1) and we color it assuming a nearest-neighbor interaction model (step 2). We then construct a color-preserving symmetric hypergraph connecting all different colors (step 3), which allows us to simplify the DD problem by just considering the corresponding quotient hypergraph (step 4). The DD sequence is then constructed by using additive codes, forming a universal decoupling group that suppresses arbitrary two-local interactions (step 5). We then tailor the sequence for specific qubit architectures (step 6).   }
    \label{fig: flowchart-summary}
\end{figure*}
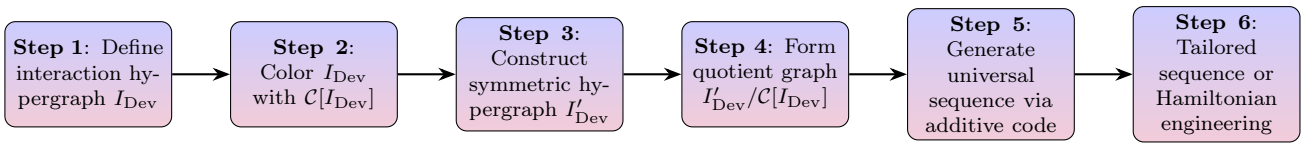
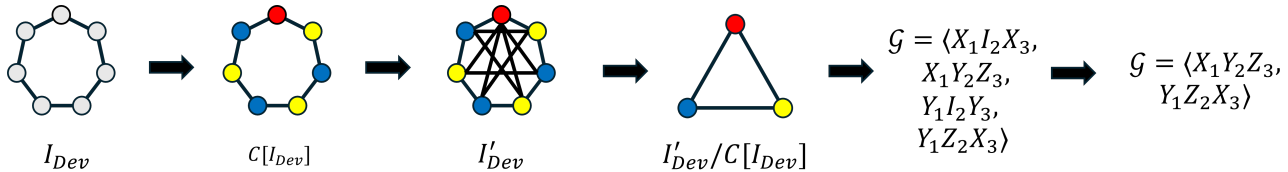
In this section, we systematically reformulate and extend the ideas introduced in Refs.~\cite{brown2024efficient, evert2024syncopated,tong2024empirical} using the language of graph theory. This perspective enables a natural generalization of previous results from $2$-local interaction models to general $k$-local interaction models. In Sec.~\ref{section: hypergraph coloring}, we describe the system Hamiltonian in terms of an interaction hypergraph and show that any $k$-local interaction induces a natural partition of physical qubits into equivalence color classes. In Sec.~\ref{section: symmetry and quotient graph}, we demonstrate that this partition gives rise to a permutation symmetry that can be exploited to reduce the original interaction hypergraph to a simpler quotient hypergraph, by factoring out the symmetry group. As a result, the task of designing a DD sequence for the original interaction hypergraph is mapped to the simpler problem of designing a DD sequence for the quotient hypergraph.

\subsection{Hypergraph coloring}
\label{section: hypergraph coloring}

To describe a general Hamiltonian of $N$ interacting qubits, we introduce the interaction hypergraph $I_{\rm Dev}=(V_{I},E_I)$. The vertex set $V_{I} = \{v_1,\dots,v_N\}$ represents physical qubits. A hyperedge $e_i \in E_I$ connecting $k$ vertices $(v_{i_1},\dots,v_{i_k})$ corresponds to a general noise model in which arbitrary Pauli interactions from $\{I, X,Y,Z \}^{\otimes k}$ may jointly act on these $k$ qubits, unless otherwise specified. This corresponds to the first step of Fig.~\ref{fig: flowchart-a}. A Pauli string associated with such a hyperedge is called $k$-local if it has support on exactly $k$ distinct vertices. Therefore, the interaction hypergraph 
$I_{\rm Dev}$ characterizes a $k$-local Hamiltonian if and only if all of its hyperedges involve at most $k$ vertices. The standard graph is simply 2-local hypergraph. Under this definition, the Hamiltonians considered in Refs.~\cite{brown2024efficient,evert2024syncopated} are 2-local and the interaction hypergraph $I_{\rm Dev}$ coincides with the device connectivity graph.

We now generalize the results of Refs.~\cite{brown2024efficient,evert2024syncopated} to arbitrary $k$-local Hamiltonians. In the first-order Magnus expansion given in Eq.~\eqref{eq: average Hamiltonian}, a vertex $v_i$ only interacts with other vertices that share a common hyperedge. This motivates a natural partition of the interaction hypergraph $I_{\rm Dev}$. We assign a distinct color $c_i$ to each vertex $v_{i} \in V_{I}$, such that no two vertices $v_i,v_j$ in a hyperedge share the same color (step 2 of Fig.~\ref{fig: flowchart-a}). We denote this hypergraph coloring as $C[I_{\rm Dev}]$, and refer to the number of colors used (i.e. the chromatic number) as $\chi_I$. 

We remark that this hypergraph coloring problem can be addressed efficiently using well-established graph-coloring algorithms \cite{mostafaie2020systematic}. However, standard graph-coloring algorithms are typically designed for graphs and not hypergraphs. To apply these algorithms to a hypergraph coloring problem, we formally construct an auxiliary graph by adding an edge between each pair of vertices $v_i$ and $v_j$ that share a hyperedge in $I_{\rm Dev}$. A proper coloring of this auxiliary graph corresponds to a valid coloring of the original hypergraph $I_{\rm Dev}$..  
 
In the next section, we show that the color partition $C[I_{\rm Dev}]$ respects an underlying hypergraph symmetry. Using this symmetry, the problem of designing a DD sequence for the interaction hypergraph $I_{\rm Dev}$ is significantly simplified.

\subsection{Symmetry and quotient hypergraph}

We now show how the color partition $C[I_{\rm Dev}]$ can be exploited to simplify the problem. The strategy is as follows. We first construct an expanded hypergraph $I_{\rm Dev}^{'} = (V_I,E_{I}^{'})$ from $I_{\rm Dev}$ such that $I_{\rm Dev} \subseteq I_{\rm Dev}^{'}$, and the color partition $C[I_{\rm Dev}]$ remains a valid vertex coloring for $I_{\rm Dev}^{'}$. Because $I_{\rm Dev} \subseteq  I_{\rm Dev}^{'}$, any decoupling group $\mathcal{G}$ designed for $I_{\rm Dev}^{'}$ is also valid decoupling group for $I_{\rm Dev}$. We then show that designing a decoupling group for $I_{\rm Dev}^{'}$ is equivalent to designing one for the significantly simpler quotient hypergraph $I_{\rm Dev}^{'}/C[I_{\rm Dev}]$, which contains only $\chi_I$ vertices. In this way, the original problem is reduced to constructing a decoupling group for the quotient hypergraph $I_{\rm Dev}^{'}/C[I_{\rm Dev}]$.

More precisely, the expanded hypergraph $I_{\rm Dev}^{'}$ is constructed from $I_{\rm Dev}$ and its coloring $C[I_{\rm Dev}]$ as follows. The hypergraph $I_{\rm Dev}^{'}$ shares the same vertex set $V_{I}^{'} = V_{I}$ as $I_{\rm Dev}$. We traverse the colored hypergraph $I_{\rm Dev}$ and record all tuples $(c_{i_1},\dots,c_{i_k})$ such that there exists a hyperedge in $I_{\rm Dev}$ connecting vertices with those colors. The new hyperedge set $E_{I}'$ is then constructed by appending to the original set $E_{I}$ additional hyperedges of the form $(v_{i_1},\dots,v_{i_k}) \notin E_{I}$, where the vertices  $(v_{i_1},\dots,v_{i_k})$ have distinct colors that match one of the recorded color tuples $(c_{i_1},\dots,c_{i_k})$. This procedure ensures that the color partition $C[I_{\rm Dev}]$ remains valid for the hypergraph $I_{\rm Dev}^{'}$, see step 3 of Fig.~\ref{fig: flowchart-a}.

\begin{figure}
    \centering
    \includegraphics[width=0.75\linewidth]{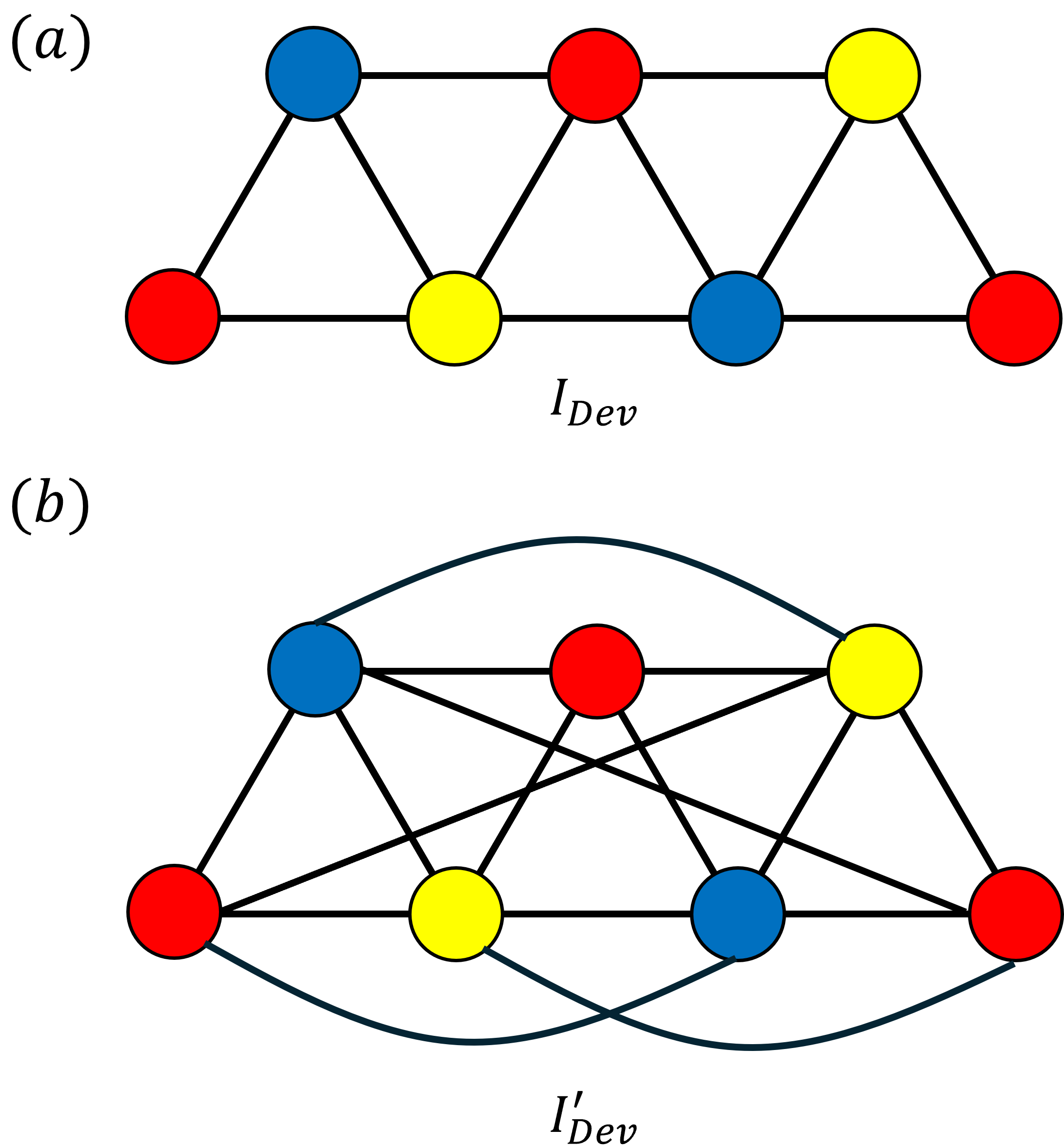}
    \caption{ \justifying \textbf{Colored bilinear array graph.} (a) Interaction hypergraph $I_{\rm Dev}$ of $7$ physical qubits with two-local interactions arranged in a bilinear array, along with the minimal color partition $C[I_{\rm Dev}]$. (b) Expanded hypergraph $I_{\rm Dev}^{'}$ constructed from $I_{\rm Dev}$ and $C[I_{\rm Dev}]$. }
    \label{fig: G and expanded G}
\end{figure}

To better illustrate this construction, here we apply it to an explicit example. We consider a system of seven qubits arranged in a bilinear array with two-local interactions. Figure~\ref{fig: G and expanded G}(a) illustrates the corresponding interaction hypergraph $I_{\rm Dev}$ along with its vertex coloring $C[I_{\rm Dev}]$. A general Hamiltonian for this system is 
\begin{equation}
    H = \sum_{i} \vec{h}_i \cdot \vec{\sigma}_i + \sum_{\langle ij \rangle} \vec{\sigma}_i \cdot A_{ij}  \vec{\sigma}_j.
\end{equation}
where $\vec{h}_i$ are local fields and $A_{ij}$ are real coupling matrices that specify pairwise interactions. For this example, the expanded hypergraph $I_{\rm Dev}^{'}$ constructed from the above procedure is shown in Fig.~\ref{fig: G and expanded G}(b), with new edges connecting vertices of different colors. 

The expanded hypergraph $I^{'}_{\rm Dev}$ may appear more complex than the original interaction hypergraph $I_{\rm Dev}$, suggesting that designing a suitable decoupling group $\mathcal{G}$ for $I^{'}_{\rm Dev}$ could be more challenging. 
However, $I^{'}_{\rm Dev}$ inherits a strong symmetry from the color partition $C[I_{\rm Dev}]$. In particular, $I^{'}_{\rm Dev}$ is invariant under permutations of vertices that share the same color, a symmetry we (with a slight abuse of notation) also denote by $C[I_{\rm Dev}]$. This invariance significantly simplifies the construction of the decoupling group: it implies that it suffices to consider decoupling groups $\mathcal{G}$ that themselves respect the symmetry $C[I_{\rm Dev}]$ \cite{zanardi1999symmetrizing}. 
Practically, this means applying identical control pulses to all qubits within the same color class.

\begin{figure}
    \centering
    \includegraphics[width=0.8\linewidth]{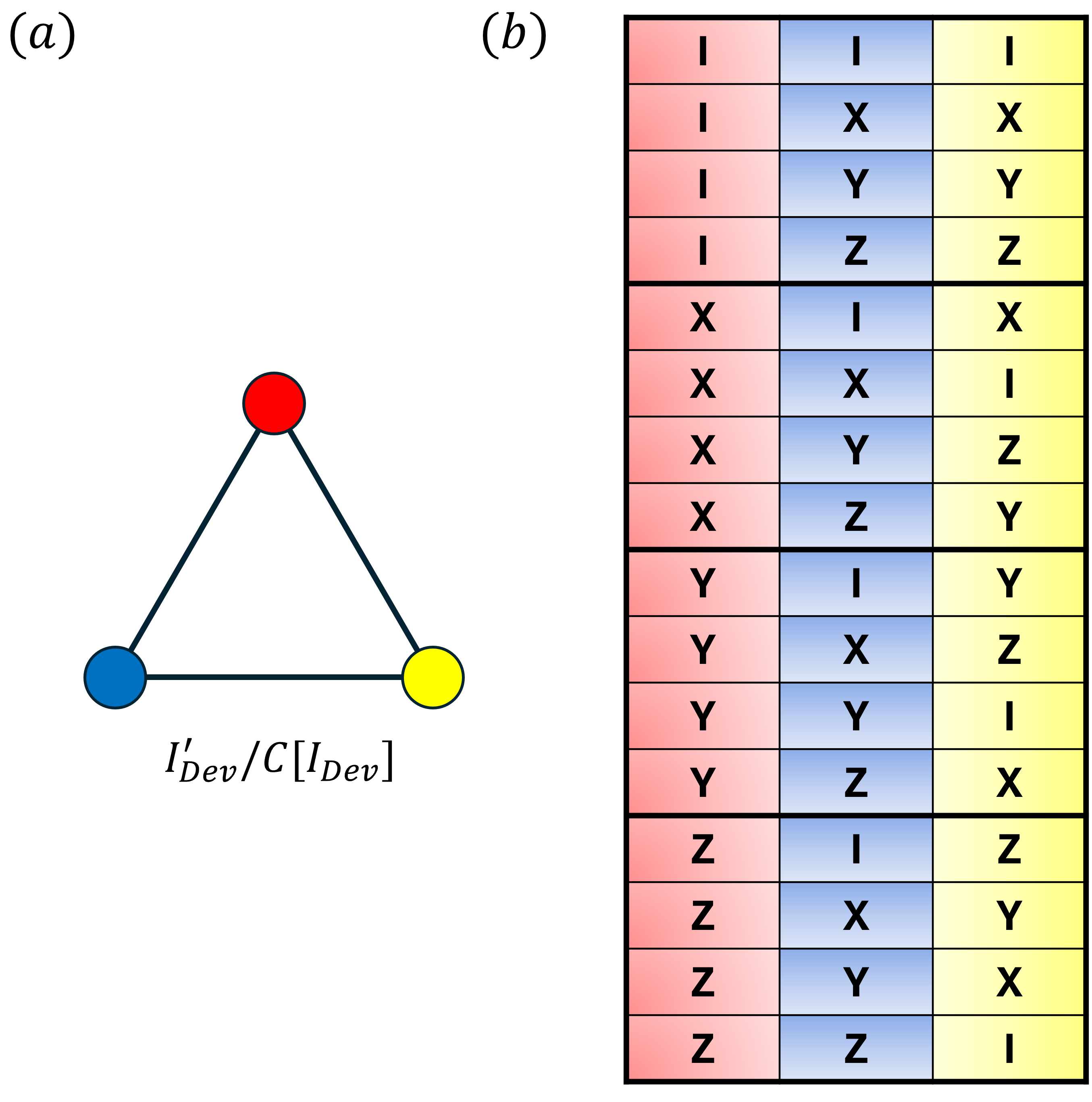}
    \caption{\justifying \textbf{Dynamical decoupling from orthogonal arrays.} (a) The quotient hypergraph $I_{\rm Dev}^{'}/C[I_{\rm Dev}]$ formed by collapsing all physical qubits of the same color in $I_{\rm Dev}^{'}$ into a single equivalence class. (b) The orthogonal array  $\text{OA}(L=16,\chi_I=3,S=4,k=2)$ which generates a universal DD sequence for two-local Hamiltonian defined by the quotient hypergraph $I_{\rm Dev}^{'}/C[I_{\rm Dev}]$. Each column is colored according to the associated color class $c_i$.  }
    \label{fig: expander quotient graph}
\end{figure}

As a complementary view on this symmetry, rather than directly working with the symmetric decoupling group $\mathcal{G}$, we can instead analyze the quotient hypergraph $I^{'}_{\rm Dev}/C[I_{\rm Dev}]$. The quotient hypergraph $I^{'}_{\rm Dev}/C[I_{\rm Dev}]$ is formed by collapsing all vertices $v_i \in V_{I}^{'}$ with the same color $c_i$ into a single super-vertex $c_i \in V_{I}^{'}/C[I_{\rm Dev}]$ in the quotient hypergraph. We will refer to the super-vertex $c_i$ in the quotient hypergraph $I^{'}_{\rm Dev}/C[I_{\rm Dev}]$ as the color class $c_i$. A hyperedge $(c_{i_1}, \dots, c_{i_k})$ exists in $I^{'}_{\rm Dev}/C[I_{\rm Dev}]$ if and only if there is at least one hyperedge $(v_{i_1},\dots,v_{i_k}) \in E_{I}^{'}$, such that the vertices $v_{i_j}$ are colored $c_{i_j}$ respectively. Figure~\ref{fig: expander quotient graph} illustrates this transformation from the expanded hypergraph $I_{\rm Dev}^{'}$ (shown in Fig.~\ref{fig: G and expanded G}) to its corresponding quotient hypergraph $I_{\rm Dev}^{'}/C[I_{\rm Dev}]$. We emphasize that the same quotient hypergraph also captures different topologies, including the ring topology in Fig.~\ref{fig: flowchart-summary}.

It follows directly that if there is a DD sequence that universally suppresses all interactions in the quotient hypergraph $I_{\rm Dev}^{'}/C[I_{\rm Dev}]$, then this sequence can be lifted to a valid decoupling sequence for the expanded hypergraph $I_{\rm Dev}^{'}$, and consequently on the interaction hypergraph $I_{\rm Dev}$. The lifting procedure consists of applying the same control pulses in parallel to all vertices in $I_{\rm Dev}^{'}$ that share the same color $c_i$. This approach is valid for both bang-bang and bounded-strength control schemes.

We note that the explicit construction of the expanded interaction hypergraph $I_{\rm Dev}^{'}$ in principle is not required to design the DD group. Rather, it provides a conceptual tool that reveals the underlying symmetry structure of the colored interaction hypergraph. The expanded hypergraph also provides a key general insight, namely any two interaction hypergraphs that share the same expanded hypergraph $I_{\rm Dev}^{'}$ can be universally decoupled using the same decoupling group $\mathcal{G}$, see the examples in Figs.~\ref{fig: flowchart-summary} and~\ref{fig: G and expanded G}. In fact, an even stronger statement holds: if two interaction hypergraphs share the same quotient hypergraph $I_{\rm Dev}^{'}/C[I_{\rm Dev}]$ and interaction model, they admit the same decoupling group capable of universally suppressing their system interactions. 

In the following section, we demonstrate how to obtain time-efficient decoupling groups for the quotient hypergraph $I^{'}_{\rm Dev}/C[G_{\rm Dev}]$ using constructions from classical coding theory.

\section{Time-efficient decoupling groups from classical additive codes}

In the previous section, we demonstrated how the problem of constructing a decoupling group $\mathcal{G}$ for an arbitrary $k$-local interaction hypergraph $I_{\rm Dev}$ can be reduced to the simpler task of constructing a decoupling group for the quotient hypergraph $I_{\rm Dev}^{'}/C[I_{\rm Dev}]$. In this section, we explain how to build efficient decoupling groups for $I_{\rm Dev}^{'}/C[I_{\rm Dev}]$ using tools from classical coding theory. In Sec.~\ref{section: orthognal array and DD}, we introduce the connection between decoupling groups $\mathcal{G}$ and a combinatorial object known as orthogonal arrays as found in previous works \cite{Rotteler_2006,Bookatz_2016}. These objects enable constructions of decoupling groups that achieve favorable trade-offs between the DD sequence length $L$ and the chromatic number $\chi_I$. In Sec.~\ref{section: Orthogonal array and classical additive codes}, we show how to construct the required orthogonal arrays from classical additive codes. In Sec.~\ref{section: dynamical decoupling as error detection codes}, we further exploit the structure of classical additive codes to show that the decoupling group can be completely characterized by its generators, thereby reducing the design of the full decoupling group to the simpler task of designing its generators. Finally, in Sec.~\ref{section: selective DD sequences}, we present a simple numerical algorithm for finding decoupling groups that selectively cancel certain interactions while preserving others.

\label{section: decoupling group from additive code}
\subsection{Orthogonal arrays and dynamical decoupling}
\label{section: orthognal array and DD}
Time-efficient DD sequences can be constructed using combinatorial design objects. For example, Ref.~\cite{brown2024efficient} employs Hadamard matrices to construct decoupling sequences for $2$-local Hamiltonians. We generalize this approach to arbitrary $k$-local Hamiltonians by using more general combinatorial structures known as orthogonal arrays \cite{Bookatz_2016,Rotteler_2006,Stollsteimer2001}. In the following analysis, we consider the worst-case scenario where all possible $k$-local interactions between the color classes $c_i$ are present in the quotient hypergraph $I_{\rm Dev}^{'}/C[I_{\rm Dev}]$. 

An orthogonal array $\text{OA}(L,\chi_{I},S,k)$ is an $L\times \chi_{I}$ array with entries taken from a finite set $S$ and strength $k$, such that every possible $k$-tuple from $S^{\otimes k}$ appears uniformly across any choice of $k$ columns \cite{hedayat2012orthogonal}. 
In this work, we restrict ourselves to the set $S = \{ I,X,Y,Z\}$ of single-qubit Pauli operators. Each column of the orthogonal array is associated with a color class $c_i$, and each row defines a group element in the decoupling group $\mathcal{G}$. If the noise model is biased along a specific axis of the Bloch sphere (e.g. the $Z$-axis) or if we want to maintain certain interactions (e.g. $XX$ interactions), we can reduce the required size $L$ of the orthogonal array by selecting a smaller set (e.g. $S = \{I,X\}$) instead of the full single-qubit Pauli group. We remark that for any choice of parameters $\chi_I$, $S$, and strength $k$, there always exists an orthogonal array whose length $L$ is determined by these parameters \cite{hedayat2012orthogonal}.

In Fig.~\ref{fig: expander quotient graph}, we show the explicit orthogonal array $\text{OA}(L=16,\chi_I=3,|S|=4,k=2)$, which generates a universal decoupling sequence for 2-local Hamiltonians defined on the quotient hypergraph  $I_{\rm Dev}^{'}/C[I_{\rm Dev}]$ \cite{neilsloaneOrthogonalArrays}. One can verify that for any pair of columns, the rows of the $\text{OA}$ exhaustively span all combinations of $\{ I,X,Y,Z \}^{\otimes 2}$ exactly once. Moreover, each Pauli operator appears exactly four times in every column, ensuring balanced coverage and uniformity.

By construction, the action of the decoupling group $\mathcal{G}$ derived from an $\text{OA}$ with strength $k$ implements a group twirl over the $k$-fold Pauli group $\{ I,X,Y,Z\}^{\otimes k}$ on any subset of $k$ color classes.  As a consequence, all $k$-local interaction terms are dynamically suppressed, following directly from Schur's lemma [Eq.~\eqref{eq: Schur lemma}]. Furthermore, all interactions with support on less than $k$ color classes are also suppressed as a consequence of this construction. Therefore, to universally decouple all $k$-local interactions on the quotient hypergraph $I_{\rm Dev}^{'}/C[I_{\rm Dev}]$, it suffices to construct the orthogonal array $\text{OA}(L,\chi_I,S,k)$. 

We note that there is a comprehensive online library cataloging efficient (i.e good ratio between $\chi_I$ and $L$) orthogonal arrays across a wide range of parameters \cite{neilsloaneOrthogonalArrays}. In the next section, we describe how to construct length-efficient orthogonal arrays by utilizing their connection to classical error correction codes \cite{Rotteler_2006,Bookatz_2016}. Since classical codes are extensively studied, we can utilize well-established classical codes to obtain time-efficient DD sequences suppressing arbitrary $k$-local interactions \cite{Bookatz_2016}.

\subsection{Orthogonal array and classical additive codes}
\label{section: Orthogonal array and classical additive codes}
The connection between orthogonal arrays and classical linear codes is well established \cite{DELSARTE1973407, hedayat2012orthogonal,Rotteler_2006,Bookatz_2016}. In this work, we focus on the projective Pauli group, which admits a natural isomorphism to the finite field $\mathbb{F}_4$.
A finite field $\mathbb{F}_{q}$ is a set containing $q$ elements in which addition, subtraction, multiplication, and division (except by zero) are all well-defined and satisfy the field axioms. The isomorphism between the projective Pauli group and $\mathbb{F}_4$ is defined by the correspondence
\begin{equation}
    \label{eq: Pauli to elements of F4}
    I \to 0,~ X \to 1,~Z\to \omega,~Y \to 1+ \omega,
\end{equation}
where $\omega$ is a primitive element of $\mathbb{F}_4$ satisfying $\omega^2 = \omega+1$. The addition and multiplication tables for $\mathbb{F}_4$ are
\begin{equation}
    \begin{array}{c|cccc}
        + & 0 & 1 & \omega & 1+\omega \\
        \hline
        0 & 0 & 1 & \omega & 1+\omega \\
        1 & 1 & 0 & 1+\omega & \omega \\
        \omega & \omega & 1+\omega & 0 & 1 \\
        1+\omega & 1+\omega & \omega & 1 & 0 \\
    \end{array}
\end{equation}
and 
\begin{equation}
    \begin{array}{c|cccc}
        \times & 0 & 1 & \omega & 1+\omega \\
        \hline
        0 & 0 & 0 & 0 & 0 \\
        1 & 0 & 1 & \omega & 1+\omega \\
        \omega & 0 & \omega & 1+\omega & 1 \\
        1+\omega & 0 & 1+\omega & 1 & \omega \\
    \end{array} \ .
\end{equation}

This correspondence allows us to represent Pauli operators algebraically as elements of $\mathbb{F}_4$, providing a convenient framework for constructing orthogonal arrays from quaternary codes.  Furthermore, when the noise is biased (e.g. along the $Z$-axis), we may restrict to a subset of the Pauli group, in which case classical codes over the binary field $\mathbb{F}_2$ become relevant.

We consider here classical additive codes, which, as we will show in Sec.~\ref{section: dynamical decoupling as error detection codes}, provide a natural framework for interpreting a decoupling sequence as an error detection code. An additive code $\mathcal{C}$ is a subset of $\mathbb{F}_{q}^{n}$ (a vector of length $n$ in which its entries are from $\mathbb{F}_q$) that is closed under addition. While every linear code is an additive code, the converse is not true: additive codes are not necessarily closed under scalar multiplication unless the field is $\mathbb{F}_2$. 

We characterize the code $\mathcal{C}$ by the set of parameters $(n,|\mathcal{C}|,d)_q$, where $n$ is the length of the codeword, $|\mathcal{C}|$ is the number of distinct code words, and $d$ is the code distance. The entries of each codeword are elements of the finite field $\mathbb{F}_q$. The distance $d$ is defined as the minimum Hamming distance between any two distinct codewords $\mathbf{u},\mathbf{v} \in \mathcal{C}$. Since $\mathcal{C}$ is closed under addition, this is equivalent to the minimum Hamming weight of any nonzero codeword 
\begin{equation}
    d = \min_{\mathbf{u} \neq 0 \in \mathcal{C} } \rm{wt_H}(\mathbf{u}).
\end{equation}

The vector space $\mathbb{F}_{4}^{n}$ can be endowed with an inner product $\langle \cdot,\cdot\rangle$. A natural choice for additive codes is the trace Hermitian inner product $\langle \cdot ,\cdot \rangle_{\rm tr}:\mathbb{F}_{4}^{n} \to \mathbb{F}_2$ \cite{ezerman2011additive}, defined as
\begin{equation}
    \langle \bf{u},\bf{v} \rangle_{tr}~= \sum_{i} (u_i v_i^2 + v_i u_i^2). 
\end{equation}
Under the mapping between single-qubit Pauli operators and elements of $\mathbb{F}_4$ given in Eq.~\eqref{eq: Pauli to elements of F4}, this inner product directly gives the commutation relations of Pauli operators. That is, it evaluates to zero if the corresponding Pauli operators commute and to one if they anticommute. 

Using the trace Hermitian inner product, we define the dual code $\mathcal{C}^{\perp}$ of an additive code $\mathcal{C}$ as
\begin{equation}
    \label{eq: def dual code}
    \mathcal{C}^{\perp} = \{ \bf{v} \in \mathbb{F}_{q}^{n}~|~ \langle \bf{u}, \bf{v} \rangle_{tr} = 0, \quad \forall \bf{u}\in \mathcal{C} \}. 
\end{equation}
We denote the minimum distance of the dual code by $d^{\perp}$. Importantly, the dual code $\mathcal{C}^{\perp}$ is always an additive code, regardless of whether $\mathcal{C}$ is linear or additive. However, it is known that if $\mathcal{C}$ is an additive code of size $|\mathcal{C}|=2^{l}$ for some positive integer $l \leq 2n$, then its dual code has size $|\mathcal{C}^{\perp}|=2^{2n-l}$. This follows from the fact that $\mathbb{F}_{4}^{n}$ is isomorphic to $\mathbb{F}_{2}^{2n}$ under addition.

A seminal result by Delsarte~\cite{DELSARTE1973407} establishes a fundamental connection between orthogonal arrays and classical codes:
\begin{theorem}
    \label{theorem: OA and classical code}
    If $\mathcal{C}$ is a code $(n,|\mathcal{C}|,d)_q$ with dual distance $d^{\perp}$, then the codewords of $\mathcal{C}$ form the rows of an $\text{OA}(|\mathcal{C}|,n,q,d^{\perp}-1)$ with entries from $\rm \mathbb{F}_q$. Conversely, the rows of a linear $\text{OA}(|\mathcal{C}|,n,q,k)$ over $\rm \mathbb{F}_q$ form a linear code $(n,|\mathcal{C}|,d)_q$ over $\rm \mathbb{F}_q$ with dual distance $d^{\perp}\geq k+1$. If the orthogonal array has strength $k$ but not $k+1$, $d^{\perp}$ is precisely $k+1$. 
\end{theorem}
This theorem highlights the direct connection between the strength of an $\text{OA}$ and the dual distance $d^{\perp}$ of the code $\mathcal{C}$. Consequently,  to construct an $\text{OA}$ of strength $k$, corresponding to a universal DD sequence that suppresses $k$-local interactions in the original Hamiltonian, it suffices to use the dual code $\mathcal{C}^{\perp}$ of a code $\mathcal{C}$ with distance $d = k + 1$.

In Ref.~\cite{Bookatz_2016}, Bose–Chaudhuri–Hocquenghem (BCH) codes \cite{bose1960class,gorenstein1961class} were used  to construct efficient universal DD sequences for $k$-local Hamiltonians acting on physical qubits, not color classes. By applying their construction to the quotient hypergraph $I_{\rm Dev}^{'}/C[I_{\rm Dev}]$, we find that the length $L$ of a single DD cycle required to universally decouple $k$-local Hamiltonians scales as
\begin{equation}
    \label{eq: scaling for bang-bang}
     O(\chi_I^{k-1} )
\end{equation}
for bang-bang control, and as 
\begin{equation}
    \label{eq: scaling for bounded-control}
    O(\chi_{I}^{k-1} \log \chi_I)
\end{equation}
for bounded-strength control (see Appendix~\ref{section: bounded sequence theory}).

In Sec.~\ref{section: application to QC}, we will present different constructions based on alternative classical codes that generate more time-efficient decoupling groups for 2-local and 3-local Hamiltonians. While the asymptotic scaling remains the same as in Eqs.~\eqref{eq: scaling for bang-bang} and~\eqref{eq: scaling for bounded-control}, our constructions lead to improved constant prefactors compared to those based on BCH codes for the 2-local and 3-local Hamiltonians. 

\subsection{Dynamical decoupling as error detection codes}
\label{section: dynamical decoupling as error detection codes}

Historically, self-dual additive codes over $\mathbb{F}_4$ (i.e., $\mathcal{C} = \mathcal{C}^{\perp}$) have played a central role in the construction of stabilizer quantum error-correcting codes \cite{calderbank1997quantumerrorcorrectioncodes, ezerman2011additive, zeng2011transversality, haah2016algebraic}. Many properties of stabilizer codes can be analyzed through their stabilizer generators. In a closely analogous manner, we further develop the connection between dynamical decoupling and classical additive codes by showing that key properties of a DD sequence can be analyzed through the generators of its decoupling group. This perspective not only provides a compact characterization of DD sequences, but also enables systematic constructions of selective DD sequences, as we explain in this section.

From the definition of the dual code $\mathcal{C}^{\perp}$ in Eq.~\eqref{eq: def dual code}, it follows that Pauli strings corresponding to codewords in $\mathcal{C}^{\perp}$ remain invariant under the decoupling sequence generated by $\mathcal{C}$. In fact, $\mathcal{C}^{\perp}$ defines the entire invariant subspace of the decoupling sequence. Consequently, to determine whether an interaction is suppressed by the decoupling sequence, one only needs to check whether the associated codeword lies in $\mathcal{C}^{\perp}$. At first glance, this may appear computationally expensive because it requires checking the trace Hermitian inner product against all codewords in $\mathcal{C}$. However, it is not necessary to check the entire code. Instead, it is sufficient to compute the inner products with respect to the generators of $\mathcal{C}$, which fully characterize the code.

Because $\mathcal{C}$ is an additive code, it is generated by a finite set of independent generators $\Gamma = \{\gamma_1, \dots, \gamma_m \}$, where $|\Gamma| = \log_2 |\mathcal{C}|$. Importantly, if a vector $\mathbf{u} \in \mathbb{F}_4^{n}$ anticommutes with at least one of the generators $\gamma_i$, then the Pauli string associated with $\mathbf{u}$ will be suppressed by the decoupling sequence generated from the code $\mathcal{C}$ \cite{vezvaee2025demonstrationhighfidelityentangledlogical}.

This can be seen explicitly through a simple example. Suppose $\mathcal{C}$, and equivalently the decoupling group $\mathcal{G}$, has two generators $\gamma_1$ and $\gamma_2$. The decoupling group is then given by
\begin{equation}
\mathcal{G} =\{I, \gamma_1, \gamma_2, \gamma_1\gamma_2\} \ .
\end{equation}
We assume without loss of generality that $\mathbf{u}$ anticommutes with $\gamma_1$ but commutes with $\gamma_2$. Under conjugation by the elements of $\mathcal{G}$, the sign of the Pauli string corresponding to $\mathbf{u}$ evolves as
\begin{equation}
    \{ +1,-1,+1,-1\} \ . 
\end{equation}
When twirling over the group $\mathcal{G}$, all these contributions cancel out, and the corresponding interaction is completely suppressed by the decoupling sequence. The generalization to an arbitrary set of generators $\Gamma$ follows directly from the same principle: if $\mathbf{u}$ anticommutes with any generator in $\Gamma$, then averaging over the decoupling group results in the complete cancellation of the corresponding Pauli term.

We have shown that DD sequences constructed from classical additive codes behave analogously to quantum stabilizer codes. However, we emphasize two key differences. First, unlike quantum stabilizer codes, we do not require the generators of the decoupling group $\mathcal{G}$ to commute with each other. Second, in the context of dynamical decoupling, there is no active "correction" step. Instead, the twirling procedure passively removes all components of the Hamiltonian that anticommute with at least one of the generators of $\mathcal{G}$. In this sense, the role of the decoupling sequence is similar to that of an \textit{error detection code} rather than an error correction code. 

The connection between designing decoupling sequences and designing error detection codes is a key insight of our work. Specifically, if the dominant residual interactions in the system are known, we can forego a universal decoupling group $\mathcal{G}$, which suppresses all interactions, and instead remove certain generators from $\mathcal{G}$ while still suppressing the dominant residual terms \cite{paz2013optimally}. The resulting DD sequence length becomes significantly shorter. We will make extensive use of this idea in Sec.~\ref{section: crosstalk superconducting qubit} and~\ref{sec: application-spins}, where we tailor our general method to specific cases.

As an example, the orthogonal array presented in Fig.~\ref{fig: expander quotient graph} that universally decouples all 2-local interactions is generated by the decoupling group
\begin{equation}
    \mathcal{G} = \langle X_1 I_2 X_3,X_1 Y_2 Z_3, Y_1 I_2 Y_3, Y_1 Z_2 X_3  \rangle
\end{equation}
However, if we know that the residual two-qubit interactions are restricted to terms of the form $X_i X_j,Y_i Y_j,$ or $Z_i Z_j$, as in spin qubits \cite{burkard2023semiconductor,Xue2022,noiri2022fast,mills2022two}, the reduced decoupling group  
\begin{equation}
    \label{eq: reduced decoupling group}
    \mathcal{G}_{\rm red} = \langle X_1 Y_2 Z_3, Y_1 Z_2 X_3  \rangle
\end{equation}
still suppress all the relevant residual interactions. Explicitly, the elements of $\mathcal{G}_{\rm red} $ are
\begin{equation}
    \mathcal{G}_{\rm red} = \{ I_1 I_2 I_3, X_1 Y_2 Z_3,Y_1 Z_2 X_3, Z_1 X_2 Y_3  \} \ .
\end{equation} 
We note that this group corresponds to running the standard $\rm XY4$ sequence in parallel on all three qubits, with the $\rm XY4$ cycles on the second and third qubits being cyclically shifted relative to each other. Moreover, the reduced group $\mathcal{G}_{\rm red} $ still universally decouples all single-qubit terms. As a result, the DD sequence requires only $2^2 =4$ time steps instead of $2^4 = 16$, achieving a fourfold reduction in cycle length. 

Another interesting consequence of our error-detection perspective is that it enables the construction of decoupling sequences that selectively suppress certain interactions while preserving others \cite{nguyen2025single, evert2024syncopated}. For instance, in systems where anisotropic exchange interactions arise, such as spin qubits with strong spin-orbit coupling \cite{Geyer2024,nguyen2025single,saezmollejo2024exchangeanisotropiesmicrowavedrivensinglettriplet}, the decoupling sequence generated by Eq.~\eqref{eq: reduced decoupling group} preserves specific components of the antisymmetric Dzyaloshinskii–Moriya (DM) interaction. In particular, it maintains terms of the form:
\begin{equation}
\{ X_{1} Y_2, Y_1 Z_2, Y_2 Z_3, Z_2 X_3, X_1 Z_3, Y_1 X_3 \}.
\end{equation}
By alternating between DD sequences derived from $\mathcal{G}_{\rm red}$ and suitably permuted variants of it, one can fully preserve the complete DM interaction, thus enabling new possibilities for quantum simulation \cite{fariña2025siteresolvedmagnontriplondynamics,Hsiao}.

We will discuss more examples  of sequences simulating certain Hamiltonians in Sec.~\ref{section: digital-analog simulation}. In the next section, we introduce a simple search algorithm to construct selective DD sequences.

Finally, we remark that if one uses a quantum stabilizer code (i.e a self-dual additive code) to construct a DD sequence in parallel with employing the same code for active quantum error correction, it is possible to improve the overall quantum code performance by suppressing high-weight errors. This idea was explored in Refs.~\cite{paz2013optimally,vezvaee2025demonstrationhighfidelityentangledlogical}, where the authors designed decoupling sequences based on the stabilizers and logical operators of a quantum code to suppress high-weight error strings. We do not directly compare our sequences with those proposed in Refs.~\cite{vezvaee2025demonstrationhighfidelityentangledlogical,paz2013optimally}, as they serve fundamentally different purposes. The sequences in their work scale exponentially with the quantum code (system) size but are used alongside a quantum error-correcting code, such that errors that accumulate due to the long sequence length can be actively corrected. In contrast, our sequences are designed to protect bare physical qubits without additional quantum error correction. An interesting direction for future work is combining our approach with that of Refs.~\cite{paz2013optimally,vezvaee2025demonstrationhighfidelityentangledlogical}, to explore trade-offs between sequence length and robustness against errors in the presence of a quantum stabilizer code.

\subsection{Selective DD sequences}
\label{section: selective DD sequences}

We now apply our formalism to selectively suppress only particular interactions, rather than to completely decouple the system from the environment or freeze the system dynamic. To do so, we assume that the system Hamiltonian $H_{S}$ can be decomposed into two parts:
\begin{equation}
    H_{S} =  H_{S}^{\parallel}+ H_{S}^{\perp} \ ,
\end{equation}
where $H_{S}^{\parallel}$ contains the interaction terms we wish to preserve and $H_{S}^{\perp}$ contains the terms we aim to suppress. This task is particularly relevant for applications such as Hamiltonian simulation, where it is desirable to selectively engineer specific interactions \cite{Choi2020}.

Based on the connection between DD sequences and error-detection codes, we can construct a simple algorithm that finds the decoupling group performing this task. Our method explicitly searches for generators of the DD group that suppresses the undesired terms in $H_{S}^{\perp}$ while preserving the desired terms in $H_{S}^{\parallel}$. 

As a first step, we identify a set of generators that commute with all terms in $H_{S}^{\parallel}$. This is efficiently achieved by representing each Pauli string in $H_{S}^{\parallel}$ as a binary vector in $\mathbb{F}_{2}^{2n}$ via the symplectic representation
\begin{equation}
    P = \left(\prod_{i=1}^{n} X_i^{x_i} Z_i^{z_i}\right)    ~\to ~ \begin{pmatrix}
        \vec{x} ~ | ~\vec{z}
    \end{pmatrix}^{T} \quad \text{ with } \vec{x},\vec{z} \in \mathbb{F}_{2}^{n}. 
\end{equation}
In the symplectic representation, the commutation relation between two Pauli strings $\bf{u}$ and $\bf{v}$ is determined by their symplectic inner product
\begin{equation}
    \bf{u}^{T} \Omega \bf{v}, \quad  \text{ where } \quad  \Omega = \begin{pmatrix}
        0_{n \times n} & I_{n \times n}  \\
        I_{n \times n} & 0_{n \times n}
    \end{pmatrix}.
\end{equation}
The two strings commute if and only if this inner product is zero.

Let $\{ \textbf{h}_{i}^{\parallel} \}_{i=1}^{M}$ be the binary vector representations of the Pauli strings in $H_{S}^{\parallel}$, where $M$ is the number of strings. The set of binary vectors (and corresponding Pauli strings) that commute with all elements of $H_{S}^{\parallel}$ forms the null space of the matrix
\begin{equation}
    \label{eq: nullspace of HS}
    \begin{pmatrix}
        (\textbf{h}_1^{\parallel})^{T} \\ \vdots \\
        (\textbf{h}_M^{\parallel})^{T}
    \end{pmatrix} \bf{\Omega}.
\end{equation}
This null space can be efficiently computed via standard Gaussian elimination over $\mathbb{F}_2$. 

Let $\{ \bf{n}_i \}$ denote the binary vector generators of the null space from Eq.~\eqref{eq: nullspace of HS}. To construct a minimal decoupling sequence that suppresses $H_{S}^{\perp}$, we form a bipartite graph whose two partitions consist of the null space generators $\{ \bf{n}_i \}$ and the Pauli terms $\{ \bf{h}_{i}^{\perp} \}$ from $H_{S}^{\perp}$. An edge is drawn between $\bf{n}_i$ and $\bf{h}^{\perp}_j$ if and only if they anti-commute, i.e.
\begin{equation}
    \bf{n}^{T}_i~\bf{\Omega} ~ \bf{h}_{j}^{\perp} = 1 \mod 2. 
\end{equation}
Constructing this graph requires computing all pairwise symplectic inner products, with a total complexity of $ O(|\{ \bf{n}_i \} | |\{ \bf{h}_{i}^{\perp} \}|)$. 

Given the bipartite graph, we numerically identify the minimal set of generators $\{ \bf{n}_i \}$ that preserves $ H_{S}^{\parallel}$ while decoupling $H_{S}^{\perp}$. This reduces to the classical minimum set cover problem: finding the smallest subset of $\{ \bf{n}_i \}$ such that every $\bf{h}_{i}^{\perp}$ anticommutes with at least one selected generator. While the minimal set cover is generally a NP-hard problem, this problem remains tractable in simple cases, where we assume the chromatic number $\chi_I$ is small, an assumption typically valid for planar devices with short range interactions. We remark that a minimum covering set does not always exist (i.e., some terms in $H^{\perp}_{S}$ may be product of terms in $H^{||}_{S}$). In such cases, one can instead employ the integer programming method introduced in Ref.~\cite{Choi2020}.

We emphasize that this algorithm is primarily a proof of concept rather than a fully optimized solution. Nonetheless, it shows that the error-detection perspective can drastically reduce the search space by focusing only on the generators of the decoupling group instead of the entire sequence, generally enabling a more efficient and automated construction of tailored decoupling sequences. This viewpoint could also potentially benefit machine learning–based approaches \cite{tong2024empirical,rahman2024learning} to designing robust DD sequences.

\section{Applications to quantum computing}
\label{section: application to QC}
In the previous sections, we explained how to construct time-efficient DD sequences for arbitrary $k$-local Hamiltonians using tools from classical coding theory. In this section, we present concrete applications of our framework based on well-known families of classical codes. We postpone the discussion of sequence robustness against pulse imperfections to Sec.~\ref{section: robust pulse sequences}. In Sec.~\ref{section: crosstalk superconducting qubit}, we show how to construct DD sequences tailored to superconducting qubit architectures using Reed–Muller codes. In Sec.~\ref{sec: application-spins}, we develop DD sequences that suppress general two- and three-local errors using projective geometry codes, and demonstrate how to adapt these constructions to spin-qubit platforms. Finally, in Sec.~\ref{section: digital-analog simulation}, we show how our framework can be used to simulate Kitaev's honeycomb model starting from the isotropic exchange Hamiltonian native to spin-qubit platforms.

\begin{figure}
    \centering
    \includegraphics[width=0.8\linewidth]{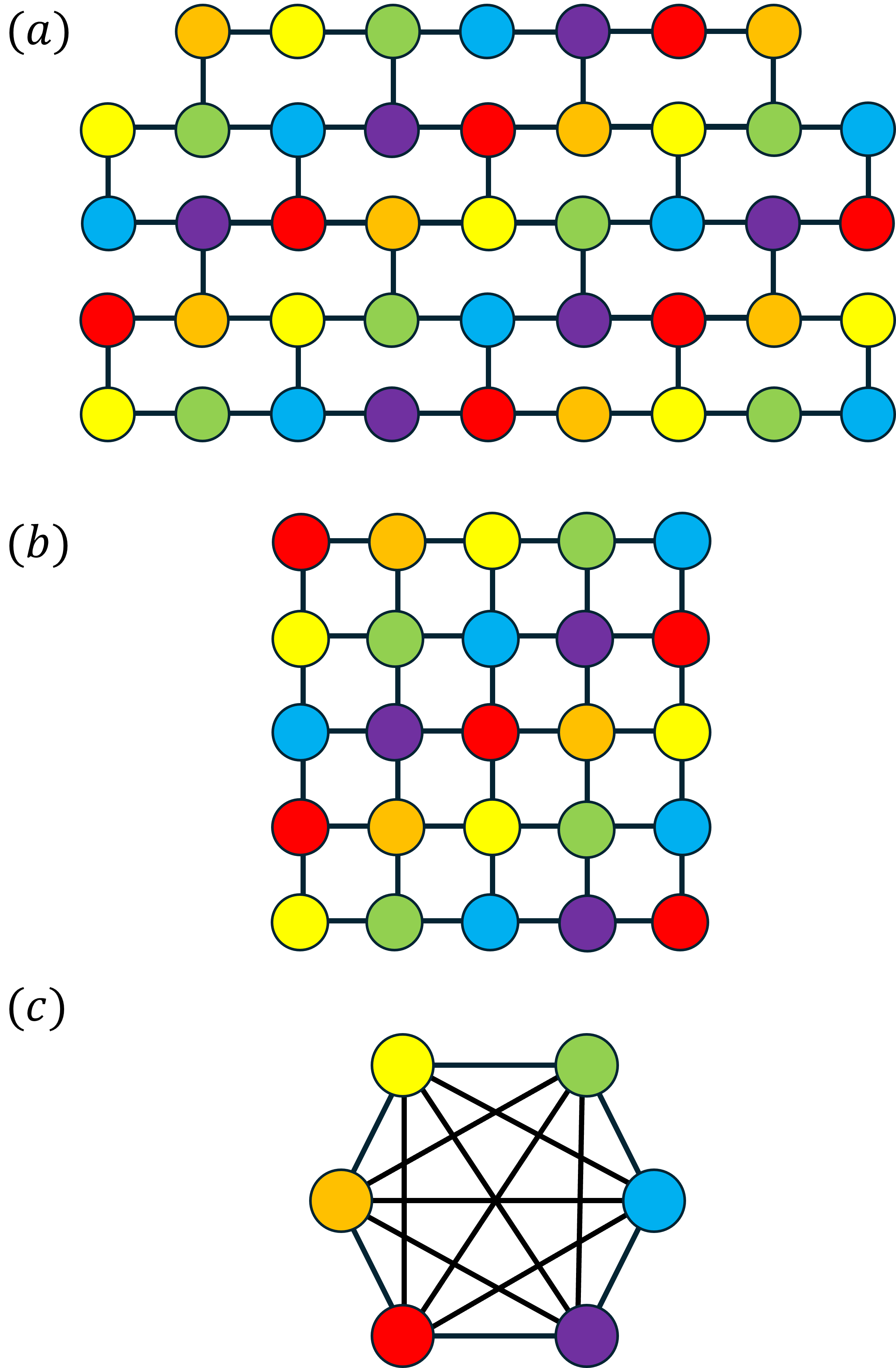}
    \caption{\justifying \textbf{Colored graphs of common superconducting qubit architectures.} We assume a noise model where each qubit experiences both longitudinal coherence errors ($Z$ errors) and transversal coherence errors ($X$ and $Y$ errors), along with residual $ZZ$ interactions between nearest- and next-nearest-neighbors, and $ZZZ$ interactions among any three consecutive qubits. The colorings also apply when residual $ZZ$ interactions extend to next-nearest neighbors. (a) Heavy-hex lattice with $\chi_I = 6$ colors. (b) Square lattice with $\chi_I = 6$ colors. (c) The resulting quotient hypergraph is identical for both architectures.  }
    \label{fig: hex-square-folded}
\end{figure}

\subsection{Residual $ZZ$ and $ZZZ$ coupling in superconducting qubits}
\label{section: crosstalk superconducting qubit}

\subsubsection{Suppressing $ZZZ$ coupling}
In superconducting platforms, suppressing residual $ZZ$ and $ZZZ$ couplings between neighboring qubits is a key challenge for scalability, as these interactions are a major source of coherent errors~\cite{Pederson2019,Menke2022,berke2022transmon}. In this section, we show that by combining the hypergraph coloring procedure introduced earlier with binary Reed–Muller codes, one can construct efficient DD sequences tailored to superconducting qubits. Specifically, we introduce a bang-bang control sequence with length scaling as $O(\chi_I)$. In Appendix~\ref{appendix: superconducting qubit (bounded control)}, we extend this result to bounded-strength control, yielding a DD sequence with length scaling as $O(\chi_I \log \chi_I)$. 

As a concrete example, Fig.~\ref{fig: hex-square-folded}(a) and Fig.~\ref{fig: hex-square-folded}(b) show the color partitions using $\chi_I =6$ for the heavy-hex lattice and the square lattice, used by IBM \cite{ibm_heavy_hex_2021} and Google \cite{Acharya2025} respectively. In this section, we assume a noise model in which each qubit is subject to both longitudinal and transversal coherence errors, along with residual $ZZ$ interactions between neighboring qubits and $ZZZ$ interactions across any three consecutive qubits. The presented color partitions also apply when considering residual $ZZ$ interactions between both nearest and next-nearest neighbors. Fig.~\ref{fig: hex-square-folded}(c) shows the corresponding quotient hypergraph induced by the color partition. We emphasize that both the heavy-hex and square lattices yield the same quotient hypergraph under this partitioning. In the following, we briefly review the Reed–Muller code and outline how to construct tailored DD sequences for superconducting qubits based on it.

The Reed-Muller code is one of the oldest and simplest families of linear binary codes with well-understood parameters and dual structure \cite{abbe2020reed}. A Reed-Muller code $\rm RM(r,m)$ is a binary code with parameters 
\begin{equation}
    (n,|\mathcal{C}|,d)_q = (2^{m},2^{\sum_{i=0}^{r} \binom{m}{i} },2^{m-r})_2
\end{equation}
The dual of $\rm RM(r,m)$ is also a Reed-Muller code, specifically $\rm RM(m-r-1,m)$. The generators $\{\gamma_i \}$ of the Reed-Muller code $\rm RM(r,m)$ can be constructed using Boolean polynomials of degree at most $r$. Each generator $\gamma_i$ corresponds to a monomial of the form
\begin{equation}
    \gamma_i \to f_{\gamma_i}(\boldsymbol{x}) = x_{i_1} x_{i_2} \dots x_{i_s},
\end{equation}
where $\boldsymbol{x} = (x_1,\dots,x_m) \in \mathbb{F}_{2}^{m}$, $ 0 \leq s \leq r$, and the total degree of $f_{g_i}$ is at most $r$. There are exactly 
\begin{equation}
    \sum_{i=0}^{r} \binom{m}{i} 
\end{equation}
linearly independent monomials, corresponding to the dimension of the code. Each generator $\gamma_{i}$ is a binary vector of length $2{^m}$ obtained by evaluating the associated polynomial $f_{\gamma_i}(\boldsymbol{x}) $ at all $2^{m}$ possible bit strings $\boldsymbol{x} \in \mathbb{F}_{2}^{m}$. 

We will use the Reed-Muller code $\mathcal{C} = \rm RM(1,m)$ to construct our DD sequence. The dual code of $\rm RM(1,m)$ is $\mathcal{C}^{\perp} = \rm RM(m-2,m)$, consequently, it has the dual distance 
\begin{equation}
    d^{\perp} = 2^{m-(m-2) } = 4. 
\end{equation}
If we replace each codeword in $\mathcal{C}$ by substituting $0$ with the identity $I$ and $1$ with $X$, then Theorem~\ref{theorem: OA and classical code} ensures that the resulting DD sequence suppresses all $Z$-type interactions with weight up to $d^{\perp}-1= 3$. 

The generators of $\rm RM(1,m)$ correspond to all linear Boolean functions over $\mathbb{F}_{2}^{m}$, i.e., the constant function $f_{\gamma_0} = 1$ and the monomials $f_{\gamma_i} = x_{i}$ with $i=1,\dots,m$. As an example, Table~\ref{tab: generator-RM13-binary} lists the generators of $\rm RM(1,3)$ as rows, corresponding to the chromatic number $\chi_I = 2^{3}=8$. The generator $\gamma_0$ (first row) associated with the constant function $f_{\gamma_0} = 1$ can suppress all single-qubit $Z_i$ terms as well as any three-qubit $Z_i Z_j Z_k$ terms. In addition, the other generators $\gamma_{i \geq 1}$ suppress two-qubit $Z_i Z_j$ interactions between qubits with distinct colors. Importantly, since the heavy-hex and square lattices in Fig.~\ref{fig: hex-square-folded} require only $\chi_I= 6$ colors, the DD sequence derived from Table~\ref{tab: generator-RM13-binary} is directly applicable to these architectures.

\begin{table}[h]
    \centering

    \begin{subtable}[t]{0.32\textwidth}
    \centering
    \begin{tabular}{|c|c|c|c|c|c|c|c|}
        \hline
         $C_1$ & $C_2$  & $C_3$  &  $C_4$  & $C_5$  & $C_6$  & $C_7$  & $C_8$ \\
        \hline
         1 & 1  & 1  & 1  & 1  & 1  & 1  & 1 \\
         \hline
         1 & 1  & 1  & 1  & 0  & 0  & 0  & 0 \\
         \hline
         1 & 1  & 0  & 0  & 1  & 1  & 0  & 0 \\
         \hline
         1 & 0  & 1  & 0  & 1  & 0  & 1  & 0  \\
         \hline
    \end{tabular}
    \caption{\justifying Binary generator matrix of $\mathrm{RM}(1,3)$.}
    \label{tab: generator-RM13-binary}
    \end{subtable}
    \hfill

    \begin{subtable}[t]{0.32\textwidth}
    \centering
    \begin{tabular}{|c|c|c|c|c|c|c|c|}
        \hline
         $C_1$ & $C_2$  & $C_3$  &  $C_4$  & $C_5$  & $C_6$  & $C_7$  & $C_8$ \\
        \hline
         X & X  & X  & X  & X  & X  & X  & X \\
         \hline
         X & X  & X  & X  & Z  & Z  & Z  & Z \\
         \hline
         X & X  & I  & I  & X  & X  & I  & I \\
         \hline
         Y & Z  & Y  & Z  & Y  & Z  & Y  & Z  \\
         \hline
    \end{tabular}
    \caption{\justifying Universal modified generator with Pauli substitutions.}
    \label{tab: generator-RM13-universal}
    \end{subtable}
    \hfill

    \begin{subtable}[t]{0.32\textwidth}
    \centering
    \begin{tabular}{|c|c|c|c|c|c|c|}
        \hline
         $C_1$ & $C_2$  & $C_3$  &  $C_4$  & $C_5$  & $C_6$  & $C_7$   \\
         \hline
         X & X  & X  & X  & Z  & Z  & Z  \\
         \hline
         X & X  & I  & I  & X  & X  & I   \\
         \hline
         Y & Z  & Y  & Z  & Y  & Z  & Y  \\
         \hline
    \end{tabular}
    \caption{\justifying Punctured universal modified generator after removing $C_8$.}
    \label{tab: generator-RM13-punctured}
    \end{subtable}

    \caption{\justifying \textbf{Suppressing $ZZ$ and $ZZZ$ interactions.}  Summary of the generator matrices based on the Reed-Muller $\mathrm{RM}(1,3)$ code. (a) Binary generator. (b) Universal modified generator with Pauli substitutions for universal decoupling. (c) Punctured version optimized for two-body interactions with one fewer color. }
    \label{tab: generator-RM13-grouped} 
\end{table}

The DD sequence constructed from the $ \rm RM(1,m)$ code using the mapping $0 \to I$ and $1 \to X$ can only suppress $Z$-type interactions. However, to universally suppress all single-qubit terms in addition to residual $Z_i Z_j$ and $Z_i Z_j Z_k$ couplings, a simple modification suffices. The key idea is that the mapping can be generalized by assigning "$0$" to $I$ or $Z$, and "$1$" to either $X$ or $Y$. The resulting decoupling sequence still removes all $Z$-type interactions of weight up to three.

To universally decouple single-qubit terms, we require that the support of the generators on each color class must include at least two distinct Pauli operators from $\{X,Y,Z \}$. In other words, each column in Table~\ref{tab: generator-RM13-binary} must contain at least two different Pauli matrices. This condition can always be satisfied for the generators of $\rm RM(1,m)$ by appropriately applying the substitution rule discussed above. Table~\ref{tab: generator-RM13-universal} provides an example of the modified substitution applied to Table~\ref{tab: generator-RM13-binary}. Consequently, the length $L$ of the decoupling sequence constructed based on the Reed-Muller codes is given by 
\begin{equation}
    L = 2^{\lceil \log_2(\chi_I) \rceil  + 1} \sim O(\chi_I) \ . 
\end{equation}

In practical terms, this means that for the near-term architectures shown in Fig.~\ref{fig: hex-square-folded}, a tailored bang-bang decoupling sequence of length $L=16$ is sufficient to protect superconducting qubit systems against 3-local interactions. In contrast, as we will show later, using a universal DD sequence that suppresses all 3-local interactions requires a significantly longer sequence of length $L=64$, a fourfold increase.

\subsubsection{Tailored sequence to only $ZZ$ couplings}

Our DD sequence based on Reed-Muller codes can universally suppress single-qubit terms, as well as two-qubit  $Z_i Z_j$ and three-qubit $Z_i Z_j Z_k$ interactions. However, if three-qubit interactions are negligible or absent, we can further improve the length of the sequence. The improvement originates from two factors. First, restricting to two-body interactions reduces the chromatic number $\chi_I$ of the interaction hypergraph. Second, we can drop one generator from the $\rm RM(1,m)$ code, effectively halving the length of the decoupling sequence.

The generator we remove is $\gamma_0$ which, as previously discussed, suppresses all weight-three $Z$-type terms. As a consequence, the final color class $C_{2^{m}}$ (corresponding to the last column of Table~\ref{tab: generator-RM13-universal}) can no longer suppress single-qubit $Z$ terms. Therefore, we must also remove this color class. Removing color class $C_{2^{m}}$ is equivalent to constructing a punctured Reed-Muller code. However, since we also discard $\gamma_0$ generator, the resulting code is a modified punctured $\rm RM(1,m)$ code with dual distance $d^{\perp}=3$ instead of $d^{\perp}=4$. Consequently, the length of the DD sequence is 
\begin{equation}
    L = 2^{ \lceil \log_2(\chi_I + 1) \rceil  } \sim O(\chi_I) \ .
\end{equation}

The resulting decoupling sequence universally suppresses single-qubit terms and two-qubit $Z_i Z_j$ interactions, while requiring one fewer generator at the cost of one fewer color. In Table~\ref{tab: generator-RM13-punctured}, we present the generators of the decoupling sequence generated from Table~\ref{tab: generator-RM13-universal} by removing the first row and the last column. This modified punctured $\rm RM(1,m)$ construction is, up to minor adjustments, equivalent to the single-axis Hadamard construction introduced in Ref.~\cite{brown2024efficient}.

This means that to protect superconducting qubit systems with the architecture shown in Fig.~\ref{fig: hex-square-folded} against both nearest-neighbor and next-nearest-neighbor $ZZ$ couplings, a tailored bang-bang DD sequence of length $L=8$ suffices. In contrast, as we will show later, a universal DD sequence that suppresses all interactions up to the same range requires a significantly longer sequence of length $L=32$, again a fourfold increase.

In this section, we have shown how to construct tailored bang-bang DD sequences for suppressing residual interactions in superconducting qubits by using binary Reed–Muller codes. In Appendix~\ref{appendix: superconducting qubit (bounded control)}, we extend this construction to bounded-strength control sequences. In the next section, we turn to the spin-qubit platform and introduce tailored decoupling sequences based on quaternary codes.

\subsection{Exchange interactions in spin qubits}
\label{sec: application-spins}
\begin{figure}
    \centering
    \includegraphics[width=0.9\linewidth]{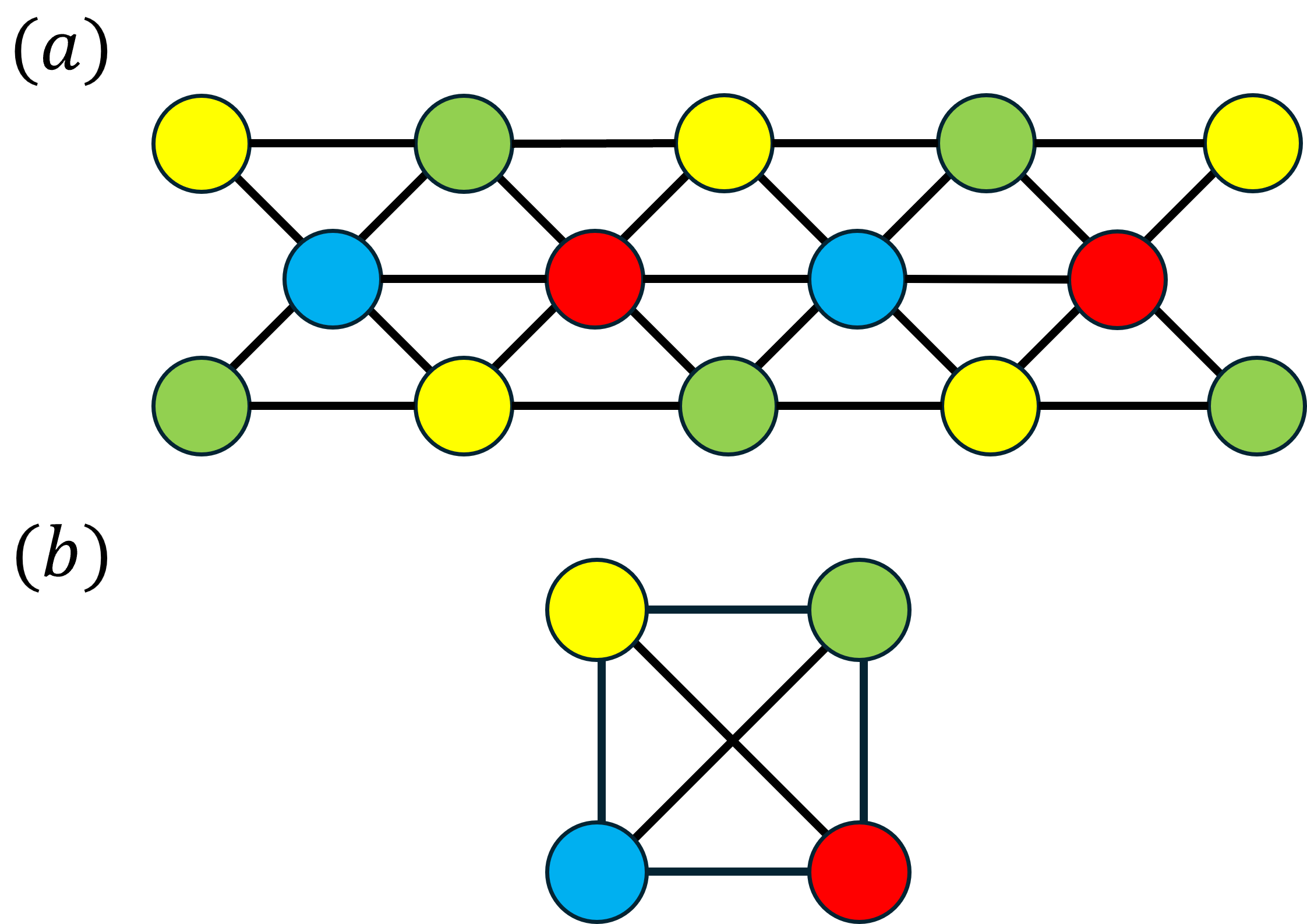}
    \caption{\justifying \textbf{Colored trilinear array architecture in the spin-qubit platform.} We assume a noise model where each qubit undergoes both longitudinal and transversal coherent error, with Heisenberg exchange between neighbors and scalar chirality interactions among any three qubits forming a triangle. (a) The color partition $C[I_{\rm Dev}]$ uses $\chi_I = 4$ colors, tailored to this interaction model. (b) The resulting quotient hypergraph is a square with diagonal links.}
    \label{fig: trilinear-array}
\end{figure}
\begin{table*}[htb!]
\centering
\renewcommand{\arraystretch}{1.2}

\begin{subtable}[t]{\textwidth}
\centering
\begin{tabular}{c c}
    \begin{tabular}{|c c | c c | c c | c c | c c|}
    \toprule
    \multicolumn{2}{|c|}{$C_1$} &
    \multicolumn{2}{c|}{$C_2$} &
    \multicolumn{2}{c|}{$C_3$} &
    \multicolumn{2}{c|}{$C_4$} &
    \multicolumn{2}{c|}{$C_5$} \\
    \midrule
    1 & 0 & 0 & 0 & 1 & 0 & 1 & 0 & 1 & 0 \\
    0 & 1 & 0 & 0 & 1 & 1 & 0 & 1 & 1 & 1 \\
    0 & 0 & 1 & 0 & 0 & 1 & 1 & 0 & 1 & 1 \\
    0 & 0 & 0 & 1 & 1 & 0 & 0 & 1 & 0 & 1 \\
    \bottomrule
    \end{tabular}
    \hspace{1.5cm}
    \begin{tabular}{|c|c|c|c|c|}
    \hline
    $C_1$ & $C_2$  & $C_3$  & $C_4$  & $C_5$ \\
    \hline
    $1$ & $0$  & $1$  & $1$  & $1$ \\
    $\omega$ & $0$  & $1+\omega$  & $\omega$  & $1+\omega$ \\
    $0$ & $1$  & $\omega$  & $1$  & $1+\omega$ \\
    $0$ & $\omega$ & $1$  & $\omega$  & $\omega$ \\
    \hline
    \end{tabular}
    \hspace{1.5cm}
    \begin{tabular}{|c|c|c|c|c|}
    \hline
    $C_1$ & $C_2$  & $C_3$  & $C_4$  & $C_5$ \\
    \hline
    $X$ & $I$  & $X$  & $X$  & $X$ \\
    $Z$ & $I$  & $Y$  & $Z$  & $Y$ \\
    $I$ & $X$  & $Z$  & $X$  & $Y$ \\
    $I$ & $Z$ & $X$  & $Z$  & $Z$ \\
    \hline
    \end{tabular}
\end{tabular}
\caption{}
\end{subtable}

\vspace{1em}

\begin{subtable}[t]{\textwidth}
\centering

\begin{tabular}{|c c c|c c c|c c c|c c c|c c c|}
\hline
$C_1$ & $C_2$ & $C_3$ & $C_4$ & $C_5$ & $C_6$ & $C_7$ & $C_8$ & $C_9$ & $C_{10}$ & $C_{11}$ & $C_{12}$ & $C_{13}$ & $C_{14}$ & $C_{15}$ \\
\hline
$X$ & $Z$ & $Y$ &
$I$ & $I$ & $I$ &
$X$ & $Z$ & $Y$ &
$X$ & $Z$ & $Y$ &
$X$ & $Z$ & $Y$ \\
$Z$ & $Y$ & $X$ &
$I$ & $I$ & $I$ &
$Y$ & $X$ & $Z$ &
$Z$ & $Y$ & $X$ &
$Y$ & $X$ & $Z$ \\
$I$ & $I$ & $I$ &
$X$ & $Z$ & $Y$ &
$Z$ & $Y$ & $X$ &
$X$ & $Z$ & $Y$ &
$Y$ & $X$ & $Z$ \\
$I$ & $I$ & $I$ &
$Z$ & $Y$ & $X$ &
$X$ & $Z$ & $Y$ &
$Z$ & $Y$ & $X$ &
$Z$ & $Y$ & $X$ \\
\hline
\end{tabular}
\caption{}
\end{subtable}

\caption{\justifying \textbf{Universal suppression of two-local and Heisenberg exchange interactions.} These tables illustrate how to derive DD groups from additive projective geometry codes based on a 1-spread in \( \mathrm{PG}(3,2) \). (a)
Left: A (complete) 1-spread in \( \mathrm{PG}(3,2) \), where each column represents a line described by two projective points in \( \mathbb{F}_2^4 \). Center: The corresponding projective points in \( \mathrm{PG}(3,4) \) constructed from the 1-spread in $\rm PG(3,2)$ using Eq.~\eqref{eq: converting from F2 to F4}. Right: The additive generators of a DD group that suppresses universally suppress two-local interactions constructed from projective points in $\rm PG(3,4)$.  (b) Generators of DD group tailored for spin-qubit systems based on a 1-spread in \( \mathrm{PG}(3,2) \). Each projective point in \( \mathrm{PG}(3,4) \) is expanded into its three equivalent representatives in \( \mathbb{F}_4^4 \), tripling the number of color classes. }
\label{tab: example of additive geometry code from PG(3,2)}
\end{table*}

Dense spin-qubit architectures are challenged by residual exchange interaction \cite{Geyer2024,bosco2024exchangeonlyspinorbitqubitssilicon,nguyen2025single} that cannot be completely turned off. The dominant residual interactions for spin-qubit systems with small spin-orbit interaction are the two-qubit Heisenberg exchange interactions
\begin{equation}
\sigma_{i} \cdot \sigma_j,
\end{equation}
and the scalar chirality interaction \cite{Sen1995} (for perpendicular magnetic fields)
\begin{equation}
\sigma_i \cdot (\sigma_j \times \sigma_k).
\end{equation}

The class of three-local Hamiltonians is particularly important, as it enables the implementation of single-step, high-fidelity three-qubit gates \cite{nguyen2025single, Gullans2019, Jiaan2024}. As a concrete example, we consider a trilinear array architecture \cite{john2024two} designed to support such gates, as illustrated in Fig.~\ref{fig: trilinear-array}(a) \cite{nguyen2025single}. We assume a noise model in which each qubit experiences both longitudinal and transversal coherent error, along with Heisenberg exchange interactions between neighboring qubits and scalar chirality interactions among any trio forming a triangle. Under this interaction model, the array admits a color partition with $\chi_I = 4$ colors. The resulting quotient hypergraph is shown in Fig.~\ref{fig: trilinear-array}(b).

To construct a tailored DD sequence suppressing all interactions up to weight three, we can generate an orthogonal array using the projective geometry code $\mathcal{C}$ and its dual (also known as the simplex code) as $\mathcal{C}^{\perp}$. The code is chosen because it enables large chromatic number $\chi_I$ using minimal number of generators, resulting in a highly time-efficient sequence. 

This section is organized as follows. We begin by briefly reviewing the construction of linear and additive codes based on projective geometry with distances $d=3$ and $d=4$. We then show how to modify these codes to tailor them to the spin-qubit platform.

\subsubsection{Linear projective geometry code}
\label{sec:LPGC}
A projective geometry space $\rm PG(n,q)$ is the set of equivalence classes in the vector space $\mathbb{F}_q^{n+1}/\{ \boldsymbol{0} \}$  of which two vectors $\bf{u},\bf{v} \in \mathbb{F}_{q}^{n+1}$ belong to the same equivalence class $[\bf{u}]$ if they differ by scalar multiplication, i.e. $\bf{u} = \alpha \bf{v}$ for some $\alpha \in \mathbb{F}_q$. Consequently, one can view each point $[\bf{u}] \in \rm PG(n,q)$ as a collection of points in $\mathbb{F}_{q}^{n+1}$ starting from $\bf{u}$ and passing through the origin. 

The projective space that we will use to construct the decoupling group is $\rm PG(n,4)$. The decoupling group is constructed using the linear check matrix $H$ of the linear code $\mathcal{C}$. As a reminder, the codeword $\bf{c} \in \mathcal{C}$ is the kernel of the check matrix, i.e. 
\begin{equation}
    H\bf{c} = \bf{0}.
\end{equation}

As a first step, we show how to construct a decoupling group for universal cancellation of all two-body interactions from the projective space. We begin by choosing the columns of the check matrix $H$ to correspond to points $[\bf{u}]$ in the projective space $\rm PG(n,4)$; that is, to equivalence classes of nonzero vectors in $\mathbb{F}_{4}^{n+1}$. The total number of such equivalence classes is given by
\begin{equation}
    \frac{4^{n+1}-1}{3}.
\end{equation}
From the construction of the check matrix, it follows that the code distance is $d \geq 3$. To see this, we proceed by contradiction and assume that there exists a codeword $\bf{c}$ with Hamming weight two. This implies the existence of two columns $[\bf{u}],[\bf{v}]$ that are linearly dependent, i.e.,
\begin{equation}
    [\bf{u}] = \alpha [\bf{v}],
\end{equation}
for some $\alpha \in \mathbb{F}_4$. However, by definition of projective space, all distinct points $[\bf{u}] \in \rm PG(n,4)$ are pairwise linearly independent. Thus, no such weight-two codeword can exist, and the minimum distance of the code must be at least three. 

Because the check matrix $H$ of the linear code $\mathcal{C}$ is the linear (not additive) generator matrix for the dual code $\mathcal{C}^{\perp}$, using Theorem~\ref{theorem: OA and classical code}, we can construct an $\text{OA}$ from $\mathcal{C}^{\perp}$ with parameters 
\begin{equation}
    \text{OA}\Big(4^{n+1},\chi_I = \frac{4^{n+1}-1}{3},4,2 \Big). 
\end{equation}
As a result, the decoupling sequence derived from this orthogonal array universally decouples all two-local interaction terms.

We now show how to construct DD sequences that universally decouple all three-local interactions using linear projective geometry code. The key requirement is that the columns of the check matrix $H$ must form a cap set. A cap set (also known as an arc in projective geometry) in $\rm PG(n,4)$  is a collection of points $[\bf{u}]$ such that no three are collinear. This condition ensures that the corresponding linear code $\mathcal{C}$ has minimum distance $d=4$, thereby detecting all weight-three errors. 

Let $n_{H}$ denote the number of columns in the check matrix, i.e. the size of the cap set. Then, the orthogonal array constructed from the dual code $\mathcal{C}^{\perp}$ has parameters
\begin{equation}
    \text{OA}(4^{n+1},\chi_I = n_H,4,3). 
\end{equation}

An important remaining question is how to construct a cap set in $\rm PG(n,4)$ with the largest possible size. In general, this is a challenging problem in projective geometry. However, explicit constructions of maximal cap sets are known for small values of $n$ \cite{FU2015}. For example, the largest cap set sizes for $n=2,3,4$ are $n_H = 6,17,41$ respectively. For higher dimensions $n$, near-optimal constructions are known, although it is unlikely  that physically relevant graphs of solid-state devices will require these large chromatic number $\chi_I$.

\subsubsection{Additive projective geometry code}
In the preceding construction, we employed linear projective geometry codes to design universal DD groups for two-local and three-local interactions. We now show that if we use additive projective geometry codes \cite{BLOKHUIS2004161}, we can construct more time-efficient  sequences. The key insight is that the field $\mathbb{F}_{4}$, when considered under addition, is isomorphic to $\mathbb{F}_2 \times \mathbb{F}_2$. Exploiting this structure, Ref.~\cite{BLOKHUIS2004161} demonstrates that the check matrix $H$ of an additive code with parameters $(\chi_I,2^{k},d)_4$ can be constructed from a collection of $\chi_I$ lines in the projective space $\rm PG(2\chi_I-k-1,2)$, such that any $d-1$ lines in the set are linearly independent.

To universally decouple two-local interactions, we require the code $\mathcal{C}$ to have distance $d=3$. Consequently, we need to choose a collection of $\chi_I$ lines in the projective space $\rm PG(2\chi_I-k-1,2)$ such that any $2$ lines in the set are linearly independent. This set is known in projective geometry as a (partial) 1-spread. The maximum logical dimension $k$ is given by 
\begin{equation}
    k = 2\chi_I-h
\end{equation}
where $h$ is the smallest integer number such that $2^{h} \geq 3\chi_I+1$ for even $h$, and $2^{h} \geq 3 \chi_I+5$ for odd $h$. As a result, the orthogonal array constructed from $\mathcal{C}^{\perp}$ with $H$ as the additive generator matrix has parameter
\begin{equation}
    \text{OA}( 2^{ \lceil \log_2 (3\chi_I+1) \rceil \backslash \lceil \log_2 (3\chi_I+5) \rceil  },\chi_I ,4,2).
\end{equation}

As a simple example, Table~\ref{tab: example of additive geometry code from PG(3,2)}(a) lists a full 1-spread of the projective space $\rm PG(3,2)$, where each projective line is represented by two projective points. Each line is further associated with a color class $c_i$. In Table~\ref{tab: example of additive geometry code from PG(3,2)}(b), we map each projective line in 
$\rm PG(3,2)$ to a projective point in $\rm PG(3,4)$ via a simple embedding from $\mathbb{F}_2 \times \mathbb{F}_2$ to $\mathbb{F}_4$
\begin{equation}
    \label{eq: converting from F2 to F4}
    (0,0) \to 0,~(1,0) \to1,~(0,1) \to \omega,~(1,1) \to 1+\omega.
\end{equation}
The resulting rows in the table form generators of a decoupling group capable of suppressing arbitrary two-local interactions.

We note that the above construction uses the same number of color classes and length in the decoupling sequence as the multi-axis protocol introduced in Ref.~\cite{brown2024efficient}. This is not a coincidence. The multi-axis protocol can be understood as implicitly employing an additive code over $\mathbb{F}_4$. This connection arises because the Hadamard matrix defines a binary linear code (the Hadamard code), whose rows or columns correspond to codewords. By grouping these rows into Schur sets \cite{leung2002simulation} and mapping from $\mathbb{F}_2 \times \mathbb{F}_2$ to $\mathbb{F}_4$, one preserves the additive structure of the binary code. Moreover, Ref.~\cite{votto2024universal} shows in their work (under the name of Walsh functions) that with minor modifications, the same construction can be extended to enable universal quantum simulation. Consequently, our framework naturally includes the protocols from Ref.~\cite{brown2024efficient,votto2024universal}, providing an alternative way to understand their underlying structure. 

To universally decouple three-local interactions, we did not find any explicit constructions using additive projective geometry codes in the literature. However, we expect that additive codes could potentially yield slightly more efficient decoupling sequences for three-local interactions as well. We leave the development of such constructions as an interesting direction for future research.

\begin{figure*}[ht]
    \centering

    \begin{subfigure}[b]{0.48\linewidth}
        \centering
        \caption{}
        \includegraphics[width=\linewidth]{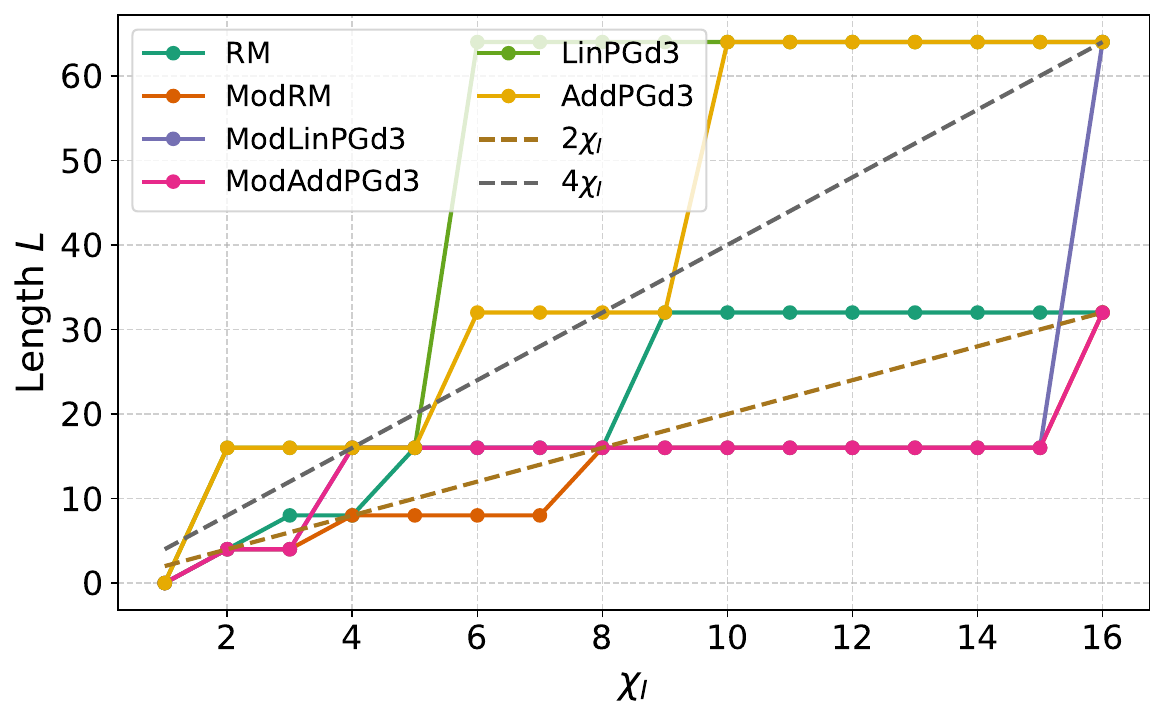}
        \label{fig: code_length_rm_pg}
    \end{subfigure}
    \hfill
    \begin{subfigure}[b]{0.48\linewidth}
        \centering
        \caption{}
        \includegraphics[width=\linewidth]{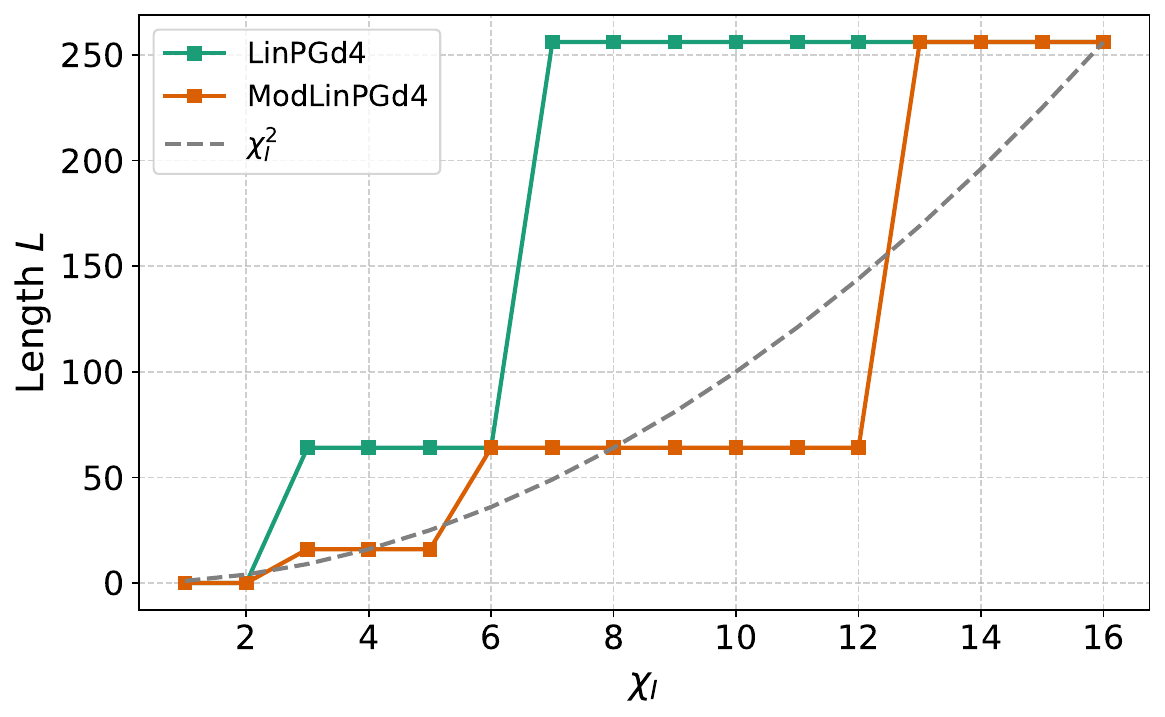}
        \label{fig: code_length_quadratic}
    \end{subfigure}

    \caption{\textbf{Scaling of decoupling sequences.}  Scaling of sequence length $L$ with the chromatic number $\chi_I$ for the code constructions listed in Table~\ref{tab:compact-decoupling}. (a) For comparison, the linear baseline $L = 4\chi_I$ is shown in black. All codes with distance $d = 3$ exhibit linear scaling in $\chi_I$, while those with $d = 4$ show quadratic scaling, consistent with the scaling reported in Ref.~\cite{Bookatz_2016}. }
    \label{fig: code-scaling}
\end{figure*}

\begin{table*}[t]
\centering
\renewcommand{\arraystretch}{1.2}
\begin{tabular}{lcccc}
\hline
\textbf{Scheme} & \textbf{\# Gen.} & \textbf{2-local} & \textbf{3-local} & \textbf{Bounded} \\
\hline
Mod. RM & $\lceil \log_2(\chi_I + 1) \rceil$ & $ZZ$ & -- & Yes \\
RM & $\lceil \log_2(\chi_I) \rceil + 1$ & $ZZ$ & $ZZZ$ & Yes \\
Lin-PG ($d=3$) & $2\log_4(3\chi_I + 1)$ & All & -- & Yes \\
Lin-PG ($d=4$) & $6,8,10$* & All & All & Yes \\
Add-PG ($d=3$) & $\log_2(3\chi_I+1) \backslash \log_2(3\chi_I+5)$ & All & -- & Yes \\
Mod. Lin-PG ($d=3$) & $2\log_4(\chi_I + 1)$ & Hexch. & -- & No \\
Mod. Lin-PG ($d=4$) & $4,6,8,10$** & Hexch. & Chir. & No \\
Mod. Add-PG ($d=3$) & $\log_2(\chi_I+1)\backslash \log_2(\chi_I+5)$ & Hexch. & -- & No \\
\hline
\end{tabular}
\caption{\textbf{Comparison of decoupling sequences.}  “\# Gen.” indicates the number of additive code generators. “Hexch.” = Heisenberg exchange; “Chir.” = scalar chirality; “All” = universal decoupling. Asterisked values * and ** denote number of additive generators for $\chi_I = 6,17,41$ and $\chi_I = 5,12,34,82$ respectively. “Bounded” indicates whether the scheme is compatible with bounded control (all work with bang-bang control).}
\label{tab:compact-decoupling}
\end{table*}

\subsubsection{Tailored sequences for spin-qubit}

We have demonstrated that projective geometry codes enable efficient decoupling of both two-local and three-local interactions, applicable to bang-bang and bounded-control sequences alike. We now present a simplified, tailored construction that universally suppresses single-qubit errors while specifically removing the Heisenberg exchange and scalar chirality interactions, the naturally occurring residual many-body interactions in spin-qubit platforms.

Let us first focus on the two-body Heisenberg exchange interactions. The main idea is to modify the columns of the check matrix such that no codeword corresponds to a term in the Heisenberg exchange interaction. Specifically, no weight-two codeword of the form $[1,1],[\omega,\omega],[1+\omega,1+\omega]$ is allowed. In the linear projective geometry code construction, only one representative for each projective equivalence class $[\bf{u}]$ is used as a column in the check matrix. As an example, the three equivalent representations of $[(1,\omega)]$ are
\begin{equation}
    [(1,\omega)] = \{ (1,\omega),(\omega,1+\omega),(1+\omega,1) \}
\end{equation}
The key observation is that if we include all projective points in the equivalence class $[\bf{u}]$ as separate columns of the check matrix $H$, no Heisenberg exchange terms arise as codewords. This effectively increases the number of allowed colors $\chi_I$ in the decoupling sequence by three times.

A similar strategy also applies to additive projective geometry codes: for each line in the projective space, there exist three distinct representations that also avoid Heisenberg exchange terms as codewords.  As an illustration, Table~\ref{tab: example of additive geometry code from PG(3,2)}(c) presents a tailored DD group designed to suppress Heisenberg exchange interactions, derived from the construction in Table~\ref{tab: example of additive geometry code from PG(3,2)}(b). Notably, this construction increases the number of color classes from 5 to 15 while retaining the same number of generators. In the case of the multi-axis protocol introduced in Ref.~\cite{brown2024efficient}, this modification is equivalent to cyclically permuting the rows in the Schur set. 

When applying this straightforward extension to Fig.~\ref{fig: trilinear-array}(b), under the assumption of residual Heisenberg exchange interactions, the tailored decoupling sequence still requires four generators, matching the number needed for a universal sequence that suppresses all two-local interactions. Therefore, for $\chi_I =4,5$ this basic approach does not yield a shorter sequence. However, in Appendix~\ref{appendix: tailored sequence for two-local spin}, we present optimized tailored sequences for these specific cases that require only three generators, offering a reduction in sequence length.

Through exhaustive numerical search, we confirm that this extended construction yields a near-optimal number of color classes for decoupling at fixed sequence lengths $L = 16,32,$ and $64$. However, we note that shorter sequences, such as those of length $L = 4$ (i.e. Eq.~\eqref{eq: reduced decoupling group}) and $L=8$ are also possible, albeit supporting fewer color classes. Such compact sequences can be systematically constructed by iteratively removing one generator from DD group with longer sequence length $L$ at a time and reducing the number of color classes, while ensuring that the resulting DD group still suppresses arbitrary single-qubit errors and detects Heisenberg exchange interactions.

When including the three-body scalar chirality interaction, a similar strategy can be applied to increase the number of allowed colors in the decoupling sequence. However, the additional constraints imposed by this interaction limit the modification of the projective geometry code’s columns to only doubling the color count $\chi_I$. 

Let us apply this idea to the quotient hypergraph in Fig.~\ref{fig: trilinear-array}(b). We begin by constructing a cap set in $\rm PG(2,4)$, which corresponds to a universal DD sequence that decouples all three-local interactions among six qubits. This cap set is given by the additive generators of the self-dual hexacode \cite{conway2013sphere}
\begin{equation}
    \begin{pmatrix}
        1 & 0 & 0 & 1 & 1 & \omega \\
        0 & 1 & 0 & 1 & \omega & 1 \\
        0 & 0 & 1 & \omega & 1 & 1 \\
        \omega & 0 & 0 & \omega & \omega & 1+\omega \\
        0 & \omega & 0 & \omega & 1+\omega & \omega \\
        0 & 0 & \omega & 1+\omega & \omega & \omega 
    \end{pmatrix}.
\end{equation}
To construct a tailored sequence, we remove the first three columns and focus on the remaining three. We then apply the expansion strategy by selecting one of these columns (e.g. the last) and multiplying it by $\omega$ to generate a new fourth column
\begin{equation}
    \begin{pmatrix}
         1 & 1 & \omega & 1+\omega \\
         1 & \omega & 1 & \omega \\
        \omega & 1 & 1 & \omega  \\
         \omega & \omega & 1+\omega & 1 \\
         \omega & 1+\omega & \omega & 1+\omega \\
         1+\omega & \omega & \omega & 1+\omega 
    \end{pmatrix} \to  \begin{pmatrix}
         X & X & Z & Y \\
         X & Z & X & Z \\
        Z & X & X & Z  \\
         Z & Z & Y & X \\
         Z & Y & Z & Y \\
         Y & Z & Z & Y 
    \end{pmatrix}.
\end{equation}
The resulting DD sequence suppresses both the Heisenberg exchange interactions and the scalar chirality terms. However, it stills contains six generators. By mapping the generator set to a minimum covering problem, as described in Sec.~\ref{section: selective DD sequences}, we find that only four of these generators are necessary to suppress all relevant interaction terms
\begin{equation}
     \begin{pmatrix}
         X & X & Z & Y \\
         X & Z & X & Z \\
         Z & Z & Y & X \\
         Z & Y & Z & Y 
    \end{pmatrix}.
\end{equation}

Without going into detail, we also report that applying the same expansion strategy to one of the unused original columns (e.g., the first column) allows us to increase the number of interaction colors to $\chi_I=5$ while still requiring only four generators
\begin{equation}
     \begin{pmatrix}
         Z & X & X & Z & Y \\
         Y & Z & X & X & Z \\
         Y & Z & Z & Y & X \\
         X & Y & Z & Z & Y
    \end{pmatrix}.
\end{equation}
These optimized sequences achieves the same suppression with only $L=16$ pulses, compared to a universal DD sequence of length $L=64$, yielding a fourfold reduction in sequence length.

We emphasize that the expansion approach is effective primarily for bang-bang sequences. Under bounded control, finite pulse durations can lead to Heisenberg exchange or scalar chirality interactions transforming into other three-local terms that the sequence may not suppress. In such cases, alternative constructions may be required to obtain tailored, time-efficient bounded-control DD sequences. Alternatively, one may resort to the universal DD sequences discussed in Sec.~\ref{sec:LPGC}, which, although potentially longer, suppress all three-local interactions. Finally, we note that shorter sequences can often be obtained by systematically removing generators from a universal sequence, one at a time, as demonstrated above. This process might reduces the number of supported color classes but preserving the required decoupling properties. Developing such compact, tailored sequences remains an interesting direction for future work.

For completeness, we summarize the parameters of all our constructions in Table~\ref{tab:compact-decoupling}, along with their applicability to either bang-bang sequences or bounded-control sequences. Since all the sequences we consider permit universal DD, we only specify which types of two-local and three-local interactions are eliminated by each construction. 

We also analyze the sequence lengths as a function of the chromatic number $\chi_I$. To visualize this,  in Fig.~\ref{fig: code_length_rm_pg}  we show the scaling of different DD sequences that suppress two-local interactions as the chromatic number $\chi_I$ increases; in Fig.~\ref{fig: code_length_quadratic}, we focus on sequences targeting three-local interactions. In both cases, we compare the scaling behavior of the sequence lengths against ideal linear and quadratic scaling. For two-local interactions, we find that the universal constructions scale linearly with sequence length $4\chi_I$, while the tailored constructions scale more compactly with $2 \chi_I$. In the case of three-local interactions, both universal and tailored constructions exhibit quadratic scaling $L\propto \chi_I^2$, but the tailored sequences have a smaller prefactor, leading to more efficient implementations. These scaling behaviors are consistent with those of Ref.~\cite{Bookatz_2016}. 

\FloatBarrier
\subsection{Digital-analog simulation}
\label{section: digital-analog simulation}
\begin{figure}
    \centering
    \begin{subfigure}[b]{0.6\linewidth}
        \centering
         \caption{}
        \includegraphics[width=\linewidth]{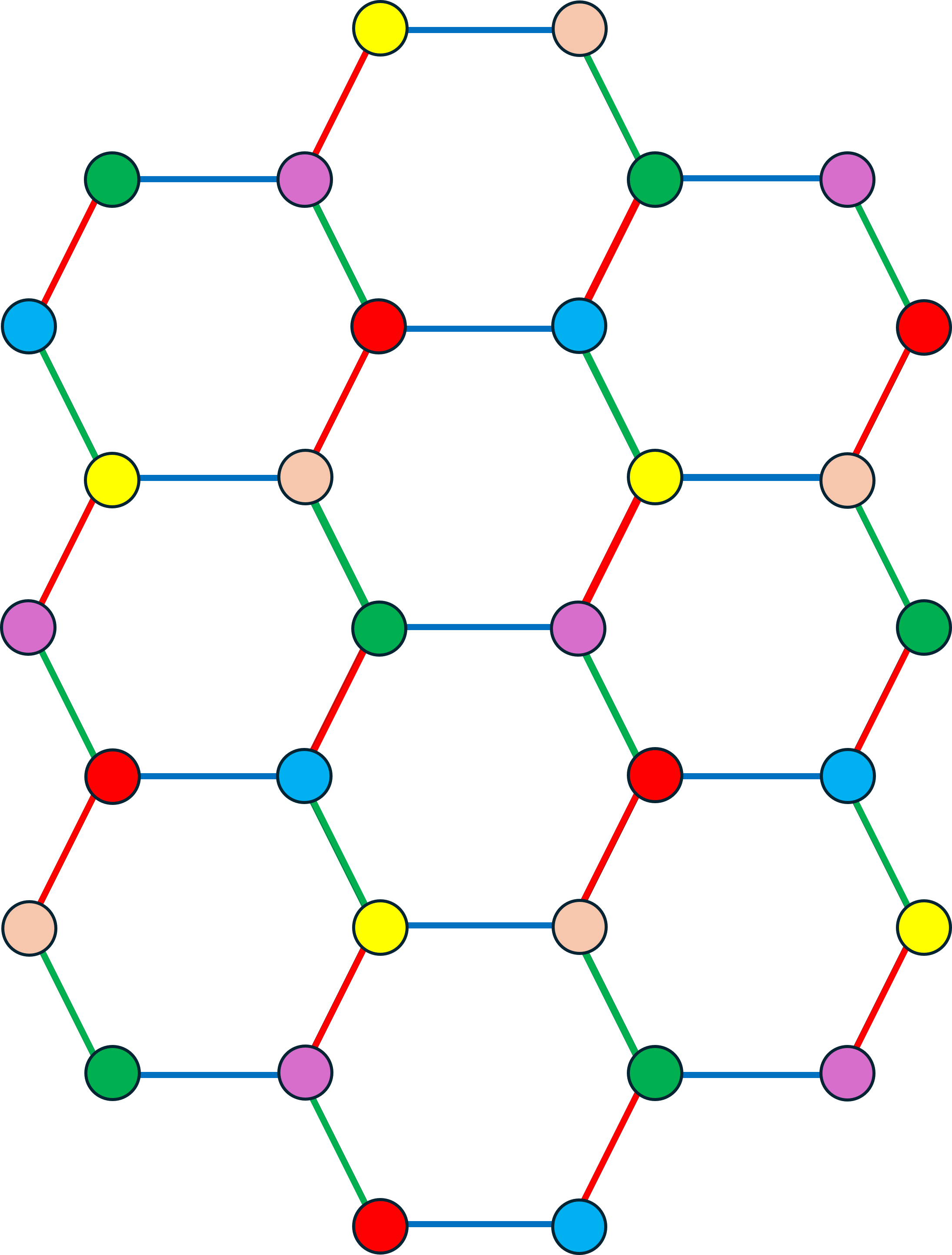}   
        \label{fig: kitaev hexagon}
    \end{subfigure}
    \hfill
    \begin{subfigure}[b]{0.6\linewidth}
        \centering
         \caption{}
        \includegraphics[width=\linewidth]{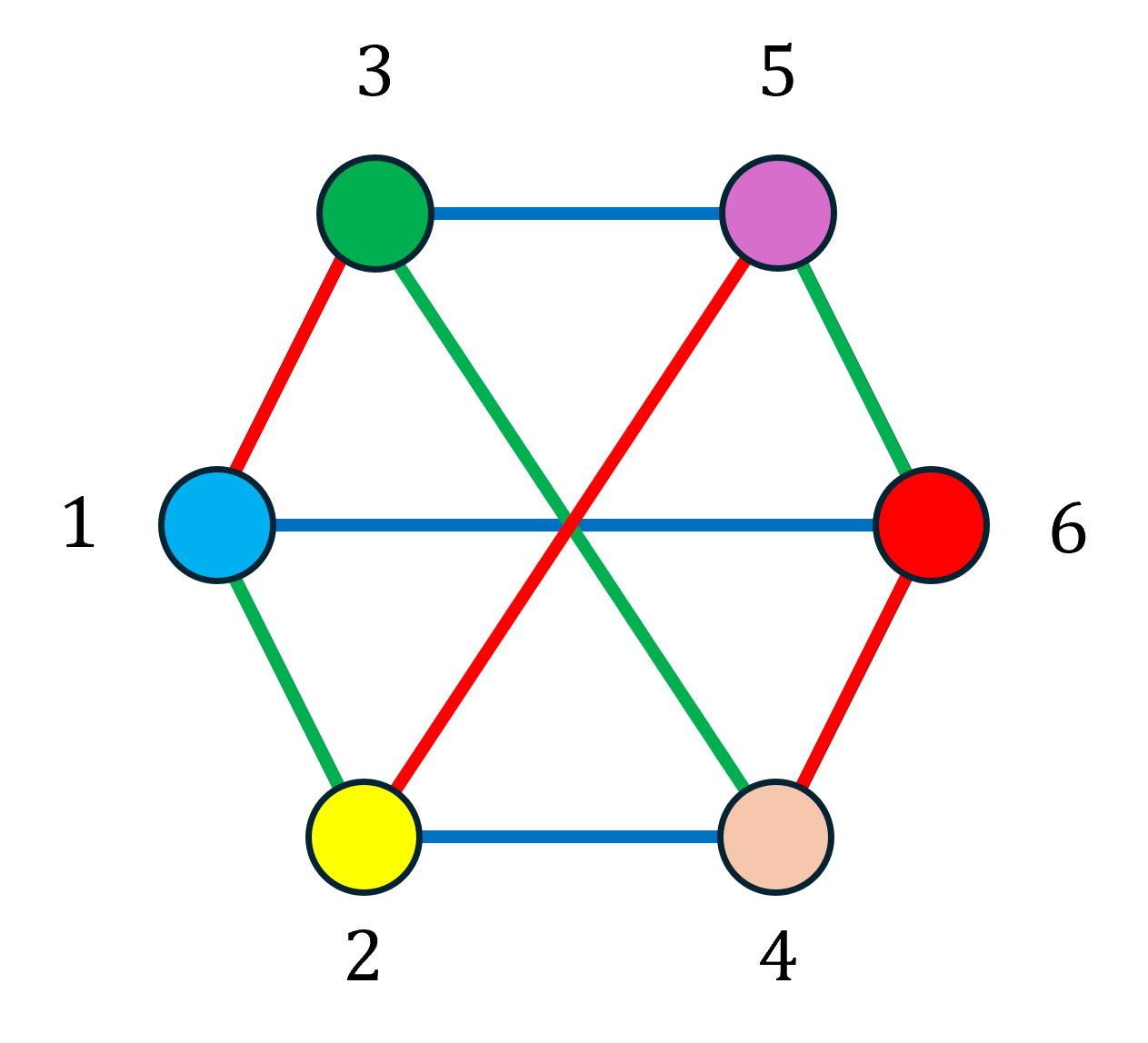}
        \label{fig: folded-kitaev}
    \end{subfigure}
    \caption{\justifying \textbf{Dynamical quantum simulation of the Kitaev model.} (a) A finite-size Kitaev honeycomb lattice, where edges are colored red, green, and blue to represent $X_i X_j$, $Y_i Y_j$, and $Z_i Z_j$ interactions, respectively. The labels $1$ through $6$ indicate the equivalence classes of qubits under the vertex color partition which respects the edges coloring, not individual physical qubits. (b) The folded Kitaev lattice (i.e., the quotient hypergraph) obtained by collapsing all qubits within the same equivalence class. Compared to the standard unit cell, the quotient hypergraph includes three additional edges (2-5), (3-4), and (1-6) that arise from the folding. }
    \label{fig: combined-kitaev}
\end{figure}

In the previous section, we showed how to construct DD sequences that suppress residual interactions within the system and its environment. In this section, we demonstrate how such sequences can also be leveraged for Hamiltonian simulation and digital-analog quantum computing.
For concreteness, we consider the Kitaev’s honeycomb lattice toy model, described by the Hamiltonian~\cite{kitaev2006anyons} 
\begin{equation}
    H_{\rm comb} = -\frac{J_{x} }{4}\sum_{\langle ij\rangle_r} X_i X_j - \frac{J_{y}}{4} \sum_{\langle ij\rangle_g} Y_i Y_j - \frac{J_{z}}{4} \sum_{\langle ij\rangle_b} Z_i Z_j,
\end{equation}
where the sum is taken over the red (r), blue (b), and green (g) colored edges in Fig.~\ref{fig: kitaev hexagon}. This model is exactly solvable by using a mapping to Majorana fermions and exhibits both gapped and gapless phases depending on whether the coupling strengths $J_x, J_y, J_z$ satisfy the triangle inequality \cite{kitaev2006anyons}. Here, we focus on simulating the system deep in the gapless phase, where $J_{x} = J_{y} = J_{z} = J$.

We assume that the native analog Hamiltonian of the spin qubit device comprises Heisenberg interactions 
\begin{equation}
    H_{\rm analog} =  \sum_{\langle ij \rangle}   \frac{J}{4}\Big(  X_i X_j +  Y_i Y_j +   Z_i Z_j \Big)\ .
\end{equation}
We aim to simulate the target Hamiltonian $H_{\rm comb}$ using $H_{\rm analog}$. First, because the native interactions are isotropic, with $XX$, $YY$, and $ZZ$ couplings have the same amplitude, we must design a DD group that selectively preserves only one type of interaction (either $XX$, $YY$, or $ZZ$) based on the edge label in the honeycomb lattice. Second, once the desired interaction type is isolated, we must flip the global sign of the Hamiltonian to convert the native antiferromagnetic coupling (positive $J$) into a ferromagnetic one (negative $J$).

As in previous sections, we begin by identifying the effective number of equivalence color classes in Kitaev’s honeycomb model. If we were to universally decouple all interactions, the honeycomb lattice would be trivially 2-colorable. However, because we aim to selectively preserve certain interactions, the effective model becomes more complex. 

In particular, due to the link-dependent interaction terms in the Hamiltonian, the symmetry of the model reduces the problem to an effective system involving six color classes. In Fig.~\ref{fig: kitaev hexagon}, we color the qubits that are equivalent under the translation symmetry of the honeycomb lattice. The resulting quotient hypergraph, formed by identifying these symmetry-equivalent sites, contains six color classes. There are three extra edges appear between opposite effective vertices, as shown in Fig.~\ref{fig: folded-kitaev}. 

To demonstrate the power of the effective graph in Fig.~\ref{fig: folded-kitaev}, we use the simple algorithm presented in Sec.~\ref{section: selective DD sequences} to find the minimal decoupling group for the bang-bang sequence. We also explicitly construct a decoupling group for the  bounded sequence, see Appendix~\ref{appendix: simulating Kitaev's honeycomb model with bounded control}. Using the same terminology as before, we associate $H_{S}^{\parallel} = H_{\rm comb}$. Consequently, the remaining interactions belong to the list $H_{S}^{\perp}$.

By finding the kernel of the linear map defined in Eq.~\eqref{eq: nullspace of HS}, we identify four operators spanning the kernel space that commute with $H_{S}^{\parallel}$. These operators are
\begin{subequations}
\begin{align}
& W_1 = Z_1 X_2 Z_3 X_4,\quad W_2 = X_3 Z_4 X_5 Z_6, \\
& W_3 = Y_1 Y_3 X_5 X_6,\quad W_4 = Y_2 X_3 X_4 Y_5  ,
\end{align}
\end{subequations}
and are closely related to a known symmetry of the Kitaev honeycomb model, namely the plaquette operators \cite{kitaev2006anyons}
\begin{equation}
    W_{p,1} = Z_1 X_2 Y_3 Y_4 X_5 Z_6, ~\quad  W_{p,2} =Y_1 Y_2 Z_3 X_4 Z_5 X_6 
\end{equation}
One can verify that $W_{1} W_{2} = W_{p,1}$ and $W_{3} W_{4} = W_{p,2}$, that is, each pair provides a decomposition of the original plaquette symmetry. The emergence of these subsymmetries is expected: since the effective Kitaev honeycomb model arises from folding the original lattice, it can exhibit additional symmetries not present in the unfolded geometry. 

If we take $W_{p,1}$ as the generator of $\mathcal{G}$, the resulting sequence effectively removes the interaction terms along the outer edges of the hexagon in Fig.~\ref{fig: folded-kitaev}, aligning the system’s dynamics with that of the Kitaev model. However, the inner-edge interactions commute with $W_p,1$, so they are not eliminated by this sequence and remain in the effective Hamiltonian. This again reflects the presence of the subsymmetries introduced by lattice folding. In a similar manner, if we take $W_{p,2}$ as the generator of $\mathcal{G}$, there are interactions on outer edges that commute with $W_{p,2}$. 

Either by numerically solving the associated minimum-set covering problem or by manually verifying it, we find that using either the pair $(W_{1},W_{2})$ or $(W_{3},W_{4})$  as generators is sufficient to construct a DD group $\mathcal{G}$ for the bang-bang sequence. Consequently, the effective Hamiltonian have the same interaction as shown in the edges of Fig.~\ref{fig: folded-kitaev}. However, the global site is opposite of what it should be, i.e $H_{S}^{\parallel} = - H_{\rm comb} $. 

To flip the global sign of the Hamiltonian, we use the following three operators
\begin{equation}
    \label{eq: sign-flip operators}
    X_{1} X_{4} X_{5},~Y_{1} Y_{4} Y_{5},~Z_{1} Z_{4} Z_{5}. 
\end{equation}
Each of these operators commutes only with edges involving the same type of Pauli interaction and anti-commutes with the rest. For instance, $X_1 X_4 X_5$ commute with $X_1 X_3$ but anti-commutes with $Z_1 Z_2$. Consequently, by sequentially conjugating the effective Hamiltonian $H_{S}^{\parallel}$ with these operators, we flip the sign of $H_{S}^{\parallel}$, yielding the target Kitaev Hamiltonian:
\begin{equation}
    \sum_{P \in \{ X,Y,Z \} } P_{1} P_{4} P_{5}~H_{S}^{\parallel }~P_{1} P_{4} P_{5} = H_{\rm comb}. 
\end{equation}
Alternatively, we can view the three operators in Eq.~\eqref{eq: sign-flip operators} as (almost) forming a group. By including the identity operator, they generate the group
\begin{equation}
\langle X_{1} X_{4} X_{5}, Z_1 Z_4 Z_5 \rangle.
\end{equation}
Twirling over the full group (including the identity) would suppress $H_{S}^{\parallel}$ entirely. However, omitting the identity from the twirling set leads to the desired sign-flipping behavior described above. In conclusion, a full DD cycle implementing this sign-flipping mechanism consists of three such conjugations, each comprising a 4-element DD sequence, resulting in a total sequence length of $L=3 \times 4 = 12$.

In this section, we have demonstrated how to construct a bang-bang sequence that simulates Kitaev’s honeycomb model using only Heisenberg exchange interactions. In Appendix~\ref{appendix: simulating Kitaev's honeycomb model with bounded control}, we extend this construction to bounded-control sequences for the same task. While the bounded-control sequence is significantly longer, it offers the advantage of allowing independent tuning of the interaction strengths $J_x,J_y$ and $J_z$ through the appropriate choice of control pulses.

\section{Designing robust sequences} 
\label{section: robust pulse sequences}
In the previous sections, we implicitly assumed that $X$, $Y$, and $Z$ $\pi$-pulses can be implemented perfectly. In experiments, however, pulse imperfections are inevitable and can significantly degrade the performance of a DD sequence if left unaddressed. This section introduces two complementary strategies to mitigate such imperfections and enhance the robustness of our DD constructions. In Sec.~\ref{section: minimize number of pulses}, we show how to reduce the total number of applied pulses in DD sequences derived from classical additive codes. By exploiting the freedom in choosing the generator set, we can substantially lower the pulse count and establish a lower bound on the minimum number of pulses required based on the code distance. In Sec.~\ref{section: mitigating imperfect pulses}, we go a step further by addressing the imperfections of the pulses. We demonstrate how proper scheduling of $(\pi)$ and $(-\pi)$ pulses, together with a mirror-sequence strategy, can suppress both propagated pulse errors and residual free-evolution errors stemming from imperfect decoupling. Finally, in Sec.~\ref{section: numerical simulation}, we perform numerical simulations of the proposed DD sequences on the bilinear array shown in Fig.~\ref{fig: G and expanded G}(a), including longitudinal and transverse coherent error as well as residual nearest- and next-nearest-neighbor interactions.

\subsection{Minimizing the number of pulses}
\label{section: minimize number of pulses}

In this section, we present a method to minimize the number of pulses applied in each round of a DD sequence. Reducing the pulse count per round offers several benefits: it relaxes demands on classical control electronics, decreases the accumulation of pulse-induced errors and crosstalk, and mitigates heating effects on the chip caused by frequent pulse application. To achieve this, we introduce the canonical implementation of DD sequences based on an additive group $\mathcal{G}$ generated by a set $\Gamma_{\mathcal{G}} = \{ \gamma_i \}$, where each applied pulse corresponds to one of the group generators $\gamma_i$. In the canonical implementation, the problem of minimizing the number of pulses per round reduces to finding a generating set for $\mathcal{G}$ with minimal Hamming weights. Using this correspondence, we show that the code distance $d$ provides a tight lower bound on the minimum number of pulses required per round.

We define $\mathrm{Cay}(\mathcal{G}, \Gamma_{\mathcal{G}})$ as the Cayley graph of the decoupling group $\mathcal{G}$ with respect to the generating set $\Gamma_{\mathcal{G}}$ \cite{Viola2003,Bookatz_2016}. Since we consider $\mathcal{G}$ to be an additive group (i.e., a subgroup of the Abelian projective Pauli group), the resulting Cayley graph is undirected. Its vertices correspond to the group elements $g_i \in \mathcal{G}$, and an edge labeled by a generator $\gamma_k \in \Gamma_{\mathcal{G}}$ connects two vertices $g_i$ and $g_{i+1}$ if and only if
\begin{equation}
    g_{i+1} = \gamma_k  g_i  \text{ or equivalently }  \gamma_k g_{i+1} =  g_i.
\end{equation}

We define a canonical implementation of a DD sequence as any closed path on the Cayley graph $\mathrm{Cay}(\mathcal{G}, \Gamma_{\mathcal{G}})$ that visits each vertex exactly once, starting and ending at the identity element. In other words, a canonical implementation corresponds to a Hamiltonian cycle on $\mathrm{Cay}(\mathcal{G}, \Gamma_{\mathcal{G}})$. Such a cycle can be specified by the ordered list of visited vertices (excluding the initial identity)
\begin{equation}
    \{ g_{1},g_2,\dots, g_{L-1},g_L = I\}.
\end{equation}

In practice, the corresponding DD implementation takes the form
\begin{equation}
      \label{eq: canonical implementation of DD}
      U_{\Delta}(g_{L-1} U_{\Delta} g_{L-1})\dots(g_2U_{\Delta} g_{2})(g_1U_{\Delta} g_{1}).
\end{equation}
where each $U_{\Delta}$ denotes a free-evolution segment. Because the sequence follows a Hamiltonian cycle in the Cayley graph starting from the identity, the vertices obey
\begin{subequations}
    \begin{gather}
        g_1 \in \Gamma_{\mathcal{G}}, \quad \text{and } \quad  g_{i+1}g_{i} \in \Gamma_{\mathcal{G}}. 
    \end{gather}
\end{subequations}
Thus, after combining adjacent pulses around each free-evolution interval in Eq.~\eqref{eq: canonical implementation of DD}, the resulting canonical DD sequence contains only pulses drawn from the generating set $\Gamma_{\mathcal{G}}$.

As an illustrative example, let us revisit the $XY4$ sequence, which can be interpreted as a DD protocol generated from the single-qubit projective Pauli group. We take the generating set to be $\{X, Y\}$ rather than $\{X, Z\}$. The resulting Cayley graph is
\begin{equation}
\begin{tikzpicture}[
  node distance=2cm,
  vertex/.style={circle, draw, minimum size=8mm, inner sep=0pt}
]

\node[vertex] (I) {I};
\node[vertex] (X) [right of=I] {X};
\node[vertex] (Z) [below of=X] {Z};
\node[vertex] (Y) [below of=I] {Y};

\draw[-] (I) edge["$X$"] (X)
          (X) edge["$Y$"] (Z)
          (Z) edge["$X$"] (Y)
          (Y) edge["$Y$"] (I);

\draw[->, thick, blue, bend right=20] (I) to (X);
\draw[->, thick, blue, bend right=20] (X) to (Z);
\draw[->, thick, blue, bend right=20] (Z) to (Y);
\draw[->, thick, blue, bend right=20] (Y) to (I);

\end{tikzpicture}.
\end{equation}
The blue directed edges depict a Hamiltonian cycle on the graph corresponding to the standard implementation of the $XY4$ sequence. Each edge label denotes the generator applied to move between vertices, showing explicitly how the alternating application of $X$ and $Y$ pulses traverses the entire decoupling group.

To recap, we have defined canonical implementations of DD sequences. We now turn to the problem of minimizing the number of pulses applied per round. In a canonical implementation, each pulse corresponds to a generator ${\gamma_i}$ of the additive group $\mathcal{G}$, so this problem reduces to minimizing the Hamming weight of these generators. As an example, consider the binary generator matrix of the $\mathrm{RM}(1,3)$ code shown in Table~\ref{tab: generator-RM13-binary}. The first row of this matrix has weight $8$, indicating that there exists a round in which control pulses must be applied simultaneously on all qubits. However, the choice of generator set $\Gamma_{\mathcal{G}}$ for the group $\mathcal{G}$ is not unique. Any linear combination of the generators can serve as an alternative valid generating set. For instance, if $\{\gamma_1, \gamma_2, \gamma_3 \}$ is a generator set, then $\{\gamma_1, \gamma_2 + \gamma_3, \gamma_1 + \gamma_3 \}$ also generates the same group. Applying this idea to Table~\ref{tab: generator-RM13-binary}, we can construct an alternative generator matrix by replacing the first row with the sum of the first and second rows. The resulting generator matrix is
\begin{equation}
    \label{eq: new generator matrix for R13}
    \begin{tabular}{|c|c|c|c|c|c|c|c|}
        \hline
         $C_1$ & $C_2$  & $C_3$  &  $C_4$  & $C_5$  & $C_6$  & $C_7$  & $C_8$ \\
        \hline
         0 & 0  & 0  & 0  & 1  & 1  & 1  & 1 \\
         \hline
         1 & 1  & 1  & 1  & 0  & 0  & 0  & 0 \\
         \hline
         1 & 1  & 0  & 0  & 1  & 1  & 0  & 0 \\
         \hline
         1 & 0  & 1  & 0  & 1  & 0  & 1  & 0  \\
         \hline
    \end{tabular}
\end{equation}
In the new generator matrix, the first generator has a Hamming weight of four, half that of the original, thereby reducing the number of simultaneous pulses required per round. This idea naturally extends to all DD sequences discussed in this work, since all our sequences are constructed from additive codes.

We now formally define the optimization problem of finding a generator matrix for the group $\mathcal{G}$ with minimal Hamming weight. Let us start with an initial choice of generator matrix, which we also denote (by abuse of notation) as  $\Gamma_{\mathcal{G}}$, whose rows $\gamma_i$ are vectors over either $\mathbb{F}_4$ or $\mathbb{F}_2$. We define a row-combination matrix $R$ as an invertible square matrix with binary entries ($0$ or $1$). An alternative, but equivalent, generator matrix $\overline{\Gamma}_{\mathcal{G}}$ for the same group $\mathcal{G}$ can then be obtained as
\begin{equation}
    \overline{\Gamma}_{\mathcal{G}} \equiv R \Gamma_{\mathcal{G}}
\end{equation}
If we define $\mathrm{Wg}(\Gamma_{\mathcal{G},i})$ as the Hamming weight of the $i$-th row of $\Gamma_{\mathcal{G}}$, the problem of minimizing the number of control pulses per round can be formulated as
\begin{equation}
    \label{eq: minimizing number of pulse per round}
    \min_{R} \max_{i} \text{Wg}(R \Gamma_{\mathcal{G},i}). 
\end{equation}
In other words, the goal is to find the row-combination matrix $R$ that minimizes the maximum weight among the generators, thereby reducing the number of simultaneous pulses required in a single round of the DD sequence.

We are not aware of any algorithms that efficiently solves Eq.~\eqref{eq: minimizing number of pulse per round}, nor is its computational complexity established. A straightforward, though potentially suboptimal, choice for the generator matrix $\Gamma_{\mathcal{G}}$ is its standard form \cite{nielsen2010quantum}. Nevertheless, a trivial lower bound on the minimum number of pulses per round can be inferred from the code properties of $\mathcal{G}$. Since $\mathcal{G}$ defines an additive code, the minimum Hamming weight of any generator is bounded by the code distance $d$
\begin{equation}
    \label{eq: trivial lower bound number applied pulses}
    \min_{R} \min_i \text{Wg}(R \Gamma_{\mathcal{G},i}) \geq d.
\end{equation}
As an example, the new generator matrix of the $\rm RM(1,3)$ code in Eq.~\eqref{eq: new generator matrix for R13} achieves this bound, with all rows having the minimum possible Hamming weight. This observation implies that a robust DD sequence should be constructed from codes that balance two competing requirements: \textit{a high dual distance $d^{\perp}$ to suppress high-weight errors, and a low distance $d$ to minimize the number of pulses per round.}

The discussion above implicitly assumes that each color class contains roughly the same number of qubits. If some color classes are significantly larger than others, Eq.~\eqref{eq: minimizing number of pulse per round} must be modified to account for the unequal distribution. In that case, the trivial lower bound in Eq.~\eqref{eq: trivial lower bound number applied pulses} becomes the product of the code distance $d$ and the size of the smallest color class.

As a final remark, the flexibility in choosing the generator set can also be exploited for other optimization tasks. For example, common native physical single-qubit gates are the $X$- and $Y$-gates, while the $Z$-gate is typically virtual. Performing a physical $Z$-gate therefore requires a combination of $X$- and $Y$-gates, which increase gate errors. Consequently, one could construct an alternative generator matrix $\Gamma_{\mathcal{G}}$ that minimizes the use of $Z$-gates, reducing the overall pulse error. We leave this direction as future work.

\subsection{Mitigating imperfect pulses}
\label{section: mitigating imperfect pulses}

In the previous section, we have introduced the canonical implementation of DD sequences and discussed how to minimize the number of pulses applied per round. In this section, we turn to the complementary question of how to make these sequences robust against pulse imperfections. We present a simple and general strategy for implementing canonical DD sequences in a way that suppresses pulse errors. The key idea is based on two techniques: first, to schedule the $(\pi)$ and $(-\pi)$ pulses so that their errors systematically cancel out (in the same spirit as the universal robust sequences of Ref.~\cite{Genov2017}); and second, to incorporate a mirror-sequence construction, analogous to how the standard $XY4$ sequence is extended to the more robust $XY8$ sequence. Together, these techniques provide a simple yet effective approach for mitigating pulse imperfections.

We start by rewriting the canonical DD sequence in Eq.~\eqref{eq: canonical implementation of DD} by grouping adjacent pulses
\begin{equation}
      \label{eq: canonical implementation of DD simplified}
      U_{\Delta} p_{L} \dots U_{\Delta} p_2 U_{\Delta} p_{1}.
\end{equation}
Each pulse $p_i$ is a tensor product of single-qubit Pauli operators
\begin{equation}
    p_i = p_{i,1}\otimes p_{i,2} \dots \otimes  p_{i,N} \in \Gamma_{\mathcal{G}}
\end{equation}
and is realized in practice by applying single-qubit gates $p_{i,k}$ in parallel across $N$ qubits. We assume that the dominant source of pulse error is systematic over- or under-rotation. The case of misaligned rotation axes is more complicated and we discuss it in Appendix~\ref{appendix: robustness under misaligned rotation axis}. Under this assumption, the noisy implementations of $\theta$-pulses for the $k$-th qubit in the $i$-th pulse are
\begin{subequations}
    \begin{gather}
            (\tilde{p}_{i,k})_{\theta } \equiv \exp\Big[-\frac{i}{2}~ \text{sign}(\theta)(\theta+ 2\epsilon_{ik})p_{i,k} \Big],
    \end{gather}
\end{subequations}
here $\epsilon_{i,k}$ captures the over- or under-rotation error. We assume that 
\begin{equation}
    |\epsilon_{i,k}| \leq \epsilon \quad \forall i,k 
\end{equation}
for small $\epsilon$. 

In this section, we consider two equivalent implementation of the $p_{i,k}$ gate, namely $\theta=\pi$ and $\theta = -\pi$ pulses. A noisy pulse $\tilde{p}_i$ can then be written as a product of the ideal pulse $p_i$ and a composite error operator $E[p_i]$ that accumulates all individual errors  from each single qubit gates
\begin{equation}
\label{eq: definition of noisy pulses}
\tilde{p}_i = E[p_i] p_i =(\prod_{k} e^{\pm i \epsilon_{ik}p_{i,k}})  p_i,
\end{equation}
where the sign of the rotation error depends on whether we apply the $(\pi)$ or $(-\pi)$-pulses. Accordingly, the noisy DD sequence becomes
\begin{equation}
    \label{eq: noisy DD sequence}
    U_{\Delta}E[p_L] p_L\dots U_{\Delta} E[p_1] p_1. 
\end{equation}
To analyze the effect of imperfect pulses, we commute the error operators $E[p_i]$ through the ideal pulses $p_i$ and the free-evolution segment $U_{\Delta}$. Our goal is to express the noisy DD sequence in Eq.~\eqref{eq: noisy DD sequence} with the following form
\begin{multline}
        \label{eq: final error sources}
   \Big( \prod_{i=1}^{L} \tilde E[p_i] \Big) U_{\Delta,L}~p_L ~U_{\Delta,L-1} p_{L-1}\dots U_{\Delta,1} p_1  \\
   \equiv   \Big( \prod_{i=1}^{L} \tilde E[p_i] \Big) e^{-i \Delta L (H_{\rm targ}+H_{\rm error} ) }.
\end{multline}
Here, $\tilde{E}[p_i]$ denotes the effective error operator obtained after commuting $E[p_i]$ through all subsequent ideal pulses $p_j$ with $j>i$, while 
\begin{equation}
    \label{eq: definition of modified effetive Hamiltonian}
    U_{\Delta,i} \equiv \exp(-i \Delta H_{\rm eff,i})
\end{equation}
represents the modified free-evolution segment that incorporates the influence of all preceding errors $E[p_j]$ with $j \le i$. 

There are two distinct error types in Eq.~\eqref{eq: final error sources}. The first arises from the propagated pulse errors $\tilde{E}[p_i]$, which accumulate as the sequence proceeds. In Sec.~\ref{section: scheduling pi and -pi pulses}, we show that these errors can be suppressed by appropriately scheduling $(\pi)$ and $(-\pi)$ pulses so that their leading-order contributions cancel. The second originates from the residual error Hamiltonian $H_{\rm error}$  due to each free-evolution segment $U_{\Delta,i}$ evolves under a slightly different effective Hamiltonian. We address the suppression of the second error type in Sec.~\ref{section: mirror sequence} using the mirror sequence. 

\subsubsection{Scheduling $(\pi)$ and $(-\pi)$-pulses}
\label{section: scheduling pi and -pi pulses}
In this subsection, we show that errors due to $\tilde{E}[p_i]$ can be suppressed by an appropriate scheduling of $(\pi)$- and $(-\pi)$-pulses within the canonical DD sequence. The key observation underlying this result is that in a canonical DD sequence, each generator $\gamma_i$ appears an even number of time. By choosing the sign of the applied pulse (for example, using a $(\pi)$-pulse the first time and a $(-\pi)$-pulse the second), we can ensure that the first-order rotation errors introduced by these imperfect pulses cancel out, leaving only higher-order residual terms.

As a first step, we need to compute the exact form of $\tilde{E}[p_i]$. Since each error operator $E[p_i]$ is composed of independent single-qubit rotation errors (see Eq.~\eqref{eq: definition of noisy pulses}), we can analyze how each individual rotation error propagates through subsequent ideal pulses $p_j$. Consider commuting a single-qubit rotation error $e^{-i \epsilon_{ik} p_{i,k}}$ through a Pauli pulse $p_j$. Depending on whether the two Pauli operators commute or anticommute, the sign of the rotation angle may flip, converting an over-rotation into an under-rotation, or vice versa
\begin{equation}
    p_{j}~e^{-i \epsilon_{ik}p_{i,k}} = \begin{cases}
        e^{-i \epsilon_{ik}p_{i,k}}~p_{j} \quad \text{ if } [p_{i,k},p_j]=0, \\
        e^{+i \epsilon_{ik} p_{i,k}}~p_j \quad \text{ otherwise}
    \end{cases}.
\end{equation}
Therefore, the effective error operator $\tilde{E}[p_i]$ can be obtained by tracking how each local error term $e^{-i \epsilon_{ik} p_{i,k}}$ transforms under all subsequent ideal pulses $p_j$ with $j>i$.

To formalize this, we define a commutation propagation function
\begin{equation}
    \text{Com}(p_{i,k},l) \equiv \begin{cases}
        (\prod_{j=i+1}^{l} p_{j})^{\dagger}  p_{i,k} (\prod_{j=i+1}^{l} p_{j})  & \text{ for } l>i \\
        p_{i,k } & \text{ for } l= i \\
        0 & \text{ otherwise } 
    \end{cases}
\end{equation}
which determines the final effective sign of a rotation error on $k$-th qubit from round $i$ after propagating through the sequence up to round $l\leq L$. With the newly defined function, the full propagated error operator is given by 
\begin{equation}
    \tilde{E}[p_i] = \prod_{k} e^{\pm i \epsilon_{ik}\text{Com}(p_{i,k},L)},
\end{equation}
where the $\pm$ sign corresponds to whether the implemented pulse on the $k$-th qubit is a $(-\pi)$- or $(\pi)$-rotation, respectively. 

Expanding to first order in $\epsilon$, the accumulated propagated pulse error can be approximated by keeping first order contribution in the Baker-Campbell-Hausdorff formula
\begin{equation}
    \label{eq: cumulative pulse error}
    \prod_{i=1}^{L} \tilde{E}[p_i]  \approx \prod_{k} \exp\Big[i \sum_{i=1}^{L} (\pm \epsilon_{ik})  \text{Com}(p_{i,k},L) +O(\epsilon^2) \Big]. 
\end{equation}
We note that when all the pulses ${ p_{i,k} }$ acting on a given qubit commute with each other, as is the case for biased decoupling sequences, where only $Z$-type or only $X$-type pulses are predominantly applied, the above expression becomes exact rather than an approximation.

To achieve first-order cancellation of the propagated errors, it is sufficient that every operator $p_{i,k}$ appearing in the exponent has a matching counterpart $p_{j,k}$ (with $j \neq i$) that is identical 
\begin{equation}
    \label{eq: condition for cancellation of propagated error}
    p_{i,k}= p_{j,k}.
\end{equation}
Assuming this condition holds, we evaluate the signs of $\text{Com}(p_{i,k},L)$ and $\text{Com}(p_{j,k},L)$. If the signs of these two terms are identical, we apply a $(\pi)$-pulse for one and a $(-\pi)$-pulse for the other. Conversely, if their signs are opposite, we apply the same pulse type, either both $(\pi)$ or both $(-\pi)$. As a concrete example, by applying this pulse schedule to the $XY4$ sequence reproduces the known results \cite{ZhiHui2012,Wang2012} that the standard implementation using a single pulse type, either all $(\pi)$ pulses or all $(-\pi)$ pulses
\begin{multline}
        \label{eq: robust XY4}
    U_{\Delta}(Y)_{\pi} U_{\Delta} (X)_{\pi} U_{\Delta} (Y)_{\pi} U_{\Delta} (X)_{\pi} \\
    \text{ or }~U_{\Delta}(Y)_{-\pi} U_{\Delta} (X)_{-\pi} U_{\Delta}  (Y)_{-\pi} U_{\Delta} (X)_{-\pi}
\end{multline}
is more robust to pulse errors than alternating between $(\pi)$ and $(-\pi)$ pulses.

Our final step is to show that in the canonical DD sequence, each generator $\gamma_{i}$ necessarily appears an even number of time, ensuring that Eq.~\eqref{eq: condition for cancellation of propagated error} can always be fulfilled. Since $\mathcal{G}$ is an additive group, its Cayley graph $\mathrm{Cay}(\mathcal{G}, \Gamma_{\mathcal{G}})$ is isomorphic to the hypercube graph $\mathbb{Z}_{2}^{|\Gamma_{\mathcal{G}}|}$. Accordingly, we can associate each vertex in $\mathrm{Cay}(\mathcal{G}, \Gamma_{\mathcal{G}})$ with a binary coordinate in $\mathbb{Z}_{2}^{|\Gamma_{\mathcal{G}}|}$, and each generator $\gamma_{i}$ with a unit basis vector $\mathbf{e}_i$. In a Hamiltonian cycle on the hypercube that starts and ends at the origin, every coordinate direction $\mathbf{e}_i$ must be traversed an even number of time, half in the positive and half in the negative direction. Therefore, each generator $\gamma_{i}$ appears an even number of times in the canonical implementation.

We remark that our strategy is simple yet broadly applicable to any DD sequence generated by an additive group. Nevertheless, once a specific DD sequence is chosen, there remains significant room for further optimization. For example, by carefully arranging the order in which the generators $\gamma_i$ appear, it may be possible to achieve error cancellation beyond the first order \cite{Childs2021}. An interesting direction would be to employ evolutionary ML methods to find the optimal Hamiltonian cycle \cite{tong2024empirical}. We leave this avenue for future work.

\subsubsection{Mirror sequence}
\label{section: mirror sequence}
In the previous subsection, we addressed how to mitigate propagated pulse errors $\tilde{E}[p_i]$. Assuming that the canonical DD sequence is implemented using the $(\pi)$ and $(-\pi)$ pulse schedule described in Sec.~\ref{section: scheduling pi and -pi pulses}, Eq.~\eqref{eq: final error sources} can be approximated as
\begin{equation}
 \label{eq: half-cycle error exponent form}
  U_{\Delta}\tilde{p}_{L} \dots U_{\Delta}\tilde{p}_2 U_{\Delta} \tilde{p}_{1} \approx e^{-i \Delta L(H_{\rm targ}+ H_{\rm error})},
\end{equation}
where the only remaining error arises from the fact that each free-evolution segment evolves under a slightly different effective Hamiltonian, giving rise to a residual error Hamiltonian $H_{\rm error}$. In this section, we introduce the mirror sequence, which is directly analogous to the construction that promotes the $XY4$ sequence to the $XY8$ sequence~\cite{Wang2012}. We show that the mirror sequence suppresses $H_{\rm error}$ to first order in $\epsilon$, at the cost of doubling the sequence length.

As a first step, we analyze the effect of pulling all the pulse errors $E[p_i]$ from inside Eq.~\eqref{eq: noisy DD sequence} to the left-hand side, as done in Eq.~\eqref{eq: final error sources}. The effect of commuting the pulse error $E[p_i]$ through the free-evolution segment $U_{\Delta}$ is given by
\begin{equation}
    \label{eq: modified free-time evolution}
    e^{-i H_\text{tot} \Delta}E[p_i]= E[p_i] e^{-i (E[p_i]^{\dagger} H_\text{tot} E[p_i])\Delta},
\end{equation}
This shows that each free-evolution segment effectively evolves under a slightly modified Hamiltonian due to pulse imperfections. Assuming that the rotation errors $\epsilon_{i,k}$ are small, the exponent on the RHS of Eq.~\eqref{eq: modified free-time evolution} can be expanded to first order in $\epsilon$ as
\begin{equation}
    \label{eq: left-over terms form modified Hamiltonian}
    E[p_i]^{\dagger} H_\text{tot} E[p_i] \approx H_{\text{tot}} + i \sum_{k} (\pm \epsilon_{i,k}) [p_{i,k},H_\text{tot}].
\end{equation}
A key observation is that the residual error on the right-hand side is additive to first order in $\epsilon$, i.e.
\begin{multline}
       E[p_j]^{\dagger} E[p_i]^{\dagger} H_\text{tot} E[p_i] E[p_j] \approx H_{\text{tot}}   \\
       + i \sum_{k} (\pm \epsilon_{i,k}) [p_{i,k},H_\text{tot}]+i \sum_{k} (\pm \epsilon_{j,k}) [p_{j,k},H_\text{tot}].
\end{multline}

As a result, using Eq.~\eqref{eq: left-over terms form modified Hamiltonian} and accounting for the sign flips that occur when propagating the pulse errors $E[p_i]$ through the ideal pulses $p_j$ (as discussed in Sec.~\ref{section: scheduling pi and -pi pulses}), we find that the effective Hamiltonian during the $l$-th free-evolution segment is approximately
\begin{equation}
    H_{\rm eff,l} \approx H_\text{tot} + \sum_{i=1}^{l} \sum_{k}(\pm \epsilon_{i,k}) [\text{Com}(p_{i,k},l),H_{tot}  ].
\end{equation}
This expression makes explicit that each effective free-evolution operator $U_{\Delta,l}$ contains contributions from imperfect pulses applied at all earlier times. Consequently, after completing one imperfect canonical DD sequence, the net evolution is governed by an effective Hamiltonian $H_{\rm eff}$ that can be decomposed into the desired target Hamiltonian $H_{\rm targ}$, which we seek to retain, and the residual error Hamiltonian $H_{\rm error}$.
\begin{multline}
        \label{eq: half-cycle effective Hamiltonian}
        H_{\rm eff}  \equiv H_{\rm targ} + H_{\rm error}   =\sum_{l=1}^{L}\frac{ g_{l} H_{\rm eff,l} g_{l}  }{L} \approx  H_{\rm targ} \\
        + \frac{1}{L} \sum_{l=1}^{L} g_{l}\Big(\sum_{i=1}^{l} \sum_{k}(\pm \epsilon_{i,k}) 
        \times [\text{Com}(p_{i,k},l),H_\text{tot}  ] \Big) g_{l}.
\end{multline}

To suppress the residual evolution generated by $H_{\rm error}$, we append a mirror sequence in which the pulses and the evolution segments appear in the reverse order
\begin{equation}
   U_{\Delta}p_L \dots U_{\Delta}p_{1} \quad  \xrightarrow{\text{Mirror sequence}} \quad  p_1 U_{\Delta} \dots p_{L} U_{\Delta}
\end{equation}
For instance, applying the mirror sequence to the $XY4$ sequence in Eq.~\eqref{eq: robust XY4} leads to the robust (and symmetric) $XY8$ sequence 
\begin{multline}
    \label{eq: XY8 mirror}
        \Big[(X_{\pi}) U_{\Delta}(Y)_{\pi} U_{\Delta} (X)_{\pi} U_{\Delta}  (Y)_{\pi} U_{\Delta} \Big]\Big[U_{\Delta}(Y)_{\pi} \\
        \times U_{\Delta} (X)_{\pi} U_{\Delta}  (Y)_{\pi} U_{\Delta} (X)_{\pi}  \Big]
\end{multline}

To explain why appending a mirror sequence can suppresses $H_{\rm error}$, we rewrite the noisy mirror DD sequence in the form
\begin{multline}
    \label{eq: reverse error decomposition of mirror sequence}
   \tilde{p}_1 U_{\Delta} \dots \tilde{p}_{L} U_{\Delta}
   \equiv    e^{-i \Delta L (H_{\rm targ}+H_{\rm error}^{\rm mirror} ) } \Big( \prod_{i=1}^{L} \tilde E[p_i] \Big),
\end{multline}
where all pulse errors $E[p_i]$ are propagated to the right-hand side, in contrast to the left-propagation used in Eq.~\eqref{eq: final error sources}. Recall that, with the appropriate scheduling of $(\pi)$- and $(-\pi)$-pulses, the accumulated pulse-error operator satisfies $\Big( \prod_{i=1}^{L} \tilde E[p_i] \Big) \approx I$ to first order in $\epsilon$. Due to the reversed direction in which the pulse errors $E[p_i]$ are commuted through the mirror sequence, it is straightforward to show that
\begin{equation}
    H_{\rm error}^{\rm mirror} = -H_{\rm error}.
\end{equation}
As a result, the residual error Hamiltonian $H_{\rm error}^{\rm mirror}$  generated by the mirror sequence cancels the residual error Hamiltonian $H_{\rm error}$ from the original sequence in Eq.~\eqref{eq: half-cycle error exponent form}. Although this derivation relies on the first-order Magnus expansion and therefore demonstrates cancellation only to first order in $\Delta$, the overall symmetric structure of the combined sequence and its mirror (see Eq.~\eqref{eq: XY8 mirror}), together with the locality of the noise model, ensures that all errors are suppressed to second order in $\Delta$. 

We emphasize that using the $(\pi)$ and $(-\pi)$ pulse schedule described in Sec.~\ref{section: scheduling pi and -pi pulses} is essential for error cancellation. Without this specific scheduling, the combined noisy DD sequence and its mirror take the form
\begin{equation}
    e^{-i \Delta L(H_{\rm targ}- H_{\rm error})}  \Big( \prod_{i=1}^{L} \tilde E[p_i] \Big)^2 e^{-i \Delta L(H_{\rm targ} +  H_{\rm error})}  
\end{equation}
which, in general, does not lead to cancellation of over- and under-rotation errors.

\subsection{Numerical simulation}
\label{section: numerical simulation}

\begin{figure}
    \centering
    \includegraphics[width=0.85\linewidth]{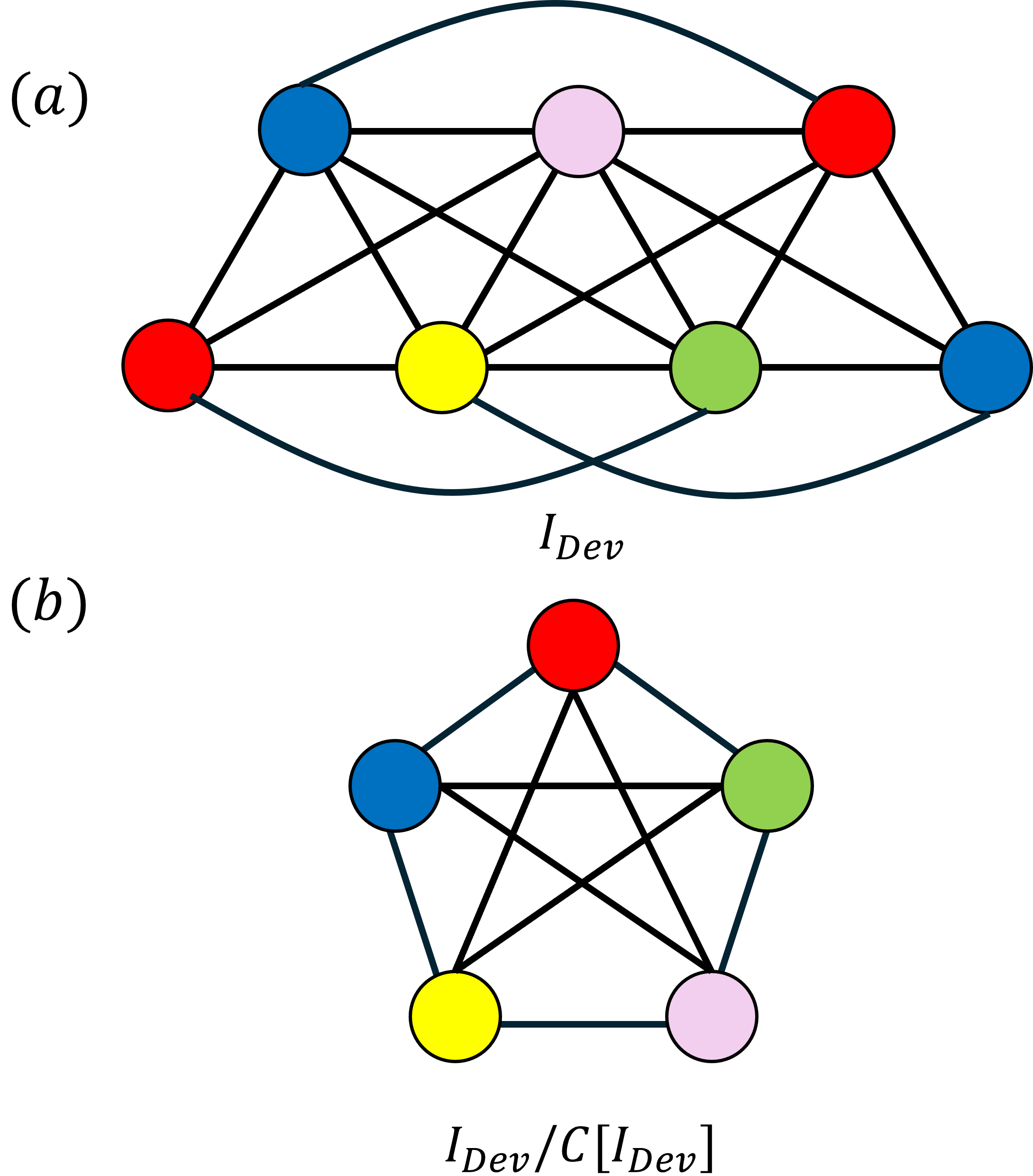}
    \caption{\justifying \textbf{Interaction and quotient hypergraph of bilinear array} (a) Interaction hypergraph $I_{\rm Dev}$ of the device presented in Fig.~\ref{fig: G and expanded G}(a) but with nearest- \textit{and} next-nearest neighbor interactions, along with the minimal color partition $C[I_{\rm Dev}]$. (b) The resulting quotient hypergraph $I_{\rm Dev}/C[I_{\rm Dev}]$ which is a simply fully connected pentagon.  }
    \label{fig: G and quotient NNN}
\end{figure}

As a proof of principle for the proposed robust sequence design, we consider the seven-qubit bilinear array shown in Fig.~\ref{fig: G and expanded G}a. In the rotating frame of the qubits, we assume the system is subject to an always-on Hamiltonian
\begin{multline}
        \label{eq: phenological error Hamiltonian}
        \sum_{i} (\delta\omega_{z,i} Z_i + \delta\omega_{x,i} X_i + \delta \omega_{y,i} Y_i) \\
        + \sum_{d(i,j)=1} J_{ij} Z_i Z_j + \sum_{d(i,j)=2} K_{ij} Z_i Z_j
\end{multline}
which models longitudinal coherent error $\{ \omega_{i,z}\}$, transversal coherent error $\{ \omega_{i,x},\omega_{i,y} \}$, residual nearest-neighbor interactions $\{ J_{ij} \}$, and next-nearest-neighbor interactions $\{ K_{ij} \}$. With the above Hamiltonian, the corresponding interaction hypergraph $I_{\rm Dev}$ and its quotient hypergraph $I_{\rm Dev}/C[I_{\rm Dev}]$ are given in Fig.~\ref{fig: G and quotient NNN}. 

We assume a simple error model in which the typical interaction strengths satisfy
\begin{equation}
\delta \omega_{z,i} \sim \delta \omega_{x,i} \sim \delta \omega_{y,i} \sim J_{ij} \sim 1~\text{MHz},
\quad K_{ij} \sim 0.1~\text{MHz}.
\end{equation}
Systematic single-qubit over-rotation or under-rotation errors $\epsilon_{i,k}$ are taken to be independently and uniformly distributed over the interval $[-5\degree,5\degree]$. These errors may vary between different realizations of the DD sequence but are fixed within each realization. After completing each DD sequence, we compute the trace fidelity with respect to the identity operation
\begin{equation}
    F \equiv \Big|  \frac{\text{Tr}(U_{\rm DD})}{2^7} \Big|^2
\end{equation}
which quantifies how effectively the sequence suppresses the unwanted dynamics. We compare the resulting fidelities for four cases: (i) no dynamical decoupling, (ii) the standard $XY4$ sequence, (iii) the $XY8$ sequence, (iv) the punctured RM(1,3) sequence listed in Table~\ref{tab: generator-RM13-punctured} (i.e the first five columns), and (v) its robust implementation. The robust RM(1,3) sequence is applied once, while the other sequences are repeated an appropriate number of times so that the total evolution time is identical across all cases (for example, repeating the $XY4$ sequence four times).

\begin{figure}
    \centering
    \includegraphics[width=0.95\linewidth]{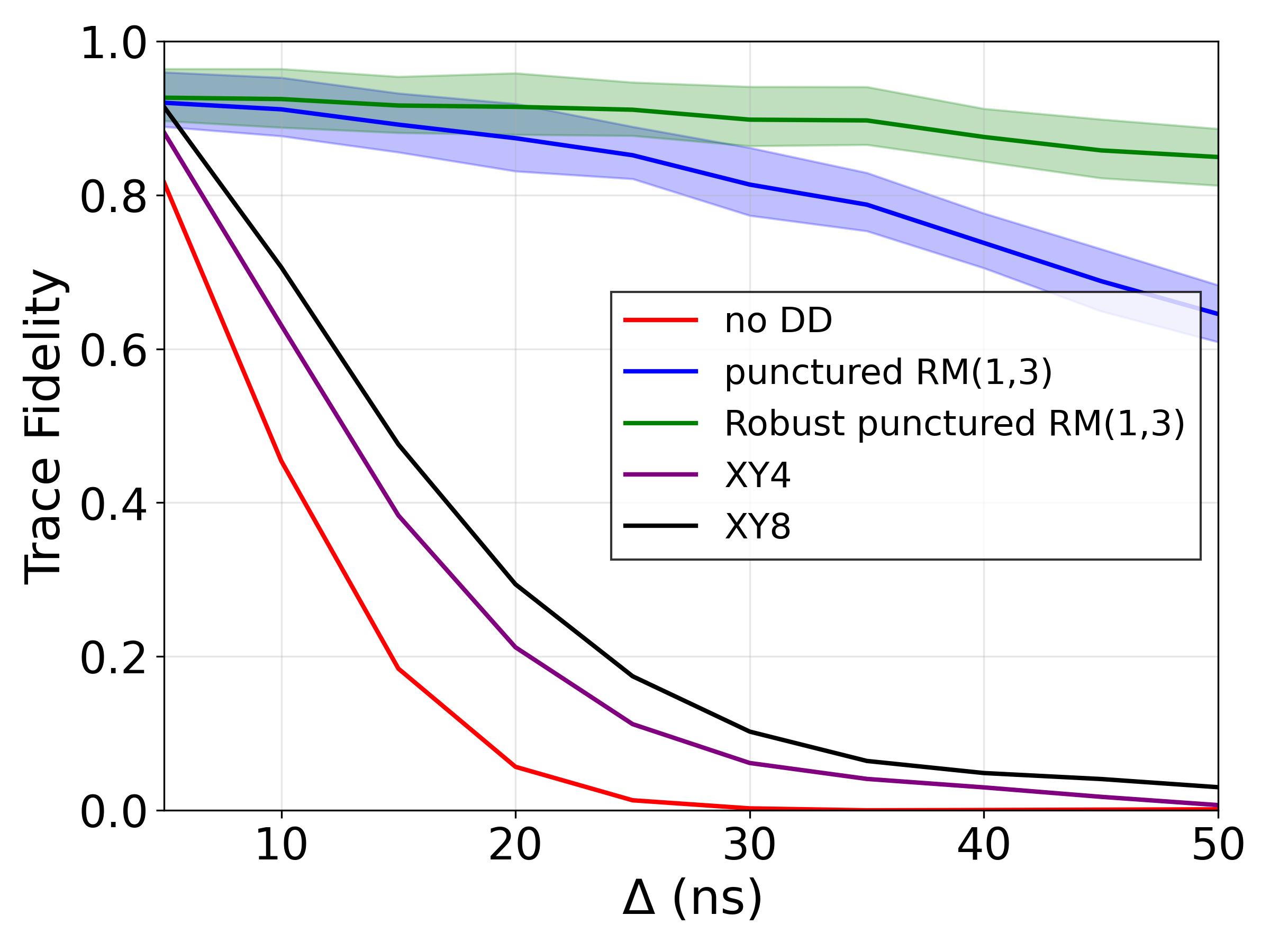}
    \caption{\justifying \textbf{Trace fidelity of different DD sequences.}
The mean value of trace fidelity as a function of the free-evolution segment duration \(\Delta \in [5,50]\,\mathrm{ns}\) (in step of $5~\rm ns$) for the seven-qubit bilinear array shown in Fig.~\ref{fig: G and expanded G}(a), subject to the error Hamiltonian in Eq.~\eqref{eq: phenological error Hamiltonian}. We compare four cases: no dynamical decoupling (red), the standard \(XY4\) sequence (purple), the punctured RM\((1,3)\) sequence (blue), and its robust implementation (green). All sequences are applied for the same total evolution time. Each data point is averaged over 200 Monte Carlo realizations. Shaded regions indicate the interquartile range (\(\pm 25\%\)) around the mean trace fidelity. Surprisingly, we find that the quartile range for the red, purple, and black curves are small and thus not visible on the plot.
}
    \label{fig: trace fidelity of DD}
\end{figure}

For each value of $\Delta$ ranging from $5~\rm ns$ to $50 \rm ns$ in step of $ 5~\rm ns$, we perform 200 Monte Carlo realizations. The results are shown in Fig.~\ref{fig: trace fidelity of DD}. Without dynamical decoupling, the accumulated errors rapidly scramble the qubit state. While the $XY4$ sequence suppresses single-qubit longitudinal and transversal coherent error, residual two-qubit interactions still lead to significant fidelity decay. We further observe that although the $XY8$ sequence improves the trace fidelity, it still exhibits an exponential decay due to these residual two-qubit interactions. In contrast, the punctured RM(1,3) DD sequence fully suppresses all error terms in the Hamiltonian of Eq.~\eqref{eq: phenological error Hamiltonian}. Finally, we observe that the robust implementation of the RM(1,3) DD sequence sequence yields a substantial improvement in trace fidelity as the segment duration $\Delta$ increases, in agreement with the arguments presented above.

In summary, by combining pulse scheduling with the mirror sequence, we achieve first-order suppression in the pulse error amplitude $\varepsilon$ and second-order suppression in the free-evolution time $\Delta$ at the cost of doubling the sequence length $L$.

\section{Conclusion and outlook}
\label{section: outlook}

In this work, we have introduced a unified framework for constructing dynamical decoupling (DD) sequences for quantum devices with arbitrary connectivity and general $k$-local noise models. We first showed that the task of suppressing $k$-local interactions can be substantially simplified by exploiting symmetries induced by noise locality. This reduction enables the construction of DD sequences whose length depends only on the device topology, rather than on the total system size. We then leveraged the connection between DD sequences and classical additive codes: when the decoupling group is a subgroup of the single-qubit Pauli group, DD sequences can be interpreted as quaternary error-detecting codes over $\mathbb{F}_4$. This perspective provides a systematic route to identifying good DD sequences from well-studied classical codes. Finally, we demonstrated how the group structure of these sequences can be exploited to construct implementations that are robust against systematic pulse over- and under-rotation errors.

As concrete demonstrations of our framework, we constructed explicit, time-efficient DD sequences tailored to superconducting and spin-qubit platforms that selectively suppress arbitrary interactions up to three-body order. Beyond error mitigation, we showed that the same framework can be used to engineer effective Hamiltonians for digital–analog quantum simulation. In particular, we presented a practical scheme for simulating the Kitaev honeycomb model on spin-qubit architectures, illustrating both the flexibility and the broader applicability of our approach.

Several interesting directions remain open for future investigation. Our current constructions are based on DD groups that are subgroups of the Pauli group, which naturally support a mapping to additive codes over $\mathbb{F}_4$. An important next step is to extend this framework to DD groups that are subgroups of the Clifford group \cite{read2025platonic,read2025factor}. Such sequences could achieve symmetrization of the system dynamics~\cite{zanardi1999symmetrizing,read2025platonic}, and could become an additional tool for Hamiltonian simulation \cite{Choi2020}. In fact, Ref.~\cite{read2025factor} is closely aligned in spirit with our work, leveraging the symmetries of specific spin–spin interactions to design efficient decoupling sequences based on single-qubit Clifford groups. Another natural and important question is to incorporate selective DD sequences into quantum gate operations. As demonstrated in our recent work~\cite{nguyen2025single}, inserting an echo pulse within a quantum gate that selectively flips interaction terms can reduce gate infidelity by at least an order of magnitude. A systematic extension of our framework to general quantum operations may be possible by combining it with dynamically corrected gate protocols~\cite{Khodjasteh2009}. Finally, our present analysis is restricted to DD sequences with an explicit group structure and equally spaced pulses. However, more general DD protocols that lack a strict group structure and employ non-equidistant pulse timings have been shown to provide higher-order error suppression with fewer pulses~\cite{Uhrig2007,yang2008universality,wang2011protection,zhang2025learningsteerquantummanybody}, as well as the ability to engineer effective interaction strengths beyond simple on–off control~\cite{hayes2014programmable,o2019hamiltonian,rajabi2019dynamical,Choi2020}. Extending our hypergraph-coloring–based approach to encompass these more general DD schemes remain an open question. 

\section{Acknowledgment}
We thank members of the Bosco, Rimbach-Russ, Terhal, and Vandersypen groups for valuable discussions. We are grateful to Pablo Cova Fariña for pointing out that the Kitaev honeycomb model can be simulated using a dynamical decoupling sequence of length $L=4$. We thank Yuning Zhang for helpful discussions on constructing robust dynamical decoupling sequences and for sharing his Mathematica code.

This research was supported by the EU through the H2024 QLSI2 project,  by the Army Research Office under Award Number: W911NF-23-1-0110, and by NCCR Spin (grant number 225153). The views and conclusions contained in this document are those of the authors and should not be interpreted as representing the official policies, either expressed or implied, of the Army Research Office or the U.S. Government. The U.S. Government is authorized to reproduce and distribute reprints for Government purposes notwithstanding any copyright notation herein.


\appendix

\section{Bounded control sequence}
\label{section: bounded sequence theory}
In this appendix, we review the bounded-control scheme for dynamical decoupling sequences \cite{Viola2003}. The key idea is that the control operations are generated by a set of generators of the decoupling group $\mathcal{G}$. The decoupling sequence is implemented by continuously and smoothly evolving between elements of $\mathcal{G}$ via these generators, rather than through instantaneous (bang-bang) pulses.

Let $\Gamma_{\mathcal{G}} = \{\gamma_1,\dots,\gamma_m \}$ denote a generating set of the decoupling group $\mathcal{G}$. We assume that each generator $\gamma_j \in \Gamma_{\mathcal{G}}$ can be implemented over a uniform time interval $\Delta$ using a time-dependent control Hamiltonian $h_{\gamma_j}(t)$ such that
\begin{equation}
    \gamma_{j} = \mathcal{T} \exp\Big(-i \int_{0}^{\Delta} dt~ h_{\gamma_j}(t)\Big).
\end{equation}

In the bounded control scheme, the control propagator $U_{c}$ follows a balanced cycle in $\mathrm{Cay}(\mathcal G,\Gamma_{\mathcal{G}})$. A balanced cycle is defined as a closed path that traverses each edge labeled by generator $\gamma_j$ exactly $\lambda_j$ times, where $\lambda_j$ is a positive integer. Since $\mathrm{Cay}(\mathcal G,\Gamma_{\mathcal{G}})$ is a regular graph, such a cycle always exists, and its length is exactly $|L| =|\mathcal{G}|(\sum_j \lambda_j)$.  For clarity, we denote $\Lambda = \sum_j \lambda_j$.

The control propagator $U_c$ starts at the identity element of the group $\mathcal{G}$ and evolves according to a path $L = \{ l_1,\dots,l_{|L|} \}$, where each $l_i \in \Gamma_{\mathcal{G}}$ specifies a generator. Due to the deterministic structure of the Cayley graph, each vertex admits a unique outgoing edge for a given generator, ensuring an unambiguous evolution along the path. The control propagator is defined piecewise over intervals of duration $\Delta$ as 
\begin{equation}
    U_{c}\Big[ (j-1)\Delta + \tau \Big]  =  u_{l_j}(\tau) U_{c}\Big[ (j-1)\Delta \Big],
\end{equation}
where 
\begin{equation}
    u_{l_j}(\tau) = \mathcal{T } \exp\Big(-i \int_{0}^{\tau} dt~ h_{l_j}(t)\Big)
\end{equation}
for $\tau \in [0,\Delta)$. 

Using Eq.~\eqref{eq: average Hamiltonian}, the average Hamiltonian $\overline{H}$ can be grouped into elements of the generating set $\Gamma_{\mathcal{G}}$, namely 
\begin{subequations}
    \begin{align}
    \overline{H} 
    & = \Pi_{\mathcal{G}}\Big( F_{\Gamma}(H) \Big)
    \end{align}
\end{subequations} 
where the channel $F_{\Gamma}(H)$ is defined as
\begin{equation}
    F_{\Gamma}(H) = \left(\sum_{\gamma_j \in \Gamma }  \frac{\lambda_j}{\Lambda \Delta} \int_{0}^{\Delta} dt~ u_{\gamma_j}^{\dagger}(t) H u_{\gamma_j}(t) \right).
\end{equation}
As a result, the total length of the bounded-control sequence increases by an additional factor proportional to the number of generators in $\Gamma_{\mathcal{G}}$.

Since $F_{\Gamma}$ is a trace-preserving quantum channel, it maps traceless Hamiltonians to traceless operators. Therefore, if $U_{g}$ defines an irreducible representation of $\mathcal{G}$, the decoupling sequence $\Pi_{\mathcal{G}}$ also decouples $F_{\Gamma}(H)$. However, in the more general case where $U_{g}$ is not irreducible, the decoupling is guaranteed only if $F_{\Gamma}(H)$ does not acquire components within the invariant subspace of $\Pi_{\mathcal{G}}$. Therefore, successful decoupling under bounded control requires designing both the decoupling group $\mathcal{G}$ and its generator set  $\Gamma_{\mathcal G}$ such that neither $H $ nor $F_{\Gamma}(H) $ has support in the invariant subspace preserved by the bang-bang decoupling $\Pi_{\mathcal{G}}$.

\section{Superconducting qubit (bounded control)}
\label{appendix: superconducting qubit (bounded control)}
In the main text, we showed how to construct a bang-bang decoupling sequence tailored to superconducting qubits using the binary Reed-Muller code. In this appendix, we extend that construction to the bounded-control setting, demonstrating how the same code framework can be adapted when only finite-time control pulses are available.

As a first step, we describe how to implement the control pulses associated with the generators of the decoupling group in the quotient hypergraph $I_{\rm Dev}^{'}/C[I_{\rm Dev}]$. Each generator $\gamma_{i} \in \mathcal{G}$ is a Pauli string of the form
\begin{equation}
    \gamma_{p_i} =  [\gamma_{p_i}]_{1} \otimes \dots [\gamma_{p_i}]_{\chi_G}, 
\end{equation}
where each element $[\gamma_{p_i}]_{k}$ acts on the $k$-th color class. To lift this operator from the quotient hypergraph back to the original interaction hypergraph, we impose that $[\gamma_{p_i}]_{k} $ acts uniformly on all qubits belonging to the color class $c_k$. That is, for each qubit $q_j \in c_k$, the same single-qubit Pauli operator $[\gamma_{p_i}]_{k} $ is applied.

We assume the control Hamiltonian used to implement $\gamma_{p_i}$ is composed of local, single-qubit Pauli operators, and is given by
\begin{equation}
    h_{\gamma_{pi}}(t) = \sum_{c_k \in C[G_{\rm Dev}]} \sum_{q_j \in c_k }  A_{k}^{q_j}(t)   [\gamma_{p_{i}}]_{k}^{(q_j)},
\end{equation}
where $[\gamma_{p_{i}}]_{k}^{(q_j)}$ denotes the local Pauli operator assigned to qubit $q_j \in c_k$ and $A_{k}^{q_j}(t)$ are the time-independent amplitude of the pulses. The pulse amplitudes are controlled such that
\begin{equation}
    \gamma_{p_i} = \mathcal{T} \exp\Big( -i \int_{0}^{\Delta}  h_{\gamma_{p_i}}(t)  \Big).
\end{equation}
The decomposition ensures that each control pulse can be executed using local, parallel single-qubit operations in a way that respects the color partition of the interaction hypergraph. 

As an example, we show that, under the substitution $0 \to I$ and $1 \to X$, the bounded-control decoupling sequence constructed from the $RM(1,m)$ code continues to suppress all $Z$-type Pauli strings of weight up to three. Since the generators are all $X$-type Pauli strings, the control Hamiltonian implements rotations around the $X$-axis. As a result, due to the finite duration of the control pulses, the channel $F_{\Gamma}$ rotates a Pauli $Z$ operator into linear combination of $Z$ and $Y$
\begin{equation}
    Z \to \cos(\alpha) Z + \sin(\alpha) Y. 
\end{equation}

The key insight is that the $Y$ Pauli operator remains detectable by the $X$-type generators, since $X$ anticommutes with $Y$. Therefore, any nontrivial operator in the image of $F_{\Gamma}$ still anticommutes with some generator in the decoupling group $\mathcal{G}$. As a result, $\mathcal{G}$ not only eliminates all $Z$-type strings of weight three or less but also cancels their rotated counterparts generated by the bounded-control channel $F_{\Gamma}$ thereby ensuring effective suppression even under bounded-control sequence.

For the universal bounded-control sequence constructed from the $\rm RM(1,m)$ code, code, we cannot simply apply the substitution 
$0$ with $I$ or $Z$ and $1$ with $X$ and $Y$ as done in the main text.This is because, under these substitutions, the channel $F_{\Gamma}$ generally maps any $Z$-type Pauli string into a linear combination of arbitrary Pauli strings of the same weight, involving any mixture of $X,Y$ and $Z$. Consequently, some of these terms may commute with all generators of the decoupling group $\mathcal{G}$ and therefore are not suppressed by the sequence. This highlights a key limitation of bounded-control schemes for universal decoupling: unless the decoupling group is carefully designed, the sequence may fail to eliminate all error terms.

To overcome this issue, we introduce an additional generator of the form
\begin{equation}
Z_{c_1} \otimes Z_{c_2} \otimes \dots \otimes Z_{c_{\chi_I}},
\end{equation}
which applies a global $Z$ operator to each color class. This operator commutes with all $Z$-type Pauli strings, meaning the corresponding control Hamiltonian does not introduce any unwanted interactions that would require further cancellation. Crucially, it anticommutes with any single-qubit $X$ or $Y$ operator acting on a color class, thereby enabling universal decoupling of single-qubit errors. As a result, the total length of the bounded-control decoupling sequence becomes
\begin{equation}
2^{\lceil \log_2(\chi_I +1)\rceil +1}\left(\lceil \log_2(\chi_I +1)\rceil +1\right) \sim O(\chi_I \log \chi_I),
\end{equation}
which scales linearly with the chromatic number $\chi_I$, with only a logarithmic overhead arising from the bounded-control implementation.

\section{Tailored sequence for two-local interactions in spin-qubit architecture with $\chi_I=4,5$}
\label{appendix: tailored sequence for two-local spin}

In this appendix, we discuss how to construct tailored sequence for two-local interactions in spin-qubit architecture with $\chi_I=4,5$ using only three generators. The construction is based on additive code in $\rm PG(2,2)$. 

We begin by labeling the seven points in the projective plane $\rm PG(2,2)$ as follows
\begin{equation}
\begin{bmatrix}
    P_1 & (1,0,0)^{T} \\
    P_2 & (0,1,0)^{T} \\
    P_3 & (0,0,1)^{T} \\
    P_4 & (1,1,0)^{T} \\
    P_5 & (1,0,1)^{T} \\
    P_6 & (0,1,1)^{T} \\
    P_7 & (1,1,1)^{T}
\end{bmatrix}.
\end{equation}
Throughout this section, we represent a line in $\rm PG(2,2)$ by an ordered pair of points $(P_i,P_j)$. The requirement that the code detects the Heisenberg exchange interaction translates into the following constraints on pairs of lines $(P_i,P_j)$ and $(P_n,P_m)$
\begin{subequations}
    \label{eq: heisenberg lines constraints}
    \begin{align}
       & P_i \neq P_n, ~P_j \neq P_, \\
       & P_i+P_j \neq P_n +P_m.
    \end{align}
\end{subequations}

These constraints can be reformulated as a combinatorial problem reminiscent of a Sudoku puzzle. Consider a $7 \times 7$ table whose rows and columns are labeled by the points $P_1$ through $P_7$. Each entry in the table is defined as the sum of the corresponding row and column labels
\begin{equation}
    \begin{tabular}{c|ccccccc}
 & P1 & P2 & P3 & P4 & P5 & P6 & P7 \\
\hline
P1 &   & P4 & P5 & P2 & P3 & P7 & P6 \\
P2 & P4 &   & P6 & P1 & P7 & P3 & P5 \\
P3 & P5 & P6 &   & P7 & P1 & P2 & P4 \\
P4 & P2 & P1 & P7 &   & P6 & P5 & P3 \\
P5 & P3 & P7 & P1 & P6 &   & P4 & P2 \\
P6 & P7 & P3 & P2 & P5 & P4 &   & P1 \\
P7 & P6 & P2 & P4 & P3 & P5 & P1 &   
\end{tabular}
\end{equation}
The problem then becomes finding the largest possible subset of table entries such that no row, column, or resulting entry value appears more than once. This ensures that the pairwise constraints on lines are satisfied. For small instances such as this $7\times 7$ table, heuristic search algorithms can be used to efficiently find such maximal subsets.

One such solution is the following set of line triples
\begin{subequations}
      \begin{align}
        &\{ (P_1,P_2,P_4),(P_2,P_3,P_6),(P_3,P_4,P_7) \\
        & ,(P_4,P_6,P_5),(P_5,P_7,P_2)  \}.
      \end{align}
\end{subequations}
Converting these lines back into their point coordinates in $\rm PG(2,2)$ and then into Pauli operators, we obtain the following set of additive generators for the tailored decoupling group:
\begin{equation}
    \begin{tabular}{|c c | c c | c c | c c | c c|}
    \toprule
    \multicolumn{2}{|c|}{$C_1$} &
    \multicolumn{2}{c|}{$C_2$} &
    \multicolumn{2}{c|}{$C_3$} &
    \multicolumn{2}{c|}{$C_4$} &
    \multicolumn{2}{c|}{$C_5$} \\
    \midrule
    1 & 0 & 0 & 0 & 0 & 1 & 1 & 0 & 1 & 1 \\
    0 & 1 & 1 & 0 & 0 & 1 & 1 & 1 & 0 & 1 \\
    0 & 0 & 0 & 1 & 1 & 0 & 0 & 1 & 1 & 1 \\
    \bottomrule
    \end{tabular} \to \begin{tabular}{|c  | c  | c | c  | c |}
    \toprule
    \multicolumn{1}{|c|}{$C_1$} &
    \multicolumn{1}{c|}{$C_2$} &
    \multicolumn{1}{c|}{$C_3$} &
    \multicolumn{1}{c|}{$C_4$} &
    \multicolumn{1}{c|}{$C_5$} \\
    \midrule
    X & I & Z & X & Y \\
    Z & X & Z & Y & Z \\
    I & Z & X & Z & Y \\
    \bottomrule
    \end{tabular}
\end{equation}
It is straightforward to verify that this decoupling group universally suppresses all single-qubit terms as well as Heisenberg exchange interactions in systems with$\chi_I = 4$ or $5$. 

In general, finding the largest set of ordered point pairs in $\rm PG(n,2)$ that satisfy the constraints of Eqs.~\ref{eq: heisenberg lines constraints} is computationally difficult. In fact, this problem can be shown to be NP-hard, as it is a specific instance of the well-known 3-dimensional matching problem. Consequently, the simple expansion strategy presented in the main text should be viewed as a heuristic algorithm. While effective for small instances, it leaves significant room for optimization in larger settings.

\section{Simulating Kitaev's honeycomb model with bounded control}
\label{appendix: simulating Kitaev's honeycomb model with bounded control}

For the bounded-control setting, however, additional terms can be generated through the channel $F_{\Gamma}$.  Accounting for these terms, we numerically find that all four operators $W_{1},W_{2},W_{3},$ and $W_4$ are required to fully decouple the undesired interactions. The resulting sequence removes all two-local terms except those in $H_{\rm comb}$. Moreover, by tuning the control Hamiltonian appropriately, it becomes possible to simulate an anisotropic Kitaev model with $J_{x} \neq J_y \neq J_z$, even if the native system Hamiltonian consists solely of Heisenberg exchange interactions. 
The total number of time slices in the bounded-control protocol is then given by
\begin{equation}
2^4 \times 4 = 64.
\end{equation}

\section{Sequence robustness under misaligned rotation axis}
\label{appendix: robustness under misaligned rotation axis}
In Sec.~\ref{section: robust pulse sequences} of the main text, we assumed that the dominant pulse error arises from systematic over- or under-rotation and presented strategies for making DD sequences robust against such errors. In this appendix, we analyze the case in which the dominant pulse error instead originates from misalignment of the rotation axis, and we explain why the proposed strategies are ineffective against this type of pulse error. 

As a concrete example, we consider a noisy implementation of the $(\tilde X)_{\theta}$ pulse given by 
\begin{equation}
    (\tilde{X})_{\theta } \equiv \exp\Big[-\frac{i}{2}\theta(\sqrt{1-\eta_y^2-\eta_z^2} X + \eta_y Y + \eta_z Z)  \Big],
\end{equation}
where $\eta_y, \eta_z \ll 1$ quantify small deviations of the rotation axis away from the ideal $X$ direction. As in the main text, we decompose the noisy pulse into an error operator $E[(X)_{\theta}]$ followed by an ideal $(X)_{\theta}$ pulse
\begin{equation}
    \label{eq: definition of pulse decomposition}
    (\tilde{X})_{\theta} = (X)_{\theta}~E[(X)_{\theta}].
\end{equation}

To determine the explicit form of the error operator $E[(X)_{\theta}]$, we employ the following math trick. Consider
\begin{equation}
\label{eq: set up for differential equation}
\exp(i \lambda H)\exp[-i\lambda (H + \eta)] = F(\lambda),
\end{equation}
with $F(0)=I$. Differentiating both sides with respect to $\lambda$ yields
\begin{subequations}
\begin{align}
\frac{d F(\lambda)}{d \lambda}
&= \exp(i \lambda H)(-i \eta)\exp[-i\lambda (H + \eta)] \\
&= \exp(i \lambda H)(-i \eta)\exp(-i \lambda H),F(\lambda),
\end{align}
\end{subequations}
where, in going from the first to the second line, we have inserted the identity $I=e^{-i \lambda H} e^{i \lambda H}$ and used Eq.~\eqref{eq: set up for differential equation}. To first order in $||\eta||$, the solution at $\lambda=1$ is given by the first-order Magnus expansion,
\begin{equation}
F(1) \approx \exp\Big[-i \int_{0}^{1} d\lambda
\big(\exp[i \lambda H]\eta\exp[-i \lambda H]\big)\Big].
\end{equation}
Applying this result to the present case, with the identifications
\begin{equation}
\exp[-i\lambda (H + \eta)] = (\tilde{X})_{\theta},
\quad
\exp[-i\lambda H] = (X)_{\theta},
\end{equation}
and $\theta=\pi$, we obtain the error operator
\begin{equation}
E[(X)_{\pi}] \approx
\exp\Big[-i\Big(-\eta_y Z + \eta_z Y\Big)\Big].
\end{equation}

Importantly, we find that the error operator associated with a noisy $(\tilde X)_{-\pi}$ pulse is identical to that of a $(\tilde X)_{\pi}$ pulse. To first order in the misalignment parameters $\eta$, we have
\begin{equation}
\label{eq: identical error pulse for misaligned rotation axis}
E[(X)_{-\pi}] \approx E[(X)_{\pi}] .
\end{equation}
This result can be directly verified using the identity
\begin{subequations}
\begin{align}
I &= (\tilde X)_{-\pi}(\tilde X)_{\pi} \\
&\approx \exp\Big[i\big(-\eta_y Z + \eta_z Y\big)\Big]
\exp\Big[-i\big(-\eta_y Z + \eta_z Y\big)\Big] \\
&= I ,
\end{align}
\end{subequations}
where we have kept terms only to first order in $\eta$. The fact that $E[(X)_{-\pi}]$ coincides with $E[(X)_{\pi}]$ for axis-misalignment errors has important consequences. By contrast, for over- and under-rotation errors one finds $E[(X)_{-\pi}] = E[(X)_{\pi}]^{\dagger}$, a property that allows cumulative pulse errors $\tilde E$ (given in Eq.~\eqref{eq: cumulative pulse error}) to be canceled by alternating between $(\pi)$ and $(-\pi)$ pulses, as discussed in Sec.~\ref{section: scheduling pi and -pi pulses}. This cancellation mechanism, however, is absent for misaligned rotation axis errors. Consequently, alternating between $(\pi)$ and $(-\pi)$ pulses is ineffective in this case, and suppressing the accumulated pulse error instead requires choosing a Hamiltonian cycle such that the exponent on the right-hand side of Eq.~\eqref{eq: cumulative pulse error} vanishes.

Another important distinction arises in the commutation properties of the pulse error $E[(X)_{\pi}]$ through the ideal pulse $(X)_{\pi}$. For misaligned rotation axis errors, we find
\begin{equation}
    \label{eq: commutation relation of misaligned rotation axis error}
    (X)_{\pi}~E[(X)_{\pi}] = E[(X)_{\pi}]^{\dagger}  (X)_{\pi}
\end{equation}
whereas for over- and under-rotation errors the corresponding relation is
\begin{equation}
    (X)_{\pi}~E[(X)_{\pi}] = E[(X)_{\pi}]  (X)_{\pi}.
\end{equation}
Owing to the property in Eq.~\eqref{eq: commutation relation of misaligned rotation axis error}, the mirror-sequence strategy introduced in Sec.~\ref{section: mirror sequence} can in fact be used to suppress the cumulative pulse errors $\tilde E$ arising from axis misalignment. However, this modified commutation relation also leads to a crucial limitation. Repeating the reasoning following Eq.~\eqref{eq: reverse error decomposition of mirror sequence}, one finds that the mirror sequence no longer cancels the residual Hamiltonian error $H_{\rm error}$. More explicitly, for misaligned rotation-axis errors, if the noisy DD sequence takes the form
\begin{equation}
U_{\Delta}\tilde{p}_L \dots U_{\Delta}\tilde{p}_1
= \Big( \prod_{i=1}^{L} \tilde E[p_i] \Big)
e^{-i \Delta L (H_{\rm targ}+H_{\rm error})},
\end{equation}
then the corresponding mirror sequence necessarily has the structure
\begin{equation}
\tilde{p}_1 U_{\Delta} \dots \tilde{p}_L U_{\Delta}
= e^{-i \Delta L (H_{\rm targ}-H_{\rm error})}
\Big( \prod_{i=1}^{L} \tilde E[p_i] \Big)^{\dagger}.
\end{equation}
As a result, while the cumulative propagated pulse error $\Big( \prod_{i=1}^{L} \tilde E[p_i] \Big)$ cancel out, the residual Hamiltonian errors add up, preventing full error suppression.

In summary, for misaligned rotation-axis errors, the pulse-scheduling strategies presented in Sec.~\ref{section: scheduling pi and -pi pulses} are no longer effective. The mirror-sequence extension discussed in Sec.~\ref{section: mirror sequence} can partially suppress pulse errors, but it does not eliminate them entirely. Consequently, an additional strategy is required to fully mitigate errors due to rotation-axis misalignment. We do not identify a simple solution within the scope of this work and therefore leave this problem for future investigation.

\bibliography{reference}

@article{Gullans2019,
  title = {Protocol for a resonantly driven three-qubit Toffoli gate with silicon spin qubits},
  author = {Gullans, M. J. and Petta, J. R.},
  journal = {Phys. Rev. B},
  volume = {100},
  issue = {8},
  pages = {085419},
  numpages = {7},
  year = {2019},
  month = {Aug},
  publisher = {American Physical Society},
  doi = {10.1103/PhysRevB.100.085419},
  url = {https://link.aps.org/doi/10.1103/PhysRevB.100.085419}
}

@article{Geyer2024,
author={Geyer, Simon
and Het{\'e}nyi, Bence
and Bosco, Stefano
and Camenzind, Leon C.
and Eggli, Rafael S.
and Fuhrer, Andreas
and Loss, Daniel
and Warburton, Richard J.
and Zumb{\"u}hl, Dominik M.
and Kuhlmann, Andreas V.},
title={Anisotropic exchange interaction of two hole-spin qubits},
journal={Nature Physics},
year={2024},
month={Jul},
day={01},
volume={20},
number={7},
pages={1152-1157},
issn={1745-2481},
doi={10.1038/s41567-024-02481-5},
url={https://doi.org/10.1038/s41567-024-02481-5}
}

@misc{Jiaan2024,
      title={Scalable multi-qubit intrinsic gates in quantum dot arrays}, 
      author={Jiaan Qi and Zhi-Hai Liu and Hongqi Xu},
      year={2024},
      eprint={2403.06894},
      archivePrefix={arXiv},
      primaryClass={quant-ph},
      url={https://arxiv.org/abs/2403.06894}, 
}

@article{Sen1995,
  title = {Large-U limit of a Hubbard model in a magnetic field: Chiral spin interactions and paramagnetism},
  author = {Sen, Diptiman and Chitra, R.},
  journal = {Phys. Rev. B},
  volume = {51},
  issue = {3},
  pages = {1922--1925},
  numpages = {0},
  year = {1995},
  month = {Jan},
  publisher = {American Physical Society},
  doi = {10.1103/PhysRevB.51.1922},
  url = {https://link.aps.org/doi/10.1103/PhysRevB.51.1922}
}

@article{Menke2022,
  title = {Demonstration of Tunable Three-Body Interactions between Superconducting Qubits},
  author = {Menke, Tim and Banner, William P. and Bergamaschi, Thomas R. and Di Paolo, Agustin and Veps\"al\"ainen, Antti and Weber, Steven J. and Winik, Roni and Melville, Alexander and Niedzielski, Bethany M. and Rosenberg, Danna and Serniak, Kyle and Schwartz, Mollie E. and Yoder, Jonilyn L. and Aspuru-Guzik, Al\'an and Gustavsson, Simon and Grover, Jeffrey A. and Hirjibehedin, Cyrus F. and Kerman, Andrew J. and Oliver, William D.},
  journal = {Phys. Rev. Lett.},
  volume = {129},
  issue = {22},
  pages = {220501},
  numpages = {6},
  year = {2022},
  month = {Nov},
  publisher = {American Physical Society},
  doi = {10.1103/PhysRevLett.129.220501},
  url = {https://link.aps.org/doi/10.1103/PhysRevLett.129.220501}
}

@article{Viola2003,
  title = {Robust Dynamical Decoupling of Quantum Systems with Bounded Controls},
  author = {Viola, Lorenza and Knill, Emanuel},
  journal = {Phys. Rev. Lett.},
  volume = {90},
  issue = {3},
  pages = {037901},
  numpages = {4},
  year = {2003},
  month = {Jan},
  publisher = {American Physical Society},
  doi = {10.1103/PhysRevLett.90.037901},
  url = {https://link.aps.org/doi/10.1103/PhysRevLett.90.037901}
}

@article{Xue2022,
author={Xue, Xiao
and Russ, Maximilian
and Samkharadze, Nodar
and Undseth, Brennan
and Sammak, Amir
and Scappucci, Giordano
and Vandersypen, Lieven M. K.},
title={Quantum logic with spin qubits crossing the surface code threshold},
journal={Nature},
year={2022},
month={Jan},
day={01},
volume={601},
number={7893},
pages={343-347},
issn={1476-4687},
doi={10.1038/s41586-021-04273-w},
url={https://doi.org/10.1038/s41586-021-04273-w}
}

@article{Pederson2019,
  title = {Native three-body interaction in superconducting circuits},
  author = {Pedersen, Simon Panyella and Christensen, K. S. and Zinner, N. T.},
  journal = {Phys. Rev. Res.},
  volume = {1},
  issue = {3},
  pages = {033123},
  numpages = {17},
  year = {2019},
  month = {Nov},
  publisher = {American Physical Society},
  doi = {10.1103/PhysRevResearch.1.033123},
  url = {https://link.aps.org/doi/10.1103/PhysRevResearch.1.033123}
}

@misc{bosco2024exchangeonlyspinorbitqubitssilicon,
      title={Exchange-Only Spin-Orbit Qubits in Silicon and Germanium}, 
      author={Stefano Bosco and Maximilian Rimbach-Russ},
      year={2024},
      eprint={2410.05461},
      archivePrefix={arXiv},
      primaryClass={cond-mat.mes-hall},
      url={https://arxiv.org/abs/2410.05461}, 
}

@misc{saezmollejo2024exchangeanisotropiesmicrowavedrivensinglettriplet,
      title={Exchange anisotropies in microwave-driven singlet-triplet qubits}, 
      author={Jaime Saez-Mollejo and Daniel Jirovec and Yona Schell and Josip Kukucka and Stefano Calcaterra and Daniel Chrastina and Giovanni Isella and Maximilian Rimbach-Russ and Stefano Bosco and Georgios Katsaros},
      year={2024},
      eprint={2408.03224},
      archivePrefix={arXiv},
      primaryClass={cond-mat.mes-hall},
      url={https://arxiv.org/abs/2408.03224}, 
}

@article{burkard2023semiconductor,
  title={Semiconductor spin qubits},
  author={Burkard, Guido and Ladd, Thaddeus D and Pan, Andrew and Nichol, John M and Petta, Jason R},
  journal={Reviews of Modern Physics},
  volume={95},
  number={2},
  pages={025003},
  year={2023},
  publisher={APS}
}

@article{noiri2022fast,
  title={Fast universal quantum gate above the fault-tolerance threshold in silicon},
  author={Noiri, Akito and Takeda, Kenta and Nakajima, Takashi and Kobayashi, Takashi and Sammak, Amir and Scappucci, Giordano and Tarucha, Seigo},
  journal={Nature},
  volume={601},
  number={7893},
  pages={338--342},
  year={2022},
  publisher={Nature Publishing Group UK London}
}

@article{mills2022two,
  title={Two-qubit silicon quantum processor with operation fidelity exceeding 99\%},
  author={Mills, Adam R and Guinn, Charles R and Gullans, Michael J and Sigillito, Anthony J and Feldman, Mayer M and Nielsen, Erik and Petta, Jason R},
  journal={Science Advances},
  volume={8},
  number={14},
  pages={eabn5130},
  year={2022},
  publisher={American Association for the Advancement of Science}
}

@article{john2024two,
  title={A two-dimensional 10-qubit array in germanium with robust and localised qubit control},
  author={John, Valentin and Yu, C{\'e}cile X and van Straaten, Barnaby and Rodr{\'\i}guez-Mena, Esteban A and Rodr{\'\i}guez, Mauricio and Oosterhout, Stefan and Stehouwer, Lucas EA and Scappucci, Giordano and Bosco, Stefano and Rimbach-Russ, Maximilian and others},
  journal={arXiv preprint arXiv:2412.16044},
  year={2024}
}

@article{brown2024efficient,
  title={Efficient Chromatic-Number-Based Multi-Qubit Decoherence and Crosstalk Suppression},
  author={Brown, Amy F and Lidar, Daniel A},
  journal={arXiv preprint arXiv:2406.13901},
  year={2024}
}

@article{zanardi1999symmetrizing,
  title={Symmetrizing evolutions},
  author={Zanardi, Paolo},
  journal={Physics Letters A},
  volume={258},
  number={2-3},
  pages={77--82},
  year={1999},
  publisher={Elsevier}
}

@article{ezerman2011additive,
  title={Additive asymmetric quantum codes},
  author={Ezerman, Martianus Frederic and Ling, San and Sole, Patrick},
  journal={IEEE Transactions on Information Theory},
  volume={57},
  number={8},
  pages={5536--5550},
  year={2011},
  publisher={IEEE}
}

@article{evert2024syncopated,
  title={Syncopated dynamical decoupling for suppressing crosstalk in quantum circuits},
  author={Evert, Bram and Izquierdo, Zoe Gonzalez and Sud, James and Hu, Hong-Ye and Grabbe, Shon and Rieffel, Eleanor G and Reagor, Matthew J and Wang, Zhihui},
  journal={arXiv preprint arXiv:2403.07836},
  year={2024}
}

@article{nguyen2025single,
  title={Single-step high-fidelity three-qubit gates by anisotropic chiral interactions},
  author={Nguyen, Minh TP and Rimbach-Russ, Maximilian and Vandersypen, Lieven MK and Bosco, Stefano},
  journal={arXiv preprint arXiv:2503.12182},
  year={2025},
  doi = {10.1103/kp8s-py9m},
  url = {https://link.aps.org/doi/10.1103/kp8s-py9m},
}

@article{Krodjasteh2008,
  title = {Rigorous bounds on the performance of a hybrid dynamical-decoupling quantum-computing scheme},
  author = {Khodjasteh, Kaveh and Lidar, Daniel A.},
  journal = {Phys. Rev. A},
  volume = {78},
  issue = {1},
  pages = {012355},
  numpages = {14},
  year = {2008},
  month = {Jul},
  publisher = {American Physical Society},
  doi = {10.1103/PhysRevA.78.012355},
  url = {https://link.aps.org/doi/10.1103/PhysRevA.78.012355}
}

@article{Bookatz_2016,
   title={Improved Bounded-Strength Decoupling Schemes for Local Hamiltonians},
   volume={62},
   ISSN={1557-9654},
   url={http://dx.doi.org/10.1109/TIT.2016.2535183},
   DOI={10.1109/tit.2016.2535183},
   number={5},
   journal={IEEE Transactions on Information Theory},
   publisher={Institute of Electrical and Electronics Engineers (IEEE)},
   author={Bookatz, Adam D. and Roetteler, Martin and Wocjan, Pawel},
   year={2016},
   month=may, pages={2881–2894} }

@article{viola1999dynamical,
  title={Dynamical decoupling of open quantum systems},
  author={Viola, Lorenza and Knill, Emanuel and Lloyd, Seth},
  journal={Physical Review Letters},
  volume={82},
  number={12},
  pages={2417},
  year={1999},
  publisher={APS}
}

@article{viola1998bang,
  title = {Dynamical suppression of decoherence in two-state quantum systems},
  author = {Viola, Lorenza and Lloyd, Seth},
  journal = {Phys. Rev. A},
  volume = {58},
  issue = {4},
  pages = {2733--2744},
  numpages = {0},
  year = {1998},
  month = {Oct},
  publisher = {American Physical Society},
  doi = {10.1103/PhysRevA.58.2733},
  url = {https://link.aps.org/doi/10.1103/PhysRevA.58.2733}
}

@article{leung2002simulation,
  title={Simulation and reversal of n-qubit Hamiltonians using Hadamard matrices},
  author={Leung, Debbie},
  journal={Journal of Modern Optics},
  volume={49},
  number={8},
  pages={1199--1217},
  year={2002},
  publisher={Taylor \& Francis}
}

@article{Rotteler_2006,
   title={Equivalence of Decoupling Schemes and Orthogonal Arrays},
   volume={52},
   ISSN={0018-9448},
   url={http://dx.doi.org/10.1109/TIT.2006.880059},
   DOI={10.1109/tit.2006.880059},
   number={9},
   journal={IEEE Transactions on Information Theory},
   publisher={Institute of Electrical and Electronics Engineers (IEEE)},
   author={Rotteler, M. and Wocjan, P.},
   year={2006},
   month=sep, pages={4171–4181} }

@misc{neilsloaneOrthogonalArrays,
	author = {},
	title = { {O}rthogonal {A}rrays  --- neilsloane.com},
	howpublished = {\url{http://neilsloane.com/oadir/index.html}},
}

@article{DELSARTE1973407,
title = {Four fundamental parameters of a code and their combinatorial significance},
journal = {Information and Control},
volume = {23},
number = {5},
pages = {407-438},
year = {1973},
issn = {0019-9958},
doi = {https://doi.org/10.1016/S0019-9958(73)80007-5},
url = {https://www.sciencedirect.com/science/article/pii/S0019995873800075},
author = {Philippe Delsarte}
}

@book{hedayat2012orthogonal,
  title={Orthogonal arrays: theory and applications},
  author={Hedayat, A Samad and Sloane, Neil James Alexander and Stufken, John},
  year={2012},
  publisher={Springer Science \& Business Media}
}

@misc{calderbank1997quantumerrorcorrectioncodes,
      title={Quantum Error Correction via Codes over GF(4)}, 
      author={A. R. Calderbank and E. M Rains and P. W. Shor and N. J. A. Sloane},
      year={1997},
      eprint={quant-ph/9608006},
      archivePrefix={arXiv},
      primaryClass={quant-ph},
      url={https://arxiv.org/abs/quant-ph/9608006}, 
}

@article{zeng2011transversality,
  title={Transversality versus universality for additive quantum codes},
  author={Zeng, Bei and Cross, Andrew and Chuang, Isaac L},
  journal={IEEE Transactions on Information Theory},
  volume={57},
  number={9},
  pages={6272--6284},
  year={2011},
  publisher={IEEE}
}

@article{haah2016algebraic,
  title={Algebraic methods for quantum codes on lattices},
  author={Haah, Jeongwan},
  journal={arXiv preprint arXiv:1607.01387},
  year={2016}
}

@article{FU2015,
title = {Large caps in projective space PG(r,4)},
journal = {Finite Fields and Their Applications},
volume = {35},
pages = {231-246},
year = {2015},
issn = {1071-5797},
doi = {https://doi.org/10.1016/j.ffa.2015.04.006},
url = {https://www.sciencedirect.com/science/article/pii/S1071579715000544},
author = {Qiang Fu and Ruihu Li and Luobin Guo and Gen Xu}
}

@article{BLOKHUIS2004161,
title = {Small additive quaternary codes},
journal = {European Journal of Combinatorics},
volume = {25},
number = {2},
pages = {161-167},
year = {2004},
issn = {0195-6698},
doi = {https://doi.org/10.1016/S0195-6698(03)00096-9},
url = {https://www.sciencedirect.com/science/article/pii/S0195669803000969},
author = {A Blokhuis and A.E Brouwer}
}

@article{Stollsteimer2001,
  title = {Suppression of arbitrary internal coupling in a quantum register},
  author = {Stollsteimer, Marcus and Mahler, G\"unter},
  journal = {Phys. Rev. A},
  volume = {64},
  issue = {5},
  pages = {052301},
  numpages = {8},
  year = {2001},
  month = {Oct},
  publisher = {American Physical Society},
  doi = {10.1103/PhysRevA.64.052301},
  url = {https://link.aps.org/doi/10.1103/PhysRevA.64.052301}
}

@article{tong2024empirical,
  title={Empirical learning of dynamical decoupling on quantum processors},
  author={Tong, Christopher and Zhang, Helena and Pokharel, Bibek},
  journal={arXiv preprint arXiv:2403.02294},
  year={2024}
}

@article{rahman2024learning,
  title={Learning how to dynamically decouple by optimizing rotational gates},
  author={Rahman, Arefur and Egger, Daniel J and Arenz, Christian},
  journal={Physical Review Applied},
  volume={22},
  number={5},
  pages={054074},
  year={2024},
  publisher={APS}
}

@article{kitaev2006anyons,
title = {Anyons in an exactly solved model and beyond},
journal = {Annals of Physics},
volume = {321},
number = {1},
pages = {2-111},
year = {2006},
note = {January Special Issue},
issn = {0003-4916},
doi = {https://doi.org/10.1016/j.aop.2005.10.005},
url = {https://www.sciencedirect.com/science/article/pii/S0003491605002381},
author = {Alexei Kitaev}
}

@article{Hahn1950,
  title = {Spin Echoes},
  author = {Hahn, E. L.},
  journal = {Phys. Rev.},
  volume = {80},
  issue = {4},
  pages = {580--594},
  numpages = {0},
  year = {1950},
  month = {Nov},
  publisher = {American Physical Society},
  doi = {10.1103/PhysRev.80.580},
  url = {https://link.aps.org/doi/10.1103/PhysRev.80.580}
}

@article{Meiboom1958,
    author = {Meiboom, S. and Gill, D.},
    title = {Modified Spin‐Echo Method for Measuring Nuclear Relaxation Times},
    journal = {Review of Scientific Instruments},
    volume = {29},
    number = {8},
    pages = {688-691},
    year = {1958},
    month = {08},
    issn = {0034-6748},
    doi = {10.1063/1.1716296},
    url = {https://doi.org/10.1063/1.1716296},
}

@article{maudsley1986modified,
  title={Modified Carr-Purcell-Meiboom-Gill sequence for NMR fourier imaging applications},
  author={Maudsley, AA},
  journal={Journal of Magnetic Resonance (1969)},
  volume={69},
  number={3},
  pages={488--491},
  year={1986},
  publisher={Elsevier}
}

@article{Vitali1999,
  title = {Using parity kicks for decoherence control},
  author = {Vitali, D. and Tombesi, P.},
  journal = {Phys. Rev. A},
  volume = {59},
  issue = {6},
  pages = {4178--4186},
  numpages = {0},
  year = {1999},
  month = {Jun},
  publisher = {American Physical Society},
  doi = {10.1103/PhysRevA.59.4178},
  url = {https://link.aps.org/doi/10.1103/PhysRevA.59.4178}
}

@article{duan1999suppressing,
  title={Suppressing environmental noise in quantum computation through pulse control},
  author={Duan, Lu-Ming and Guo, Guang-Can},
  journal={Physics Letters A},
  volume={261},
  number={3-4},
  pages={139--144},
  year={1999},
  publisher={Elsevier}
}

@article{szankowski2017environmental,
  title={Environmental noise spectroscopy with qubits subjected to dynamical decoupling},
  author={Sza{\'n}kowski, Piotr and Ramon, Guy and Krzywda, Jan and Kwiatkowski, Damian and others},
  journal={Journal of Physics: Condensed Matter},
  volume={29},
  number={33},
  pages={333001},
  year={2017},
  publisher={IOP Publishing}
}

@article{Alvarez2011,
  title = {Measuring the Spectrum of Colored Noise by Dynamical Decoupling},
  author = {\'Alvarez, Gonzalo A. and Suter, Dieter},
  journal = {Phys. Rev. Lett.},
  volume = {107},
  issue = {23},
  pages = {230501},
  numpages = {5},
  year = {2011},
  month = {Nov},
  publisher = {American Physical Society},
  doi = {10.1103/PhysRevLett.107.230501},
  url = {https://link.aps.org/doi/10.1103/PhysRevLett.107.230501}
}

@article{Tripathi2022,
  title = {Suppression of Crosstalk in Superconducting Qubits Using Dynamical Decoupling},
  author = {Tripathi, Vinay and Chen, Huo and Khezri, Mostafa and Yip, Ka-Wa and Levenson-Falk, E.M. and Lidar, Daniel A.},
  journal = {Phys. Rev. Appl.},
  volume = {18},
  issue = {2},
  pages = {024068},
  numpages = {24},
  year = {2022},
  month = {Aug},
  publisher = {American Physical Society},
  doi = {10.1103/PhysRevApplied.18.024068},
  url = {https://link.aps.org/doi/10.1103/PhysRevApplied.18.024068}
}

@article{qiu2021suppressing,
  title={Suppressing coherent two-qubit errors via dynamical decoupling},
  author={Qiu, Jiawei and Zhou, Yuxuan and Hu, Chang-Kang and Yuan, Jiahao and Zhang, Libo and Chu, Ji and Huang, Wenhui and Liu, Weiyang and Luo, Kai and Ni, Zhongchu and others},
  journal={Physical Review Applied},
  volume={16},
  number={5},
  pages={054047},
  year={2021},
  publisher={APS}
}

@article{West2010,
  title = {Near-Optimal Dynamical Decoupling of a Qubit},
  author = {West, Jacob R. and Fong, Bryan H. and Lidar, Daniel A.},
  journal = {Phys. Rev. Lett.},
  volume = {104},
  issue = {13},
  pages = {130501},
  numpages = {4},
  year = {2010},
  month = {Apr},
  publisher = {American Physical Society},
  doi = {10.1103/PhysRevLett.104.130501},
  url = {https://link.aps.org/doi/10.1103/PhysRevLett.104.130501}
}

@article{pasini2010optimized,
  title={Optimized dynamical decoupling for time-dependent Hamiltonians},
  author={Pasini, Stefano and Uhrig, G{\"o}tz S},
  journal={Journal of Physics A: Mathematical and Theoretical},
  volume={43},
  number={13},
  pages={132001},
  year={2010},
  publisher={IOP Publishing}
}

@article{Uhrig2007,
  title = {Keeping a Quantum Bit Alive by Optimized $\ensuremath{\pi}$-Pulse Sequences},
  author = {Uhrig, G\"otz S.},
  journal = {Phys. Rev. Lett.},
  volume = {98},
  issue = {10},
  pages = {100504},
  numpages = {4},
  year = {2007},
  month = {Mar},
  publisher = {American Physical Society},
  doi = {10.1103/PhysRevLett.98.100504},
  url = {https://link.aps.org/doi/10.1103/PhysRevLett.98.100504}
}

@article{wang2011protection,
  title={Protection of quantum systems by nested dynamical decoupling},
  author={Wang, Zhen-Yu and Liu, Ren-Bao},
  journal={Physical Review A—Atomic, Molecular, and Optical Physics},
  volume={83},
  number={2},
  pages={022306},
  year={2011},
  publisher={APS}
}

@article{yang2008universality,
  title={Universality of Uhrig Dynamical Decoupling for Suppressing Qubit Pure Dephasing<? format?> and Relaxation},
  author={Yang, Wen and Liu, Ren-Bao},
  journal={Physical review letters},
  volume={101},
  number={18},
  pages={180403},
  year={2008},
  publisher={APS}
}

@article{agarwal2020dynamical,
  title={Dynamical enhancement of symmetries in many-body systems},
  author={Agarwal, Kartiek and Martin, Ivar},
  journal={Physical Review Letters},
  volume={125},
  number={8},
  pages={080602},
  year={2020},
  publisher={APS}
}

@article{WAHUHA1968,
  title = {Approach to High-Resolution nmr in Solids},
  author = {Waugh, J. S. and Huber, L. M. and Haeberlen, U.},
  journal = {Phys. Rev. Lett.},
  volume = {20},
  issue = {5},
  pages = {180--182},
  numpages = {0},
  year = {1968},
  month = {Jan},
  publisher = {American Physical Society},
  doi = {10.1103/PhysRevLett.20.180},
  url = {https://link.aps.org/doi/10.1103/PhysRevLett.20.180}
}

@article{Choi2020,
  title = {Robust Dynamic Hamiltonian Engineering of Many-Body Spin Systems},
  author = {Choi, Joonhee and Zhou, Hengyun and Knowles, Helena S. and Landig, Renate and Choi, Soonwon and Lukin, Mikhail D.},
  journal = {Phys. Rev. X},
  volume = {10},
  issue = {3},
  pages = {031002},
  numpages = {27},
  year = {2020},
  month = {Jul},
  publisher = {American Physical Society},
  doi = {10.1103/PhysRevX.10.031002},
  url = {https://link.aps.org/doi/10.1103/PhysRevX.10.031002}
}

@misc{vezvaee2025demonstrationhighfidelityentangledlogical,
      title={Demonstration of High-Fidelity Entangled Logical Qubits using Transmons}, 
      author={Arian Vezvaee and Vinay Tripathi and Mario Morford-Oberst and Friederike Butt and Victor Kasatkin and Daniel A. Lidar},
      year={2025},
      eprint={2503.14472},
      archivePrefix={arXiv},
      primaryClass={quant-ph},
      url={https://arxiv.org/abs/2503.14472}, 
}

@article{mori2023floquet,
  title={Floquet states in open quantum systems},
  author={Mori, Takashi},
  journal={Annual Review of Condensed Matter Physics},
  volume={14},
  number={1},
  pages={35--56},
  year={2023},
  publisher={Annual Reviews}
}

@article{iserles2002expansions,
author = {Iserles, Arieh},
year = {2002},
month = {02},
pages = {},
title = {Expansions That Grow on Trees},
volume = {49},
journal = {Not. Am. Math. Soc.}
}

@article{mostafaie2020systematic,
  title={A systematic study on meta-heuristic approaches for solving the graph coloring problem},
  author={Mostafaie, Taha and Khiyabani, Farzin Modarres and Navimipour, Nima Jafari},
  journal={Computers \& Operations Research},
  volume={120},
  pages={104850},
  year={2020},
  publisher={Elsevier}
}

@article{bose1960class,
  title={On a class of error correcting binary group codes},
  author={Bose, Raj Chandra and Ray-Chaudhuri, Dwijendra K},
  journal={Information and control},
  volume={3},
  number={1},
  pages={68--79},
  year={1960},
  publisher={Elsevier}
}

@article{gorenstein1961class,
  title={A class of error-correcting codes in p\^{}m symbols},
  author={Gorenstein, Daniel and Zierler, Neal},
  journal={Journal of the Society for Industrial and Applied Mathematics},
  volume={9},
  number={2},
  pages={207--214},
  year={1961},
  publisher={SIAM}
}

@article{read2025platonic,
  title={Platonic dynamical decoupling sequences for interacting spin systems},
  author={Read, Colin and Serrano-Ens{\'a}stiga, Eduardo and Martin, John},
  journal={Quantum},
  volume={9},
  pages={1661},
  year={2025},
  publisher={Verein zur F{\"o}rderung des Open Access Publizierens in den Quantenwissenschaften}
}

@article{Genov2017,
  title = {Arbitrarily Accurate Pulse Sequences for Robust Dynamical Decoupling},
  author = {Genov, Genko T. and Schraft, Daniel and Vitanov, Nikolay V. and Halfmann, Thomas},
  journal = {Phys. Rev. Lett.},
  volume = {118},
  issue = {13},
  pages = {133202},
  numpages = {5},
  year = {2017},
  month = {Mar},
  publisher = {American Physical Society},
  doi = {10.1103/PhysRevLett.118.133202},
  url = {https://link.aps.org/doi/10.1103/PhysRevLett.118.133202}
}

@article{GULLION1990479,
title = {New, compensated Carr-Purcell sequences},
journal = {Journal of Magnetic Resonance (1969)},
volume = {89},
number = {3},
pages = {479-484},
year = {1990},
issn = {0022-2364},
doi = {https://doi.org/10.1016/0022-2364(90)90331-3},
url = {https://www.sciencedirect.com/science/article/pii/0022236490903313},
author = {Terry Gullion and David B Baker and Mark S Conradi}
}

@article{Wang2012,
  title = {Effect of pulse error accumulation on dynamical decoupling of the electron spins of phosphorus donors in silicon},
  author = {Wang, Zhi-Hui and Zhang, Wenxian and Tyryshkin, A. M. and Lyon, S. A. and Ager, J. W. and Haller, E. E. and Dobrovitski, V. V.},
  journal = {Phys. Rev. B},
  volume = {85},
  issue = {8},
  pages = {085206},
  numpages = {12},
  year = {2012},
  month = {Feb},
  publisher = {American Physical Society},
  doi = {10.1103/PhysRevB.85.085206},
  url = {https://link.aps.org/doi/10.1103/PhysRevB.85.085206}
}

@article{ZhiHui2012,
  title = {Comparison of dynamical decoupling protocols for a nitrogen-vacancy center in diamond},
  author = {Wang, Zhi-Hui and de Lange, G. and Rist\`e, D. and Hanson, R. and Dobrovitski, V. V.},
  journal = {Phys. Rev. B},
  volume = {85},
  issue = {15},
  pages = {155204},
  numpages = {15},
  year = {2012},
  month = {Apr},
  publisher = {American Physical Society},
  doi = {10.1103/PhysRevB.85.155204},
  url = {https://link.aps.org/doi/10.1103/PhysRevB.85.155204}
}

@misc{fariña2025siteresolvedmagnontriplondynamics,
      title={Site-resolved magnon and triplon dynamics on a programmable quantum dot spin ladder}, 
      author={Pablo Cova Fariña and Daniel Jirovec and Xin Zhang and Elizaveta Morozova and Stefan D. Oosterhout and Stefano Reale and Tzu-Kan Hsiao and Giordano Scappucci and Menno Veldhorst and Lieven M. K. Vandersypen},
      year={2025},
      eprint={2506.08663},
      archivePrefix={arXiv},
      url={https://arxiv.org/abs/2506.08663}, 
}

@article{Hsiao,
  title = {Exciton Transport in a Germanium Quantum Dot Ladder},
  author = {Hsiao, T.-K. and Cova Fari\~na, P. and Oosterhout, S. D. and Jirovec, D. and Zhang, X. and van Diepen, C. J. and Lawrie, W. I. L. and Wang, C.-A. and Sammak, A. and Scappucci, G. and Veldhorst, M. and Demler, E. and Vandersypen, L. M. K.},
  journal = {Phys. Rev. X},
  volume = {14},
  issue = {1},
  pages = {011048},
  numpages = {17},
  year = {2024},
  month = {Mar},
  publisher = {American Physical Society},
  doi = {10.1103/PhysRevX.14.011048},
  url = {https://link.aps.org/doi/10.1103/PhysRevX.14.011048}
}

@article{berke2022transmon,
  title={Transmon platform for quantum computing challenged by chaotic fluctuations},
  author={Berke, Christoph and Varvelis, Evangelos and Trebst, Simon and Altland, Alexander and DiVincenzo, David P},
  journal={Nature communications},
  volume={13},
  number={1},
  pages={2495},
  year={2022},
  publisher={Nature Publishing Group UK London}
}

@article{abbe2020reed,
  title={Reed--Muller codes: Theory and algorithms},
  author={Abbe, Emmanuel and Shpilka, Amir and Ye, Min},
  journal={IEEE Transactions on Information Theory},
  volume={67},
  number={6},
  pages={3251--3277},
  year={2020},
  publisher={IEEE}
}

@article{Acharya2025,
  title={Quantum error correction below the surface code threshold},
  author={{Google Quantum AI et al.}},
  journal={Nature},
  volume={638},
  number={8052},
  pages={920--926},
  year={2025},
  publisher={Nature Publishing Group UK London}
}

@misc{ibm_heavy_hex_2021,
  author       = {Paul Nation and Hanhee Paik and Andrew W. Cross and Zaira Nazario},
  title        = {The {IBM} Quantum Heavy Hex Lattice},
  year         = {2021},
  month        = {July},
  howpublished = {\url{https://www.ibm.com/quantum/blog/heavy-hex-lattice}},
}

@book{conway2013sphere,
  title={Sphere packings, lattices and groups},
  author={Conway, John Horton and Sloane, Neil James Alexander},
  year={2013},
  note={Vol.~290, Springer Science \& Business Media}
}

@article{paz2013optimally,
  title={Optimally combining dynamical decoupling and quantum error correction},
  author={Paz-Silva, Gerardo A and Lidar, Daniel A},
  journal={Scientific reports},
  volume={3},
  number={1},
  pages={1530},
  year={2013},
  publisher={Nature Publishing Group UK London}
}

@article{Parra-Rodriguez2020,
  title = {Digital-analog quantum computation},
  author = {Parra-Rodriguez, Adrian and Lougovski, Pavel and Lamata, Lucas and Solano, Enrique and Sanz, Mikel},
  journal = {Phys. Rev. A},
  volume = {101},
  issue = {2},
  pages = {022305},
  numpages = {12},
  year = {2020},
  month = {Feb},
  publisher = {American Physical Society},
  doi = {10.1103/PhysRevA.101.022305},
  url = {https://link.aps.org/doi/10.1103/PhysRevA.101.022305}
}

@article{Casas_2007,
doi = {10.1088/1751-8113/40/50/006},
url = {https://dx.doi.org/10.1088/1751-8113/40/50/006},
year = {2007},
month = {nov},
publisher = {},
volume = {40},
number = {50},
pages = {15001},
author = {Casas, Fernando},
title = {Sufficient conditions for the convergence of the Magnus expansion},
journal = {Journal of Physics A: Mathematical and Theoretical}
}

@article{Coote2025,
  title = {Resource-Efficient Context-Aware Dynamical Decoupling Embedding for Arbitrary Large-Scale Quantum Algorithms},
  author = {Coote, Paul and Dimov, Roman and Maity, Smarak and Hartnett, Gavin S. and Biercuk, Michael J. and Baum, Yuval},
  journal = {PRX Quantum},
  volume = {6},
  issue = {1},
  pages = {010332},
  numpages = {13},
  year = {2025},
  month = {Feb},
  publisher = {American Physical Society},
  doi = {10.1103/PRXQuantum.6.010332},
  url = {https://link.aps.org/doi/10.1103/PRXQuantum.6.010332}
}

@misc{read2025factor,
      title={Dynamical decoupling of interacting spins through group factorization}, 
      author={Colin Read and Eduardo Serrano-Ensástiga and John Martin},
      year={2025},
      eprint={2506.15382},
      archivePrefix={arXiv},
      primaryClass={quant-ph},
      url={https://arxiv.org/abs/2506.15382}, 
}

@article{brinkmann2016introduction,
  title={Introduction to average Hamiltonian theory. I. Basics},
  author={Brinkmann, Andreas},
  journal={Concepts in Magnetic Resonance Part A},
  volume={45},
  number={6},
  pages={e21414},
  year={2016},
  publisher={Wiley Online Library}
}

@article{BRINKMANN2025100191,
title = {Introduction to average Hamiltonian theory. II. Advanced examples},
journal = {Journal of Magnetic Resonance Open},
volume = {23},
pages = {100191},
year = {2025},
issn = {2666-4410},
doi = {https://doi.org/10.1016/j.jmro.2025.100191},
url = {https://www.sciencedirect.com/science/article/pii/S266644102500007X},
author = {Andreas Brinkmann}
}

@article{Childs2021,
  title = {Theory of Trotter Error with Commutator Scaling},
  author = {Childs, Andrew M. and Su, Yuan and Tran, Minh C. and Wiebe, Nathan and Zhu, Shuchen},
  journal = {Phys. Rev. X},
  volume = {11},
  issue = {1},
  pages = {011020},
  numpages = {49},
  year = {2021},
  month = {Feb},
  publisher = {American Physical Society},
  doi = {10.1103/PhysRevX.11.011020},
  url = {https://link.aps.org/doi/10.1103/PhysRevX.11.011020}
}

@misc{zhang2025learningsteerquantummanybody,
      title={Learning to steer quantum many-body dynamics with tree optimization}, 
      author={Jixing Zhang and Bo Peng and Yang Wang and Cheuk Kit Cheung and Guodong Bian and Andrew M. Edmonds and Matthew Markham and Zhe Zhao and Durga Bhaktavatsala Rao Dasari and Ruoming Peng and Ye Wei and Jörg Wrachtrup},
      year={2025},
      eprint={2510.07802},
      archivePrefix={arXiv},
      primaryClass={quant-ph},
      url={https://arxiv.org/abs/2510.07802}, 
}

@misc{hickman2025crosstalkrobustdynamicaldecouplingbipartitetopology,
      title={Crosstalk-Robust Dynamical Decoupling for Bipartite-Topology Quantum Processors}, 
      author={Ethan Hickman and Xiaodi Wu and Gregory Quiroz},
      year={2025},
      eprint={2506.18010},
      archivePrefix={arXiv},
      primaryClass={quant-ph},
      url={https://arxiv.org/abs/2506.18010}, 
}

@article{hayes2014programmable,
  title={Programmable quantum simulation by dynamic Hamiltonian engineering},
  author={Hayes, David and Flammia, Steven T and Biercuk, Michael J},
  journal={New Journal of Physics},
  volume={16},
  number={8},
  pages={083027},
  year={2014},
  publisher={IOP Publishing}
}

@article{o2019hamiltonian,
  title={Hamiltonian engineering with constrained optimization for quantum sensing and control},
  author={O’Keeffe, Michael F and Horesh, Lior and Barry, John F and Braje, Danielle A and Chuang, Isaac L},
  journal={New Journal of Physics},
  volume={21},
  number={2},
  pages={023015},
  year={2019},
  publisher={IOP Publishing}
}

@article{rajabi2019dynamical,
  title={Dynamical Hamiltonian engineering of 2D rectangular lattices in a one-dimensional ion chain},
  author={Rajabi, Fereshteh and Motlakunta, Sainath and Shih, Chung-You and Kotibhaskar, Nikhil and Quraishi, Qudsia and Ajoy, Ashok and Islam, Rajibul},
  journal={npj Quantum Information},
  volume={5},
  number={1},
  pages={32},
  year={2019},
  publisher={Nature Publishing Group UK London}
}

@article{votto2024universal,
  title={Universal quantum processors in spin systems via robust local pulse sequences},
  author={Votto, Matteo and Zeiher, Johannes and Vermersch, Beno{\^\i}t},
  journal={Quantum},
  volume={8},
  pages={1513},
  year={2024},
  publisher={Verein zur F{\"o}rderung des Open Access Publizierens in den Quantenwissenschaften}
}

@article{Khodjasteh2009,
  title = {Dynamically Error-Corrected Gates for Universal Quantum Computation},
  author = {Khodjasteh, Kaveh and Viola, Lorenza},
  journal = {Phys. Rev. Lett.},
  volume = {102},
  issue = {8},
  pages = {080501},
  numpages = {4},
  year = {2009},
  month = {Feb},
  publisher = {American Physical Society},
  doi = {10.1103/PhysRevLett.102.080501},
  url = {https://link.aps.org/doi/10.1103/PhysRevLett.102.080501}
}

@book{nielsen2010quantum,
  title={Quantum computation and quantum information},
  author={Nielsen, Michael A and Chuang, Isaac L},
  year={2010},
  publisher={Cambridge university press}
}

\end{document}